%% file: evolution.tex
\documentclass{aa}  

\usepackage{graphicx}
\usepackage{txfonts}
\usepackage{multirow}
\usepackage{xcolor}
\def\Prot{P_{\mathrm{rot}}}
\def\ProtJ{P_{\mathrm{rot,150}}}
\def\ProtJJ{P_{\mathrm{rot,600}}}
\def\Lx{L_{\mathrm{X}}}
\def\Rx{R_{\mathrm{X}}}
\def\Madot{\dot{M}_{\mathrm{atm}}}
\def\fas{f_{\mathrm{atm}}^{\mathrm{start}}}

\begin{document}

   \title{Constraining stellar rotation and planetary atmospheric evolution of a dozen systems hosting sub-Neptunes and super-Earths}

%   \subtitle{}
   \titlerunning{Constraining stellar rotation and planetary atmospheric evolution}
   \authorrunning{Bonfanti et al.}

   \author{A. Bonfanti\inst{1}
          \and
          L. Fossati\inst{1}
          \and
          D. Kubyshkina\inst{2,1}
          \and
          P. E. Cubillos\inst{1}
          }

   \institute{Space Research Institute, Austrian Academy of Sciences, 
              Schmiedlstrasse 6, A-8042 Graz, Austria \\
              \email{andrea.bonfanti@oeaw.ac.at}
         \and
             School of Physics, Trinity College Dublin, the University of Dublin, College Green, Dublin-2, Ireland
             }

   \date{}

% \abstract{}{}{}{}{} 
% 5 {} token are mandatory
 
  \abstract
   {Planetary atmospheric evolution modelling is a prime tool for understanding the observed exoplanet population and constraining formation and migration mechanisms, but it can also be used to study the evolution of the activity level of planet hosts.}
   {We constrain the planetary atmospheric mass fraction at the time of the dispersal of the protoplanetary disk and the evolution of the stellar rotation rate for a dozen multi-planet systems that host sub-Neptunes and/or super-Earths.}
   {We employ a custom-developed \textsc{Python} code that we have dubbed \textsc{Pasta} (\textit{P}lanetary \textit{A}tmospheres and \textit{S}tellar Ro\textit{T}ation R\textit{A}tes), which runs within a Bayesian framework to model the atmospheric evolution of exoplanets. The code combines MESA stellar evolutionary tracks, a model describing planetary structures, a model relating stellar rotation and activity level, and a model predicting planetary atmospheric mass-loss rates based on the results of hydrodynamic simulations.}
   {Through a Markov chain Monte Carlo scheme, we retrieved the posterior probability density functions of all considered parameters. For ages older than about 2 Gyr, we find a median spin-down (i.e. $P(t)\propto t^y$) of $\bar{y}=0.38_{-0.27}^{+0.38}$, indicating a rotation decay slightly slower than classical literature values ($\approx$0.5), though still within $1\sigma$. At younger ages, we find a median spin-down (i.e. $P(t)\propto t^x$) of $\bar{x}=0.26_{-0.19}^{+0.42}$, which is below what is observed in young open clusters, though within $1\sigma$. Furthermore, we find that the $x$ probability distribution we derived is skewed towards lower spin-down rates. However, these two results are likely due to a selection bias as the systems suitable to be analysed by \textsc{Pasta} contain at least one planet with a hydrogen-dominated atmosphere, implying that the host star has more likely evolved as a slow rotator. We further look for correlations between the initial atmospheric mass fraction of the considered planets and system parameters (i.e. semi-major axis, stellar mass, and planetary mass) that would constrain planetary atmospheric accretion models, but without finding any.}
   {\textsc{Pasta} has the potential to provide constraints to planetary atmospheric accretion models, particularly when considering warm sub-Neptunes that are less susceptible to mass loss compared to hotter and/or lower-mass planets. The TESS, CHEOPS, and PLATO missions are going to be instrumental in identifying and precisely measuring systems amenable to \textsc{Pasta}'s analysis and can thus potentially constrain planet formation and stellar evolution.}


   \maketitle
%
%-------------------------------------------------------------------
%
\section{Introduction}
A variety of planet-finding facilities have led to the detection of more than 4000 exoplanets and an almost equal amount of candidates\footnote{{\tt https://exoplanetarchive.ipac.caltech.edu/}}. Such a large number of planets has allowed us to form a picture of the galactic planet population and of the underlying structures, which are believed to be caused by a combination of formation, migration, and atmospheric evolution processes. For example, the so-called sub-Jovian desert \citep[e.g.][]{davis09,szabo11,mazeh16} and sub-Neptune radius gap \citep[e.g.][]{fulton17,fulton18,vanEylen18,macdonald2019} are among the most noticeable structures in the observed exoplanet population.

The radius gap and the lower edge of the sub-Jovian desert are currently believed to be the consequence of planetary atmospheric evolution, and in particular of atmospheric escape, though in case of the radius gap the nature of the main escape-driving mechanism is still unclear \citep[e.g.][]{owen17,owen18,jin18,ginzburg2018,gupta2019,gupta2020,loyd20,sandoval21}. In general, it is believed that atmospheric escape plays a pivotal role in shaping the long-term evolution of planetary atmospheres. Therefore, adequately tracking planetary atmospheric evolution gives us the possibility to further understand the observed planet population as well as gather critical information about planet formation \citep[e.g.][]{jin14,kubyshkina19AAKepler11}.

Atmospheric escape is particularly strong when planets are young and is driven by the high-energy -- X-ray and extreme UV (EUV; together XUV) -- emission of the host star \citep[blow-off; e.g.][]{watson81,lammer03} and/or by the internal atmospheric energy and low planetary gravity \citep[boil-off; e.g.][]{stokl2015,ginzburg2016,owen17,fossati17}. For computational reasons, in atmospheric evolution calculations, escape is computed employing analytical formulas \citep[e.g. the energy-limited approximation;][]{watson81,erkaev07} that, however, are unreliable, particularly in the regime that most matters for atmospheric evolution \citep[e.g.][]{kubyshkina18AAgrids,kubyshkina18ApJHBA,krenn2021}. To circumvent this problem, \citet{kubyshkina18ApJHBA} present an analytical expression that enables mass-loss rates to be computed for a variety of planets on the basis of the large grid of hydrodynamic upper atmosphere models published by \citet{kubyshkina18AAgrids}, thus enabling both boil-off and blow-off to be simultaneously accounted for. 

Estimating the evolution of a planetary atmosphere is further complicated by the fact that the evolution of the rotation rate, and thus of the high-energy emission, of late-type stars is not unique. As a matter of fact, stellar rotation rate and XUV emission are tightly bound, with faster rotating stars being XUV brighter \citep[e.g.][]{pallavicini81,pizzolato2003,johnstone15Prot150}. Along the main sequence, both rotation rate and XUV emission decrease with time, but their evolutionary path is not unique: After the saturation regime, stars born with the same properties (i.e. mass and metallicity) can have widely different initial rotation rates and XUV fluxes \citep[e.g.][]{mamajek08,denissenkov10,johnstone15Prot150,johnstone21}. The evolutionary paths of both rotation rate and XUV emission merge at an age of about 2\,Gyr \citep[e.g.][]{tu15}, meaning that for older stars it is not possible to infer the past rotation rate and XUV emission from observations alone. To overcome this problem, \citet{kubyshkina19ApJ,kubyshkina19AAKepler11} developed an algorithm that enables the evolution of a hydrogen-dominated planetary atmosphere to be tracked accounting for atmospheric escape using mass-loss rates extracted from the grid of hydrodynamic models from \citet{kubyshkina18AAgrids}. The algorithm employs a Bayesian framework and the currently observed system parameters to simultaneously constrain the past evolution of the planetary atmosphere and of the stellar rotation rate, and hence XUV emission. In practice, the algorithm returns a posterior probability density function (PDF) for the planetary initial atmospheric mass fraction and for the rotation rate of the host star at any desired age.

In its original implementation the algorithm had been applied to the HD3167, K2-32, Kepler-11, and $\nu^2$\,Lup systems, constraining their past evolution \citep{kubyshkina19ApJ,kubyshkina19AAKepler11,delrez2021}. However, the code relied on conversion parameters (e.g. between the stellar rotation rate and X-ray emission, and between X-ray and EUV flux) taken from the literature, which suffer from significant uncertainties. We present here a major update to the code, which we have dubbed \textsc{Pasta} (\textit{P}lanetary \textit{A}tmospheres and \textit{S}tellar Ro\textit{T}ation R\textit{A}tes), where we also treat these conversion parameters as free parameters and refine the gyrochronological relation. We then apply the algorithm to a large number of systems for which the masses and radii for at least one planet are known to better than 25\% and 8\%, respectively. Our code is also able to deal with multi-planet systems, which are the best targets as the multiple fitting points provided by the different planets orbiting the same star increase the constraining power of the algorithm. 

We aim at studying the stellar spin-down after carefully checking the consistency of our adopted framework with the multiple and diverse constraints coming from observations, theory, and literature. We also search for differences between the distribution of rotation rates of young planet hosts with that of young open cluster (OC) stars of comparable mass and age that may hint at the presence of star-planet interactions (SPIs) early on in the evolution of planetary systems. Finally, we look for correlations between the derived planetary initial atmospheric mass fractions and system parameters (e.g. initial atmospheric mass fraction vs. either planetary or stellar mass and semi-major axis) that would enable planetary accretion models, and thus planet formation to be empirically constrained \citep[e.g.][]{lozovsky2021}.

This paper is organised as follows. Section~\ref{sec:algorithm} describes the whole framework of the \textsc{Pasta} code, while Sect.~\ref{sec:sample} presents the sample of exoplanetary systems analysed in this work. Our results are outlined in Sect.~\ref{sec:results} and then thoroughly discussed in Sect.~\ref{sec:discussion}. Finally, our conclusions are drawn in Sect.~\ref{sec:conclusions}.

%--------------------------------------------------------------------

%
\section{The \textsc{Pasta} algorithm}\label{sec:algorithm}
We modelled the evolutionary history of both planetary atmospheres and stellar rotation rates within a Bayesian framework following a significant upgrade of \textsc{Pasta}, first presented by \citet{kubyshkina19ApJ,kubyshkina19AAKepler11}. Below, we thoroughly describe the entire algorithm and the implemented upgrades.

\subsection{Stellar and planetary models}\label{sec:models}

Modelling the evolution of an exoplanetary system requires the use of several different modelling tools and results. The various ingredients are stellar evolutionary tracks, a model of the stellar XUV flux evolution, a planetary structure model relating planetary parameters and atmospheric mass, and a model computing planetary atmospheric escape.

The stellar evolutionary tracks are those present in the MESA Isochrones and Stellar Tracks \citep[MIST;][]{choi16} grid, which has been computed employing the MESA stellar evolutionary code. Within the MIST grid, we consider tracks computed for solar metallicity stars with a mass ranging between 0.4 and 1.3 $M_{\odot}$, with steps of 0.05 $M_{\odot}$ up to 0.9 $M_{\odot}$ and steps of 0.02 $M_{\odot}$ for higher-mass stars. Each evolutionary track lists the theoretically expected values of the stellar bolometric luminosity $L_{\mathrm{bol}}$, effective temperature $T_{\mathrm{eff}}$, and radius $R_{\star}$ at any given epoch $t$. These stellar parameters are fundamental to track the equilibrium temperature of the hosted exoplanets over time (Sect.~\ref{sec:framework}).

The planetary structure model relating planetary mass $M_{\rm p}$, radius $R_{\rm p}$, equilibrium temperature $T_{\mathrm{eq}}$, and atmospheric mass $M_{\mathrm{atm}}$ is that presented by \citet{johnstone15models}. In this work, we employ the model grid described by \citet{kubyshkina18AAgrids}, which spans the 1 to 40 $M_{\oplus}$ range and thus covering super-Earths and sub-Neptunes. 

Variations in $M_{\mathrm{atm}}$ over time are computed considering the atmospheric mass-loss rate $\Madot$, which depends on the stellar XUV emission and planetary properties. \textsc{Pasta} uses the stellar rotation period $\Prot$ as proxy for the stellar high-energy flux; therefore, it is essential to reconstruct the evolution of $\Prot$ with time $t$: $\Prot (t)$, where the variable of temporal evolution $t$ may span the range from the initial reference epoch (e.g. the time of proto-planetary disk dispersal, $t_{\mathrm{disk}}$) up to the stellar age $t_{\star}$. For young stars, that is stars with an age $t_{\star}<2$ Gyr, we modelled $\Prot (t)$ as a power law of the form
\begin{equation}
\Prot (t)=P_{\mathrm{rot,}\star} \left (\frac{t}{t_{\star}} \right)^x \quad\quad t_{\star}<2\: \mathrm{Gyr}
\label{eq:ProtYoung}
\end{equation}
where $P_{\mathrm{rot,}\star}$ is the present-day rotation period of the star, while $x$ is a free parameter varying within the [0,2] range to account for the different rotation rates observed for young late-type stars \citep[see e.g.][Fig. 2]{tu15}. 

Instead, for stars older than 2 Gyr, the picture is more complicated. Despite several gyrochronological relations being available in the literature \citep[e.g.][]{barnes03,mamajek08,collierCameron09,barnes10}, all the period-age relations are calibrated upon OCs, whose age has been previously estimated through a global isochrone fitting as all cluster members are coeval. However, there is a lack of OCs for calibrating gyrochronological laws in the intermediate- and old-age domain and quite some doubts have been raised when applying gyrochronology to field stars. For example, \citet{barnes09}, \citet{brown14}, and \citet{kovacs15} found that for field stars isochronal ages are considerably greater than gyrochronological ages. In particular, \citet{kovacs15} stresses that field stars appear to have significantly lower slow-down rates compared to their cluster counterparts and \citet{vanSaders16} shows that fast rotators are unexpectedly found among stars more evolved than the Sun. A complete review of the reliable domains of application of classical gyrochronological relations is beyond the scope of this work, but the above considerations brought us to adopt a $\Prot (t)$ relation, which assume the following form when a star is older than 2 Gyr (ages in Gyr):
\begin{equation}
\Prot (t)=
\begin{cases}
P_{\mathrm{rot,}\star}\left(\frac{t}{2} \right)^x \left(\frac{2}{t_{\star}} \right)^y & t<2\: \mathrm{Gyr} \\
P_{\mathrm{rot,}\star}\left(\frac{t}{t_{\star}} \right)^y & t\ge2\: \mathrm{Gyr} \\
\end{cases}
\quad t_{\star}\ge 2\: \mathrm{Gyr,}
\label{eq:ProtOld}
\end{equation}
where the broken power-law enables us to differentiate between two evolutionary regimes, whose break is set at $t=2$ Gyr. In fact, stars even of the same spectral type may have different $\Prot$ values at $t=t_{\mathrm{disk}}$ and hence evolve along different evolutionary rotation tracks, which then tend to converge towards similar $\Prot$ values at $t$\,$\approx$\,2 Gyr, where $\Prot (t=2\, \mathrm{Gyr})$ depends on stellar mass. As a star may be characterised by highly different spin-down rates in the first part of its life ($t<2$ Gyr), the first regime of Eq. (\ref{eq:ProtOld}) is governed by the $x$ exponent spanning the [0, 2] range as it happens for Eq. (\ref{eq:ProtYoung}). Instead, the factor $\left(\frac{2}{t_{\star}} \right)^y$ allows the function to be continuous, as there is not dependence on the variable $t$. The following stellar rotation evolution ($t>2$ Gyr) is generally quieter; thus, we introduced an additional $y$ exponent, which varies within the [0.01,1] range, to model the stellar spin-down rate.
The value of the $y$ exponent is centred around $\frac{1}{2}$, which is the value adopted by the classical \citeauthor{skumanich72}-law \citep{skumanich72}. Also, we remark that this value is very close to 0.566, which is the value proposed by \citet{mamajek08} in their power-law gyrochronological relation. Equation~(\ref{eq:ProtOld}) is set up so that $\Prot (t=t_{\star}) = P_{\mathrm{rot,}\star}$ and it guarantees continuity of the function at $t=2$ Gyr. To further ensure that $\Prot (t)$ is differentiable at $t=2$ Gyr, within the range $t \in [1.5,t_{\mathrm{lim}}]$, where $2\le t_{\mathrm{lim}}=\min{\{2.5, t_{\star}\}}$, we replaced the broken power-law with a $3^{\mathrm{rd}}$-degree single spline fitted onto the power-law.

The $\Prot (t)$ function does not explicitly include tidal effects that may be induced by the planets on the host star. However the flat and broad priors on the gyrochronological exponent allow us to span between slow and fast spin-down rates. The occurring of SPI is mentioned as a possible scenario in Sect.~\ref{sec:rotperiods}.

Given $\Prot$ at any given time, the X-ray stellar luminosity $\Lx$ ($\lambda=5$-$100$ \AA) can be estimated using scaling relations. Indeed, $\Lx$, which traces stellar magnetic activity, can be expressed as a function of Rossby number
\begin{equation}
\mathrm{Ro}=\frac{\Prot}{\tau}\,,
\label{eq:Rossby}
\end{equation}
which probes the efficiency of the stellar magnetic dynamo \citep[see e.g.][]{noyes84}. $\tau$ is the convective turn-over time, which is mass-dependent and we expressed it as \citep{wright11}
\begin{equation}
\log{\tau} = 1.16 - 1.49\log{\frac{M}{M_{\odot}}} - 0.54\log^2{\frac{M}{M_{\odot}}}\,.
\label{eq:tau}
\end{equation}

For increasing stellar rotational velocities, $\Lx$ increases monotonically \citep{pallavicini81} following different mass-dependent tracks \citep{mcdonald19}. This behaviour breaks down for very fast rotators \citep[e.g.][]{micela85}, so that below a given threshold value of $\Prot$ (hence of Ro, say $\mathrm{Ro_{sat}}$) $\Lx$ saturates, that is $\Lx$ becomes $\mathrm{Ro}$-independent, while keeping the dependence on $M_{\star}$: $\Lx (\mathrm{Ro}<\mathrm{Ro_{sat}})=L_{\mathrm{X,sat}}(M_{\star})$. Therefore, following \citet{wright11}, we modelled the ratio between $\Lx$ and the stellar bolometric luminosity $L_{\mathrm{bol}}$ as
\begin{equation}
\Rx\equiv\frac{\Lx}{L_{\mathrm{bol}}}=
\begin{cases}
R_{\mathrm{X,sat}} & \mathrm{Ro}\le \mathrm{Ro_{sat}} \\
C \mathrm{Ro}^{\beta} & \mathrm{Ro}> \mathrm{Ro_{sat}} \\
\end{cases}
\label{eq:Lx}
\end{equation}
where $\mathrm{Ro_{\mathrm{sat}}}=0.13$ is the saturation threshold of Ro as quantified by \citet{wright11}, while $\beta=-2.70\pm0.13$ has been estimated by \citet{wright11} from an unbiased sample of late-type stars for which both $\Prot$ and $\Lx$ had been measured. As $R_{\mathrm{X,sat}}\equiv L_{\mathrm{X,sat}}/L_{\mathrm{bol}}$ depends on stellar mass, rather than fixing it to the best-fit value as in \citet{wright11}, we evaluated $R_{\mathrm{X,sat}}$ for different mass ranges from \citet[][Fig. 1]{mcdonald19}, adopting the values that are reported in Table \ref{tab:RxSat}. For each mass range, the $C$ coefficient in Eq.~(\ref{eq:Lx}) is then computed accordingly to guarantee the continuity of $\Rx$ at $\mathrm{Ro}=\mathrm{Ro_{sat}}$. For reference, the $C$ values for the average $\beta=-2.70$ are reported in Table~\ref{tab:RxSat}, but we remark that the default behaviour of our tool is that $\beta$ varies according to a Gaussian prior whose $\sigma=0.13$. Therefore, $C$ depends also on $\beta$.
%=====================================
\begin{table}
\caption{Saturation values ($R_{\mathrm{X,sat}}$) adopted in our work for different mass ranges, as inferred from \citet{mcdonald19}. The last column lists the values of the $C$ coefficient in Eq.~(\ref{eq:Lx}) for the average $\beta=-2.70$, $C_{\beta=-2.70}$.}
\label{tab:RxSat}
\centering
\begin{tabular}{c r r}
\hline\hline
Mass range [$M_{\odot}$] & \multicolumn{1}{c}{$R_{\mathrm{X,sat}}$} & \multicolumn{1}{c}{$C_{\beta=-2.70}$} \\
\hline
 $M<0.687$              & $7.4\times10^{-4}$ & $3.00\times10^{-6}$ \\
 $0.687\le M < 0.89$    & $4.5\times10^{-4}$ & $1.82\times10^{-6}$ \\
 $0.89\le M < 1.12$     & $2\times10^{-4}$   & $8.10\times10^{-7}$ \\
 $M\ge 1.12$            & $5.5\times10^{-5}$ & $2.23\times10^{-7}$ \\
\hline
\end{tabular}
\end{table}
%=====================================

From $L_X$, we infer the EUV ($\lambda=100$-$920$ \AA) stellar luminosity ($L_{\mathrm{EUV}}$) via the relation reported by \citet{sanzForcada11}
\begin{equation}
\begin{split}
\log{L_{\mathrm{EUV}}} & = (q_{\mathrm{SF}}\pm \sigma_{q}) + (m_{\mathrm{SF}}\pm \sigma_{m})\log{L_X} \\
                       & = (4.80\pm1.99) + (0.860\pm0.073)\log{L_X}\,,
\end{split}
\label{eq:L_EUV}
\end{equation}
where both X-ray and EUV luminosities are expressed in erg/s. The term $L_{\mathrm{EUV}}$ is the key-ingredient necessary to finally compute $\Madot$. To this end, we employed the hydro-based approximation (HBA) presented by \citet{kubyshkina18ApJHBA}, which is an analytical formulation of $\Madot$ extracted by fitting the results of the large grid of upper planetary atmosphere hydrodynamic models presented by \citet{kubyshkina18AAgrids}. The HBA speeds up decisively the computation of $\Madot$ compared to using hydrodynamic simulations and, at the same time, it removes the physical limitations of the energy-limited approximations \citep[e.g.][]{krenn2021}. In detail, \citet{kubyshkina18ApJHBA} computes $\Madot$ following two different functional behaviours, which depend on the restricted Jeans escape parameter of a planet \citep{jeans25,fossati17}
\begin{equation}
\Lambda=\frac{G M_{\rm p} m_{\rm H}}{R_{\rm p} k_{\rm B} T_{\mathrm{eq}}}\,,
\label{eq:Jeans}
\end{equation}
where $G$ is the universal gravitational constant, $k_B$ is the Boltzmann constant, and $m_H$ is the hydrogen mass. In particular, \citet{kubyshkina18ApJHBA} express the atmospheric mass-loss rate (in g/s) of a given planet as a function of semi-major axis $a$ and $R_{\rm p}$ as
\begin{equation}
\Madot = e^{\alpha_0} \left(\frac{F_{\mathrm{EUV}}}{\mathrm{erg\, cm^{-2}\, s^{-1}}} \right)^{\alpha_1}\left( \frac{a}{\mathrm{AU}} \right)^{\alpha_2} \left(\frac{R_{\rm p}}{R_{\oplus}}\right)^{\alpha_3}\Lambda^K\,,
\label{eq:Madot}
\end{equation}
where $e$ is Euler's number and $F_{\mathrm{EUV}}=\frac{L_{\mathrm{EUV}}}{4\pi a^2}$ is the EUV stellar flux at the distance of the planet, while the exponent $K$ is given by
\begin{equation}
\ln{K} = \zeta + \theta\ln{\frac{a}{\mathrm{AU}}}\,.
\end{equation}
The values of the exponents $\alpha_0$, $\alpha_1$, $\alpha_2$, $\alpha_3$, and $K$ (through $\zeta$ and $\theta$) are listed in Table \ref{tab:exponents} for convenience and vary according to whether $\ln{\Lambda}$ is smaller or larger than the threshold value $e^{\Sigma}$, where
\begin{equation}
  \Sigma=\frac{15.611 -0.578\ln{\frac{F_{\mathrm{EUV}}}{\mathrm{erg\, cm^{-2}\, s^{-1}}}} + 1.537\ln{\frac{a}{\mathrm{AU}}} + 1.018\ln{\frac{R_p}{R_{\oplus}}} }{5.564 + 0.894\ln{\frac{a}{\mathrm{AU}}}}.
  \label{eq:Sigma}
\end{equation}
%
%=====================================
\begin{table*}
\caption{Coefficients of Equation~(\ref{eq:Madot}) taken from \citet[][Table 1]{kubyshkina18ApJHBA}.}
\label{tab:exponents}
\centering
\begin{tabular}{l c c c c c c}
\hline\hline
 & $\alpha_0$ & $\alpha_1$ & $\alpha_2$ & $\alpha_3$ & $\zeta$ & $\theta$ \\
\hline
 $\Lambda < e^\Sigma$   & 32.0199   & 0.4222    & $-1.7489$ & 3.7679    & $-6.8618$ & 0.0095 \\
 $\Lambda \ge e^\Sigma$   & 16.4084   & 1.0000    & $-3.2861$ & 2.7500    & $-1.2978$ & 0.8846 \\
\hline
\end{tabular}
\end{table*}
%=====================================

Equation (\ref{eq:Madot}) enables the evolution of the atmospheric mass with time to then computed. 

\subsection{The statistical framework}\label{sec:framework}
\textsc{Pasta} works within a Bayesian framework and can take the following jump parameters: (i) the $y$ exponent of the gyrochronological relation; (ii) the stellar rotation period at an age of 150 Myr $\ProtJ$, where 150 Myr is a representative early stage of stellar evolution that enables us to directly compare the posterior PDF obtained for a given star with the respective distribution gathered from measurements of the rotation period of OC stars (e.g. \citealt{johnstone15Prot150}; see below for the reason for which we decided to use $\ProtJ$ instead of $x$ as jump parameter); (iii) the $\beta$ exponent of Eq.~(\ref{eq:Lx}); (iv) the $q_{\mathrm{SF}}$ and $m_{\mathrm{SF}}$ coefficients of Eq.~(\ref{eq:L_EUV}); (v) the epoch of dispersal of the proto-planetary disk $t_{\mathrm{disk}}$; (vi) $t_{\star}$, $P_{\mathrm{rot,\star}}$, and $M_{\star}$; and (vii) for each planet, $a$, $M_{\rm p}$, and $\fas=f_{\mathrm{atm}}(t_{\mathrm{disk}})=\left. \frac{M_{\mathrm{atm}}}{M_{\rm p}} \right|_{t=t_{\mathrm{disk}}}$, which is the atmospheric mass fraction evaluated at the beginning of the evolution, that is, for $t=t_{\mathrm{disk}}$.

\textsc{Pasta} starts the evolution after the dispersal of the proto-planetary disk, whose timescale is estimated to be a few megayears \citep[see e.g.][and references therein]{montmerle10,alexander14,kimura16,gorti16}. Therefore, although $t_{\mathrm{disk}}$ can be set as a jump parameter, for this work we fix it at $t_{\mathrm{disk}}=5$ Myr. We will explore the capability of \textsc{Pasta} to also constrain $t_{\mathrm{disk}}$ in a future work.

The default setup is imposing Normal (Gaussian) priors $\mathcal{N}(\mu, \sigma)$ on $t_{\star}$, $P_{\mathrm{rot,\star}}$, $M_{\star}$, $a$, and $M_{\rm p}$, which are obtained through observations and stellar evolutionary models. Normal priors are imposed also on $\beta$, $q_{\mathrm{SF}}$, and $m_{\mathrm{SF}}$ having $\mathcal{N}(-2.70, 0.13)$, $\mathcal{N}(4.80, 1.99)$, and $\mathcal{N}(0.860, 0.073)$, respectively. 

Instead, the parameters $y$, $\ProtJ$, and $\fas$ are set having uniform priors. In particular, a flat prior on $\ProtJ$ implies a uniform sampling of values within the specified $\ProtJ$-range and avoids biasing the sampling towards low $\ProtJ$ values, which would happen if a uniform prior on $x$ was assumed. In practice, the output of the code are posterior PDFs for $\ProtJ$ and $\fas$ that are constrained by the currently observed system parameters and according to the considered input models.

As a statistical framework, \textsc{Pasta} uses a Markov chain Monte Carlo (MCMC) scheme implemented through the \textsc{MC3} tool \citep{cubillos17}. At each chain step \textsc{Pasta} samples a set of input parameters from the priors and it first selects the proper grid of stellar models according to the step value of $M_{\star}$, so to infer $L_{\mathrm{bol}}$, $T_{\mathrm{eff}}$, and $R_{\star}$ by interpolation within the stellar evolutionary tracks at the starting epoch of evolution $t_{\mathrm{disk}}$. Then, for each considered planet in the system, \textsc{Pasta} computes the equilibrium temperature as
\begin{equation}
    T_{\mathrm{eq}} = T_{\mathrm{eff}}\sqrt{\frac{R_{\star}}{2a}}\,,
    \label{eq:Teq}
\end{equation}
which assumes zero albedo and a moderately rotating planet, so that its entire spherical surface re-irradiates the flux received by the host star. Then, $T_{\mathrm{eq}}$ and the step values of $M_{\rm p}$ and $\fas$ enable interpolation within the planetary structure model grid to obtain the initial planetary radius $\hat{R}_p(t_{\mathrm{disk}})$. At this point, from the step value of $\Prot$ computed from Eq.~(\ref{eq:ProtYoung}) or (\ref{eq:ProtOld}), \textsc{Pasta} estimates $F_{\mathrm{EUV}}$ using the scaling relations given in Eqs.~(\ref{eq:Lx}) and (\ref{eq:L_EUV}), so that $\Madot$ can be computed through Eq.~(\ref{eq:Madot}).

The code continuously increases the evolutionary age by $\Delta t$, which is adjusted such that the atmospheric mass loss is less than $5\%$ of the entire atmospheric mass $M_{\mathrm{atm}}$. At the end of each MCMC step the code has generated a planetary evolutionary track by updating $M_{\mathrm{atm}}$ (from $\Madot$ and the time step $\Delta t$), and hence the planetary radius $\hat{R}_p(t_{\mathrm{disk}}+\Delta t)$. This loop terminates (i.e. the planetary evolutionary track reaches its end) when either the evolution reaches the present day stellar age $t_{\star}$ or the atmosphere is fully lost. The theoretical present-day planetary radius $\hat{R}_p(t_{\star})$ is finally compared with the observed value $R_p$ and that specific MCMC step is accepted according to the Metropolis-Hastings acceptance rule discussed in \citet{cubillos17}.

The algorithm uses the planetary radii as observational constraints to drive the chains' convergence. An alternative option implemented in the code is to track the atmospheric mass fraction $\hat{f}_{\mathrm{atm}}(t)$, rather than $\hat{R}_p(t)$, to finally compare $\hat{f}_{\mathrm{atm}}(t_{\star})$ with its observational counterpart $f_{\mathrm{atm}}(t_{\star})$ that may be obtained through a planetary atmospheric structure model and the planetary basic parameters \citep[e.g.][]{rogers2011,lopez14,dorn2017,delrez2021}.

The final results are posterior PDFs for each considered parameter. It is crucial to check the prior-posterior consistency of those parameters for which normal priors were imposed. If those priors are respected, the posterior PDFs of the parameters for which uniform priors were set ($\ProtJ$ and $\fas$ in this case) may be considered physically reliable or, at least, consistent with the adopted framework.

As already highlighted by \citet{kubyshkina19AAKepler11}, the constraining power of the code increases with increasing number of planets in the system. This is because the code has multiple fitting points (i.e. the radius of each planet in the system), but at the same time all planets in the system orbit around the same star and thus simultaneously constrain $\ProtJ$. Furthermore, the code can also be employed to constrain planetary masses in case the values are either poorly constrained or unavailable, as shown for example by \citet{kubyshkina19ApJ} and \citet{bonfanti21}.

The reliability of \textsc{Pasta}'s results depend on the reliability and accuracy of the input system parameters, of the background models, and of the assumptions. The scheme described above is based on two main assumptions:(i) the considered planets either hosted in the past or still host a hydrogen-dominated atmosphere \citep[see][for a thorough discussion on the general validity of this assumption]{owen2020}; and (ii) planet migration happened inside the disk and the planetary orbital separation does not change following disk dispersal (i.e. $a$ is constant over time). Furthermore, the planetary structure models we consider assume that planetary cores have an Earth-like density, regardless of the planetary mass. \citet{lopez14} and \citet{petigura16} validated this assumption showing that, for planets with a hydrogen-dominated atmosphere, the planetary radius is generally independent of the assumed core composition. However, a result of this assumption is that the algorithm works appropriately only for planets with an average density $\rho_p\leq1\rho_{\oplus}$.

\section{The sample}\label{sec:sample}
We compiled the sample of systems analysed in this work on the basis of the catalogue presented by \citet{otegi20}, which contains systems for which for at least one planet the uncertainties on $M_{\rm p}$ and $R_{\rm p}$ are smaller than 25\% and 8\%, respectively. Within this sample, we further selected just systems hosting planets with $5\lesssim M_{\rm p}/M_{\oplus} \lesssim 30$ in order to remain within the boundaries of the planetary structure and escape model grids we have available. Furthermore, we considered exclusively multi-planet systems, to benefit from the simultaneous multiple constrain on $\ProtJ$.

Following the aforementioned criteria, we ended up with a dozen systems. The planetary input parameters required to run \textsc{Pasta}, and their sources, are listed in Table \ref{tab:planets} (first five columns). In most cases, we took the same sources as chosen by \citet{otegi20}, further complementing them with additional sources in case of missing information. For Kepler-11, given the several and often not consistent $M_{\rm p}$ values present in the literature, we decided to combine the solutions provided by different authors, in an attempt to increase the robustness of the considered planetary masses. In practice, we treated each estimated $M_{p\,-\sigma_{\mathrm{low}}}^{\;\;+\sigma_{\mathrm{up}}}$ as a Normal PDF with possibly asymmetric tails and we summed all PDFs referring to the same planet. We then extracted the mode and estimated the asymmetric uncertainties from the resulting distribution.
%=====================================
\input{planets}
%=====================================

Besides the planetary parameters, \textsc{Pasta} also requires $M_{\star}$, $t_{\star}$, and $P_{\mathrm{rot,}\star}$ as input. We obtained both stellar mass and age by homogeneously analysing the host stars using the isochrone placement algorithm described in \citet{bonfanti15,bonfanti16}. Basic input parameters of the isochrone placement were the stellar effective temperature $T_{\mathrm{eff,}\star}$, metallicity [Fe/H]$_{\star}$, gravity $\log{g_{\star}}$, Gaia magnitude $G_{\star}$, and the Gaia parallax $\pi_{\star}$. When available from observations, the stellar rotation period $P_{\mathrm{rot,}\star}$, the projected rotational velocity $v\sin{i}_{\star}$ and the chromospheric activity index $\log{R'_{\mathrm{HK}}}$ were further added to the basic input set of the isochrone placement to possibly remove isochronal degeneracies, to improve convergence. Following \citet{bonfanti21}, we finally enlarged the internal uncertainties on $M_{\star}$ and $t_{\star}$ by adding in quadrature 4\% and 20\%, respectively, to account for isochrones' systematics. 

Any observational $P_{\mathrm{rot,}\star}$ value imposes a further constraint during the preliminary stellar age derivation process within the isochrone placement algorithm, but it is an optional parameter. Conversely, the following application of our \textsc{Pasta} algorithm always requires an estimate for $P_{\mathrm{rot,}\star}$. Among the considered sample of systems, just Kepler-11, Kepler-411, and WASP-47 have published $\Prot$ values. For the other stars, we estimated their current rotation periods by numerically inverting the gyrochronological relation of \citet{barnes10} as their ages are known from the evolutionary isochrones and tracks via the isochrone placement technique.

All stellar input parameters are listed in Table~\ref{tab:stars}. Except for the $M_{\star}$, $t_{\star}$, and most of the $P_{\mathrm{rot,}\star}$ values, which we computed as described above, all the other listed parameters were taken from the same sources specified in Table~\ref{tab:planets}.
%=====================================
\input{stars}
%=====================================

%
\section{Results}\label{sec:results}
\textsc{Pasta} returns posterior distributions for each jump parameter. As discussed above, the main results are the posterior distributions of $\ProtJ$ and $\fas$, which were set with uniform priors, but to assess their reliability it is necessary to first check the prior-posterior agreement for those parameters for which a Normal prior based on observations was imposed.

As an example illustrating the capabilities of \textsc{Pasta}, we present in Figs.~\ref{fig:K2-285star} and \ref{fig:K2-285planets} the priors and posteriors of all considered jump parameters for the K2-285 planetary system, which contains four planets. Similar plots, but for all other considered systems, are shown in Appendix \ref{app:plots}.

\subsection{The exoplanetary system K2-285}
In the following, we aim to guide the reader through the interpretation of the plots shown Figs.~\ref{fig:K2-285star} and \ref{fig:K2-285planets}, and in turn in Appendix \ref{app:plots}. Starting from Fig.~\ref{fig:K2-285star}, the posterior distributions (blue histograms) of $t_{\star}$, $P_{\mathrm{rot},\star}$, and $M_{\star}$ (first row, second to fourth panel) nicely agree with their respective priors (red Gaussians). The very slight tension between priors and posteriors of $q_{\mathrm{SF}}$ and $m_{\mathrm{SF}}$ (third row, first and second panel) highlights the importance of setting them as jump parameters rather than fixing them (see also Sect.~\ref{sec:gyroExp}). In fact, giving enough degrees of freedom to the algorithm allows it to find the best match between theoretical and observational constraints.

The global prior-posterior agreement, despite the large number of interacting models composing the algorithm, suggests that the PDF of $\ProtJ$ unveiling the past rotation history of K2-285 (first row, leftmost panel) can be considered to be reliable within the given assumptions and gives a median value of $\bar{P}_{\mathrm{rot,}150}=15.3_{-9.4}^{+13.5}$ days. A comparison of the PDF of $\ProtJ$ with the $\ProtJ$ distribution obtained from considering stars having $|M_{\mathrm{J15}}-M_{\star,\mathrm{K2-285}}|<0.1M_{\odot}$ within the sample of \citet{johnstone15Prot150} indicates that, when it was young, K2-285 was a slow rotator.

The second row of Fig.~\ref{fig:K2-285star} contains three panels. The leftmost panel displays the estimated $L_X$ at 150 Myr, which is derived through Eq.~(\ref{eq:Lx}) and considering $\ProtJ$, while the other two panels show the PDFs of the gyrochronological exponents $x$ and $y$. Although characterised by a well defined peak, the PDF of $L_X$ presents a little bump at $\log{L_X}>29.5$, which results because $L_X$ saturates at low $\Prot$ values (saturation regime). As a matter of fact, $L_X$ and $\Prot$ are anti-correlated, but in the regime of very fast rotators (low $\Prot$), $L_X$ stops increasing and becomes equal to $L_{X,\mathrm{sat}}$, which depends only on stellar mass \citep[see e.g.][Fig.~2]{wright11}. Therefore, towards large values, the PDF of $L_X$ stops decreasing smoothly and all $L_{X,\mathrm{sat}}$ values are lumped in the last bins towards high $L_X$ values. In some cases, this bin is so pronounced to enter the highest posterior density (HPD) credible interval, but the HPD credible interval for $L_X$ should be continuous. This means that in these cases the high $L_X$ solution should be considered to be a bias driven by the lumping of high $L_X$ values in a narrow range independently of $\Prot$.

The median of the $x$ distribution $\bar{x}=0.23_{-0.16}^{+0.33}$ points towards a slow stellar spin-down if compared to the spin-down rates theoretically expected by \citet[][Fig. 2]{tu15}. In fact, the median $\bar{x}$ we derived for K2-285 corresponds to the 10$^{\mathrm{th}}$ percentile of the theoretical distribution. Moreover, the HPD credible interval of $y$ spans values below 0.50 and the median of the distribution is $\bar{y}=0.32_{-0.24}^{+0.37}$, which is smaller (though within 1$\sigma$) of the classical \citeauthor{skumanich72} exponent that is equal to $\frac{1}{2}$. We refer the reader to Sects.~\ref{sec:gyroExp} and \ref{sec:rotperiods} for a deeper discussion about the spin-down rate of our targets.

Figure \ref{fig:K2-285planets} shows the PDFs of the semi-major axis $a$ (first row), mass $M_{\rm p}$ (second row), and initial atmospheric mass fraction $\fas$ (third row) for each planet detected in the K2-285 system. The Normal priors imposed on $a$ are respected by the posteriors. Normal priors were also imposed on the masses of planets b and c (second row, two leftmost panels) as derived from radial velocity, while uniform priors were imposed on planets d and f, which have just upper mass limits \citep{palle19}.

Figure \ref{fig:K2-285planets} shows that \textsc{Pasta} is unable to constrain $\fas$ for planets b, d, and e, whose posteriors are essentially flat. In other words, any atmospheric mass fraction at $t=t_{\mathrm{disk}}$ is compatible with their present-day atmospheric content, which is negligible. Instead, for planet c \textsc{Pasta} gives a current atmospheric mass fraction of $f_{\mathrm{atm,c}}^{\mathrm{now}}$\,$\approx$\,0.025 and returns a well constrained initial atmospheric mass fraction of $\bar{f}_{\mathrm{atm,c}}^{\mathrm{start}}=0.0384_{-0.0053}^{+0.0052}$. Therefore, it is the constraint imposed by planet c that mainly drives the atmospheric evolution of the K2-285 system, defining the activity level of the host star over time. As a positive side effect of studying multi-planet system, despite the uniform priors and as a result of planet c constraining the evolution of the stellar rotation rate, \textsc{Pasta} returns a rather tight constraint on the masses of planets d and e, for which we obtain $3.59_{-0.34}^{+0.33}\, M_{\oplus}$ and $2.04_{-0.18}^{+0.20}\, M_{\oplus}$, respectively.

The K2-285 corner plot of all the relevant jump parameters within our framework is shown in Fig. \ref{fig:K2-285corner}. It emphasises the overall lack of correlation between the majority of the jump parameters with a few exceptions. For example, $q_{\mathrm{SF}}$ and $m_{\mathrm{SF}}$ are clearly anti-correlated. The reason is that the data point scatter quantitatively described by Eq.~(\ref{eq:L_EUV}) occupies a broad strip in the $\log{L_{\mathrm{EUV}}}$-$\log{L_X}$ plane, which is quantified by the high $\sigma_q$. By decreasing the intercept $q_{\mathrm{SF}}$ and increasing the slope $m_{\mathrm{SF}}$ at the same time, the best-fit line still spans the same strip.

Another anti-correlation involves $f^{\mathrm{start}}_{\mathrm{atm},c}$ and $M_c$. 
Planet c is the only one characterised by a tight constraint on its $\fas$. The higher $M_c$, the higher the gravitational potential that retains the atmosphere; therefore, \textsc{Pasta} estimates a lower value for $f^{\mathrm{start}}_{\mathrm{atm},c}$ to still match the present-day atmospheric content given the lower mass-loss rate. Finally there are positive correlations involving the masses of planets b, d, and e, which are the ones for which \textsc{Pasta} was not able to constrain their respective $\fas$. As planet d and e have no observational constraints on their masses, it is likely that the mass correlations reflect correlations present within our planetary structure models, when only $T_{\mathrm{eq}}$ drives the mass posterior PDFs.

\begin{figure*}
    \resizebox{\hsize}{!}{ \includegraphics{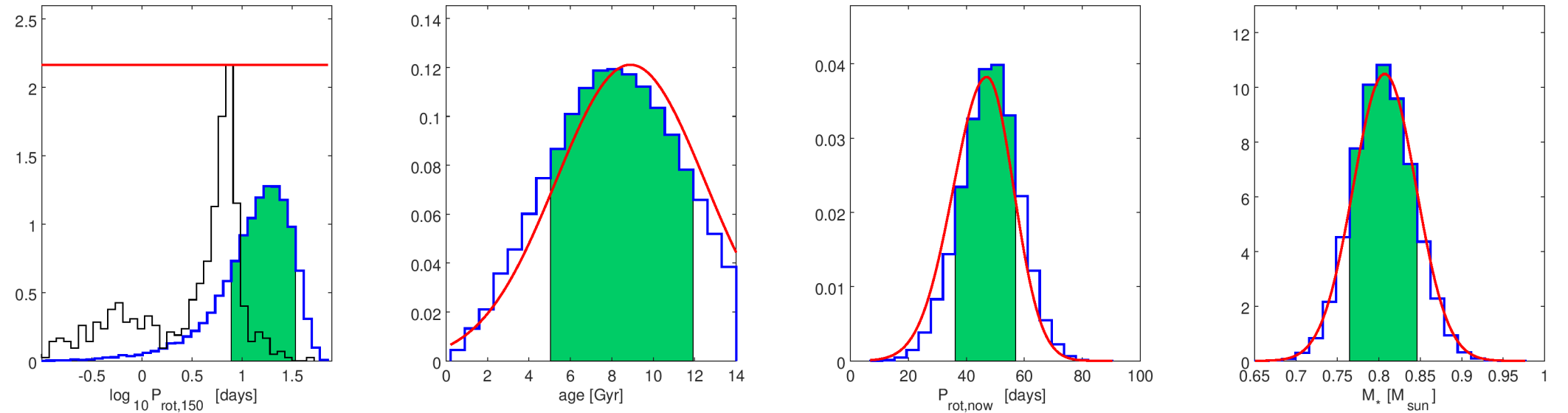}} \\
    \resizebox{\hsize}{!}{ \includegraphics{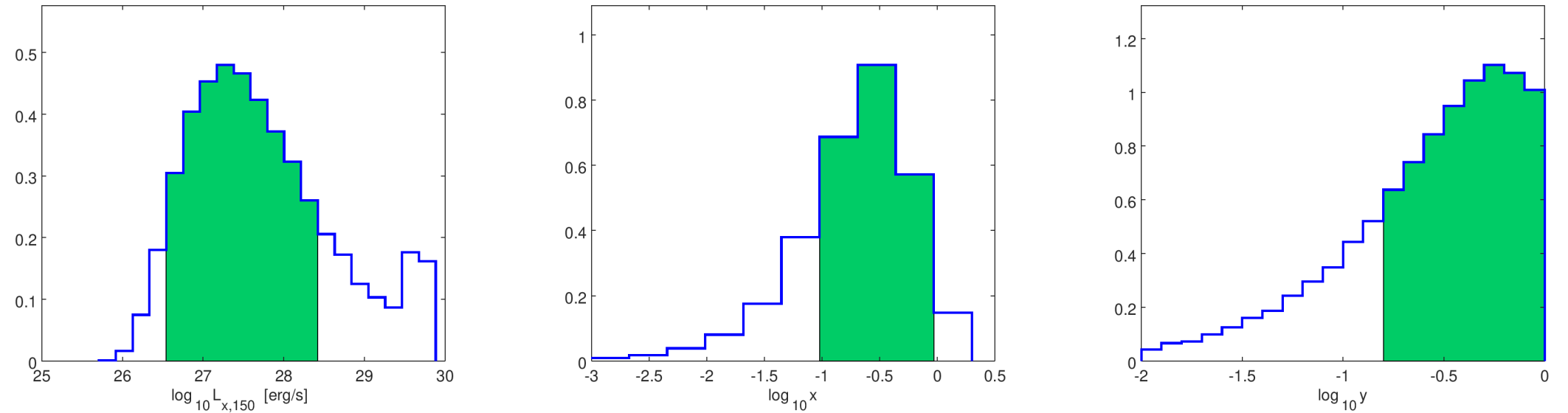}} \\
    \resizebox{\hsize}{!}{ \includegraphics{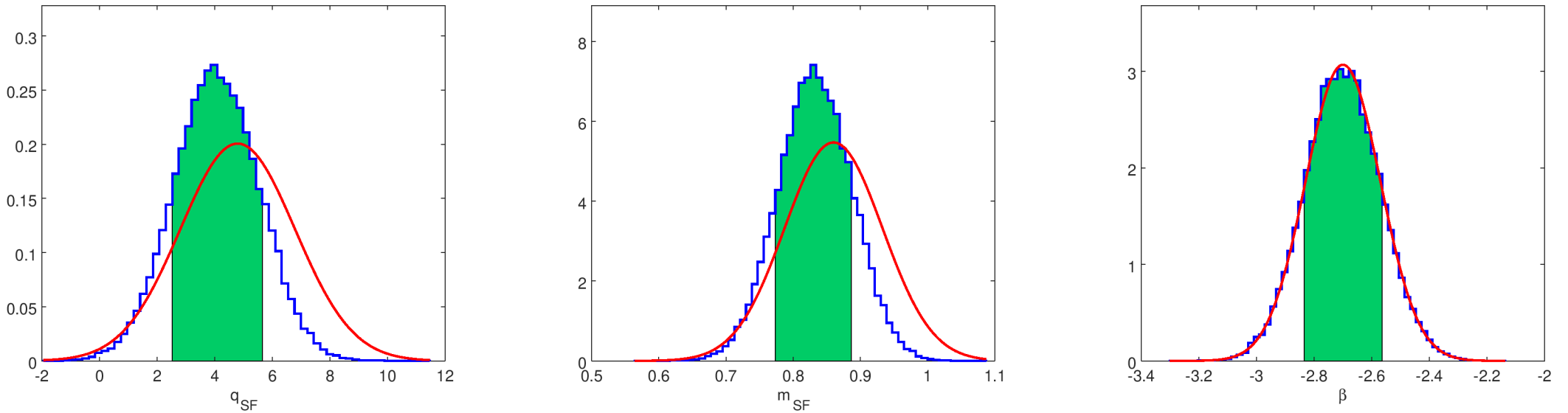}}
    \caption{Stellar properties of the K2-285 system given by \textsc{Pasta}. First row: $\ProtJ$, $t_{\star}$, $P_{\mathrm{rot},\star}$, and $M_{\star}$. Second row: $L_X$ at 150 Myr, and the $x$ and $y$ exponents adopted by the gyrochronological relations given by Eqs.~(\ref{eq:ProtYoung}) and (\ref{eq:ProtOld}). Third row: $q_{\mathrm{SF}}$ and $m_{\mathrm{SF}}$ coefficients of Eq.~(\ref{eq:L_EUV}), and the $\beta$ exponent of Eq.~(\ref{eq:Lx}). The output posterior PDFs are shown as blue histograms, with the green area representing the 68.3\% HPD credible interval. The red curves are the Gaussian or flat (uninformative) imposed priors. When absent (second row), no specific prior was imposed on the parameter; in particular, $x$ and $L_X$ are mathematically derived from the PDFs of the jump parameters. The black histogram in the top-leftmost panel represents the $\ProtJ$ distribution extracted from the sample of \citet{johnstone15Prot150} considering stars with a mass differing by less than $0.1\, M_{\odot}$ from the mass of K2-285.}
    \label{fig:K2-285star}
\end{figure*}
\begin{figure*}
    \resizebox{\hsize}{!}{ \includegraphics{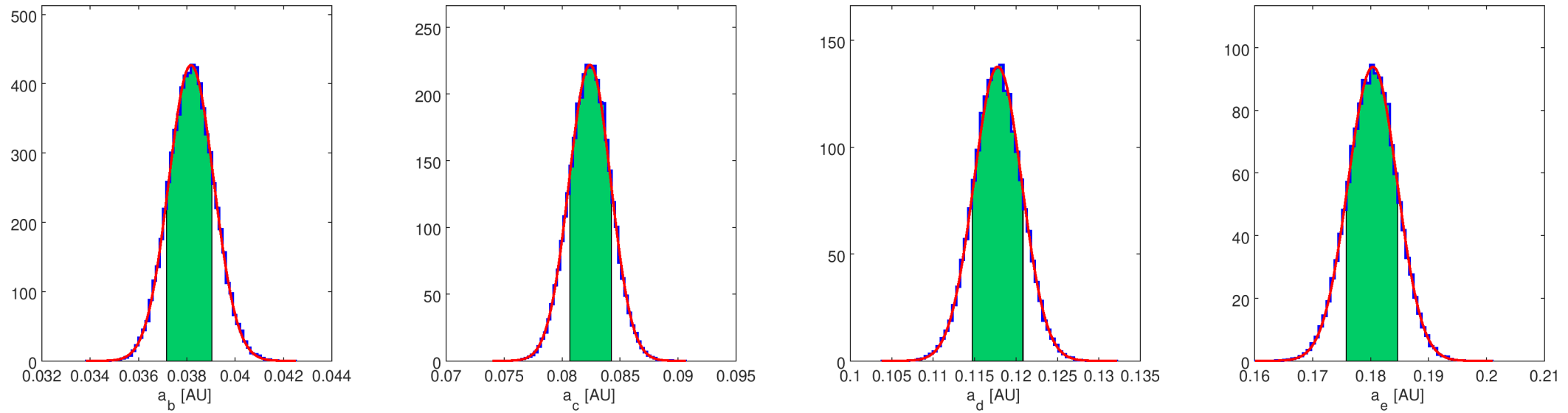}} \\
    \resizebox{\hsize}{!}{ \includegraphics{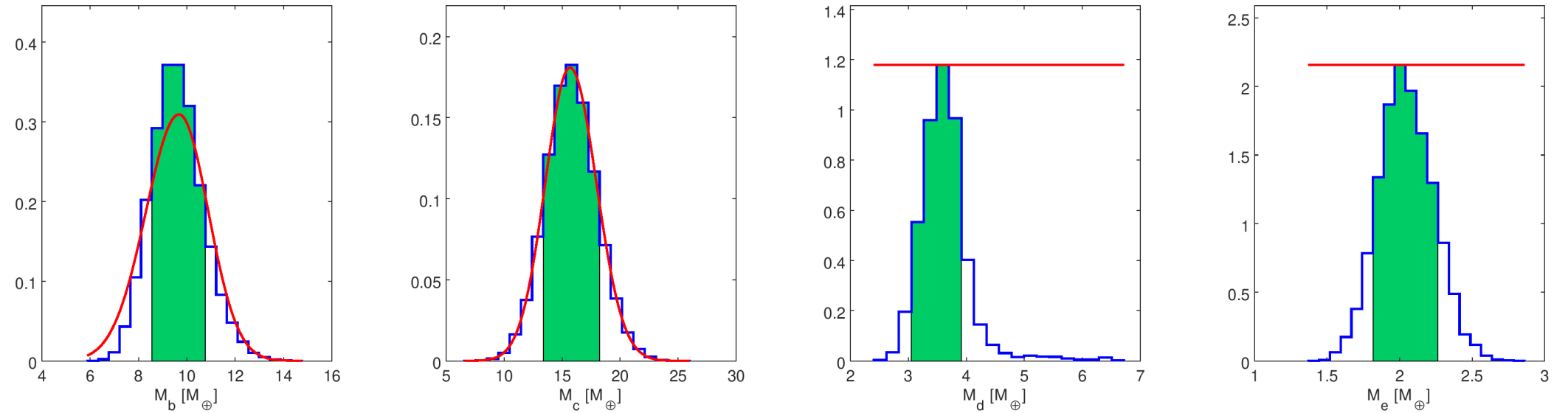}} \\
    \resizebox{\hsize}{!}{ \includegraphics{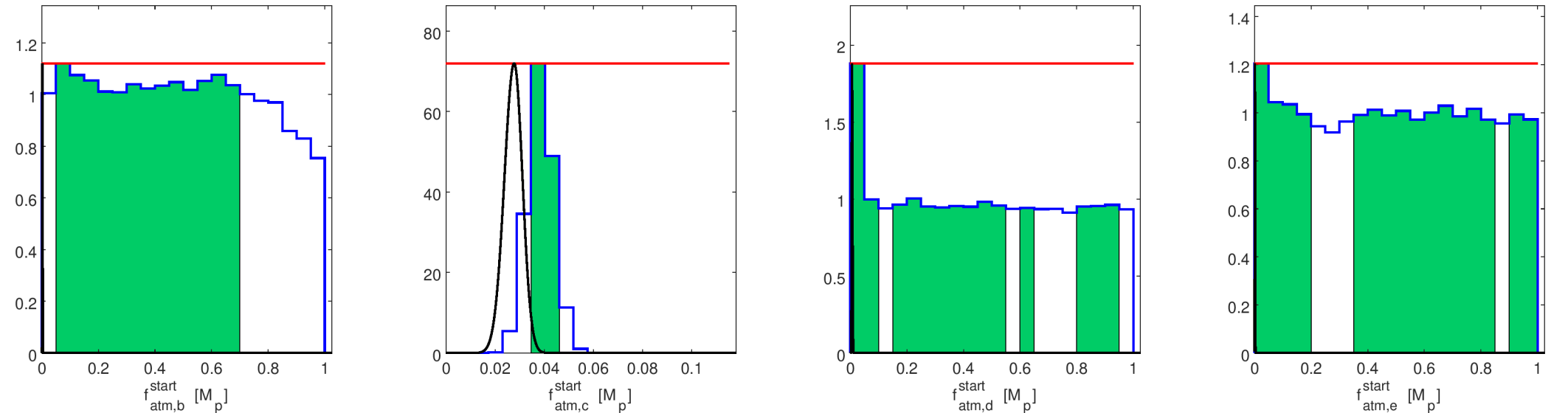}}
    \caption{PDFs given by \textsc{Pasta} for the considered planetary parameters in the K2-285 system: semi-major axis (first row), mass (second row), and $\fas$ (third row). The black curves in each panel of the third row (evident in the second panel, and just barely visible in the others as they are squeezed towards zero) indicate the predicted present-day atmospheric mass fraction. It follows that K2-285 b, d, and e have almost entirely lost their atmospheres at some point in time along the system's evolution.}
    \label{fig:K2-285planets}
\end{figure*}

\begin{figure*}
    \resizebox{\hsize}{!}{\includegraphics{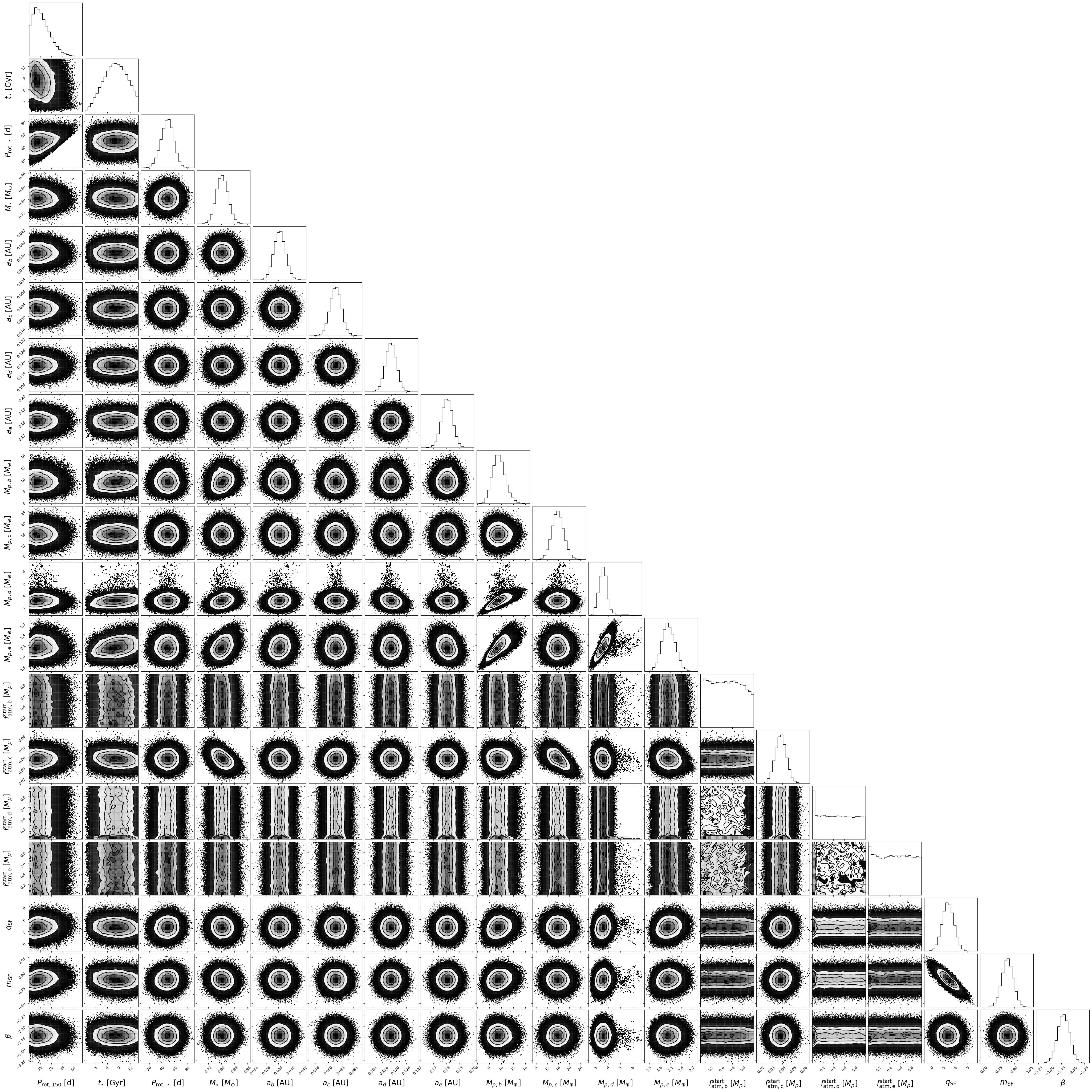}}
    \caption{Corner plot of the relevant jump parameters returned by \textsc{Pasta} for the K2-285 system.}
    \label{fig:K2-285corner}
\end{figure*}

\section{Discussion}\label{sec:discussion}

We combined the results obtained from all 12 systems to preliminary check whether the results of our framework agree with the considered empirical relations and then derive the distributions of the gyrochronological exponents $x$ and $y$ (Sect.~\ref{sec:gyroExp}). Furthermore, as \citet{johnstone15Prot150} report the observed rotation periods of OC stars of 150 and 600 Myr, we also compared the average distributions of $\ProtJ$ and $\ProtJJ$ with those derived from stars member of young OCs to look for possible traces of SPI occurring during the first stages of evolution (Sect.~\ref{sec:rotperiods}). Finally, we looked for correlations between the derived $\fas$ values and system parameters to constrain planetary atmospheric accretion models (Sect.~\ref{sec:accretion}).

\subsection{The gyrochronological exponents}\label{sec:gyroExp}

\citet{kubyshkina19ApJ} and \citet{kubyshkina19AAKepler11} presented the results of several tests performed on synthetic systems aiming at validating \textsc{Pasta}'s framework, but they did not enable a validation of the physical results. Because of the nature of the results provided by \textsc{Pasta}, it is not possible to use any real system to validate the output $\ProtJ$ and $\fas$ distributions. 
However, as our framework is built on several empirical relations and theoretical models, we checked a posteriori whether the results obtained for the different systems respect all observational, empirical, and theoretical constraints at the same time. For example, the posterior PDFs on $M_{\star}$ and $t_{\star}$ are driven by MESA theoretical models, while the priors imposed on $q_{\rm SF}$, $m_{\rm SF}$, and $\beta$ have an empirical root.

\begin{figure*}
    \centering
    \includegraphics[width=0.33\textwidth]{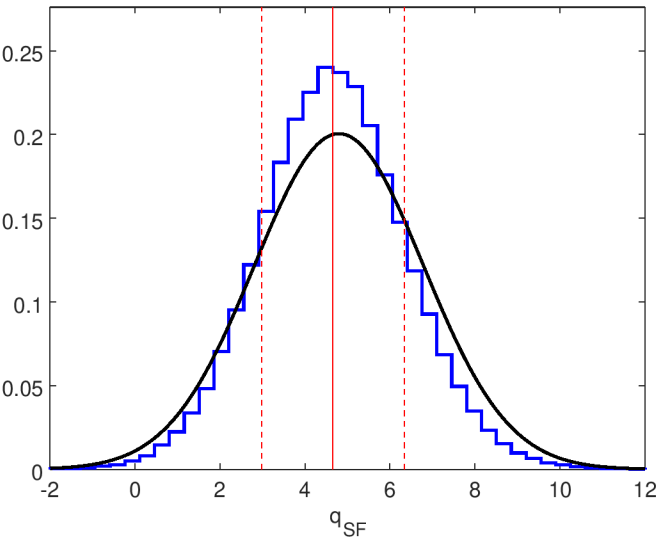}
    \includegraphics[width=0.32\textwidth]{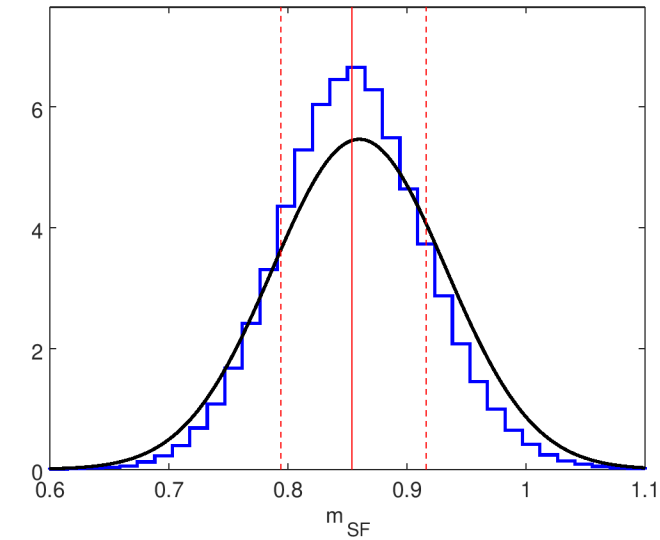}
    \includegraphics[width=0.32\textwidth]{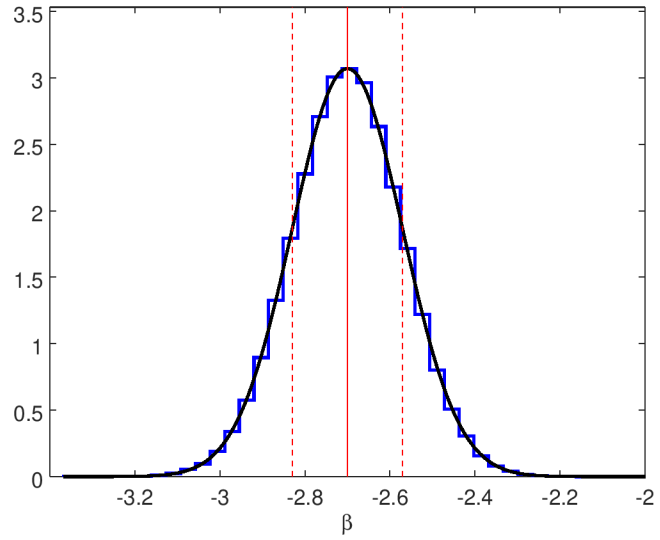}
    \caption{PDFs (blue histograms) of those jump parameters entering as coefficients in Eqs.~(\ref{eq:Lx}) and (\ref{eq:L_EUV}). \textit{Left and middle panel:} Intercept $q_{\mathrm{SF}}$ and slope $m_{\mathrm{SF}}$ of Eq.~(\ref{eq:L_EUV}). \textit{Right panel:} $\beta$ exponent of Eq.~(\ref{eq:Lx}). The median and the $\pm1$-$\sigma$ bounds of each PDF are marked as vertical solid and dashed red lines, respectively. The black Gaussians represent the priors inferred from observations for each empirical parameter.}
    \label{fig:qmBeta}
\end{figure*}

The three panels of Fig. \ref{fig:qmBeta} show the merged posterior PDFs (blue histograms) for $q_{\rm SF}$, $m_{\rm SF}$, and $\beta$, which have been obtained by combining the posterior distributions coming from the analysis of each planetary system. The agreement between posteriors and priors (black Gaussians) of these parameters in addition to the overall prior-posterior agreement of all the jump parameters seen in each analysis tells us that both the theoretical and empirical components of our framework lead to a consistent picture.

\begin{figure}
    \centering
    \includegraphics[width=\columnwidth]{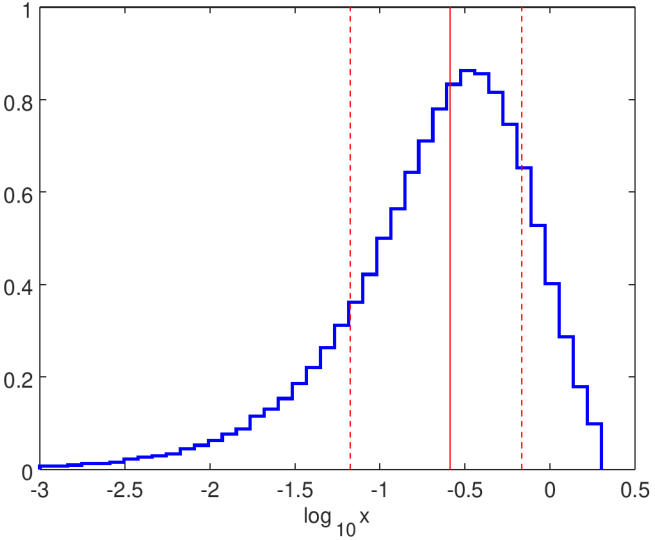}
    \includegraphics[width=\columnwidth]{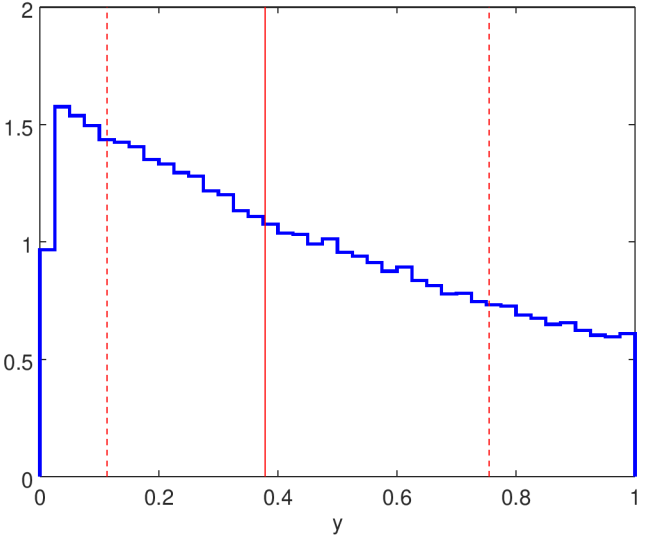}
    \caption{\textit{Top}: PDF of the $x$ exponent of our adopted gyrochronological relations expressed in Eqs.~(\ref{eq:ProtYoung}) and (\ref{eq:ProtOld}) built upon all the stellar targets analysed in this work. The vertical solid red line marks the median of the distribution, while the two vertical dashed red lines marks the $\sigma$ levels corresponding to the 15.87$^{\mathrm{th}}$ and 84.14$^{\mathrm{th}}$ percentiles. \textit{Bottom}: Same as the top panel, but for the $y$ exponent of Eq.~(\ref{eq:ProtOld}). We note that the \textit{top panel} is in log scale.}
    \label{fig:xExp}
\end{figure}

This preliminary check strengthens the statistics we derived for the gyrochronological exponents. In fact, we also merged the $x$ and $y$ posterior distributions of all our selected targets producing the histograms shown in Fig.~\ref{fig:xExp}. We recall that the $x$ exponent controls the rotation rate of the star during its first 2 Gyr of life. This rate may vary significantly from star to star, in fact rotational evolutionary tracks differing by as much as $\sim$1.5 orders of magnitude tend to converge towards the same $\Prot$ after a couple of Gyr \citep[][their Figure 2]{tu15}. Instead, the $y$ exponent controls the decay of the stellar rotation rate at older ages.

The classical \citet{skumanich72} law models the decay of $\Prot$ of main sequence stars as a function of time through a power-law whose exponent is $\frac{1}{2}$. Other examples of gyrochronological relations in the form of power-laws may be found in, for example, \citet{barnes03}, \citet{barnes07}, \citet{mamajek08}, and \citet{collierCameron09}. These works emphasise that $\Prot$ may be predicted from knowledge of both stellar age and colour index, where the latter is taken as a proxy of stellar mass. The different authors still suggest an age dependence of the form $\Prot \propto t^{\alpha}$, where $\alpha=0.5$ \citep{barnes03}, $\alpha=0.5189$ \citep{barnes07}, $\alpha=0.566$ \citep{mamajek08}, and $\alpha=0.56$ \citep{collierCameron09}. If, on the one hand, the majority of literature sources tend to prefer $\alpha$ values slightly greater than 0.5, on the other hand, \citet{meibom09} confirm the empirical $\Prot$\,$\propto$\,$t^{\frac{1}{2}}$ dependence for G-type stars, but find a slower spin-down rate (i.e. compatible with $\alpha<0.5$) for K stars. All in all, an exponent of $\sim$\,$\frac{1}{2}$ appears to be generally adequate to describe stellar spin-downs for population studies, but slight variations may be expected on a star by star basis, especially for stars with a mass different from that of G-type stars.

Figure \ref{fig:xExp} indicates that the reference (i.e. median) gyrochronological exponent for our host stars can be estimated as $\bar{y}=0.38_{-0.27}^{+0.38}$ with a general preference for slow spin-down rates, even if the 84.14$^{\mathrm{th}}$ percentile $\bar{y}_{+1\sigma}=0.76$ demonstrates the heaviness of the right tail. The difference between our inferred $\bar{y}$ value and the $y$ values reported in the literature is well below 1$\sigma$. 

The preference towards lower spin-down rates shown by our stellar sample may be the result of a selection bias. As a matter of fact, exoplanets are preferentially found around quiet stars, hence slow rotators. This is supported also by the $x$ distribution with a median of $\bar{x}=0.26_{-0.19}^{+0.42}$, which strongly favours a quiet evolution during the first stages of stellar evolution given its median. We thoroughly discuss this in Sect.~\ref{sec:rotperiods}. With reference to \citet[][Fig. 2, left panel]{tu15}, the range of our $x$ values defined by the 68.3\% confidence interval is compatible with the set of rotational evolutionary tracks spanning the same percentile range, even if our $x$ distribution is skewed towards lower values with $\bar{x}$ being approximately the rate displayed by the red track of \citet[][10$^\mathrm{th}$ percentile of the predicted rotational distribution]{tu15}.

Finally, we stress that gyrochronological relations in the literature are usually calibrated on OC stars (as they are a coeval), which are generally young. Instead, although dependent by the stellar and planetary models in use, our approach -- constrained by the present-day $M_{\mathrm{atm}}$ -- may infer the rate of rotational decay for stars of any age. The lower spin-down rates we found are consistent with the conclusions independently drawn by \citet{kovacs15} or \citet{vanSaders16} about non-cluster field stars and give an original contribution to the gyrochronological studies of planet-hosting stars.

\subsection{Rotation rates at young ages}\label{sec:rotperiods}

\begin{figure}
    \centering
    \includegraphics[width=\columnwidth]{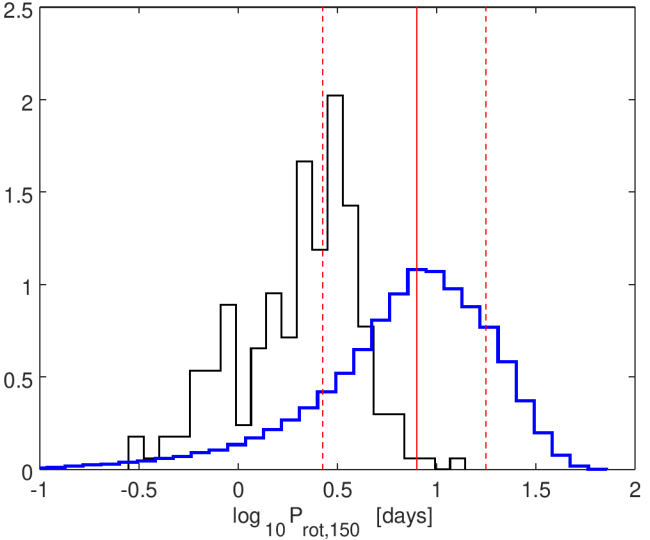}
    \includegraphics[width=\columnwidth]{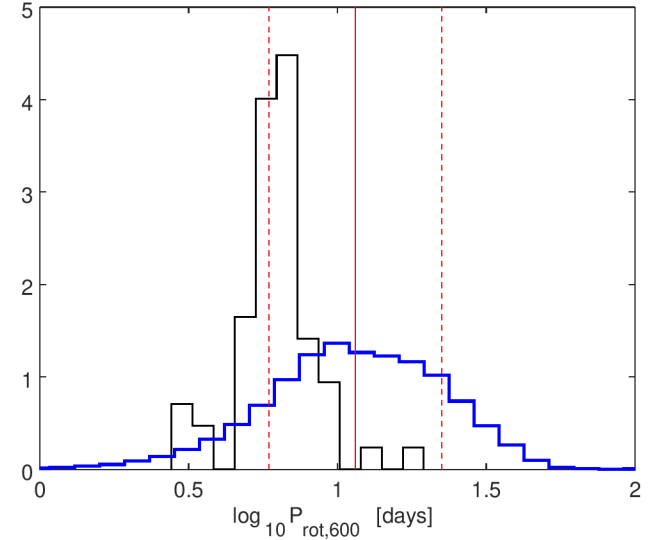}
    \caption{Top: Comparison between the $\ProtJ$ PDF obtained by combining those distributions extracted from the analysed sample (blue line) and the $\ProtJ$ distribution extracted from \citet{johnstone15Prot150}, considering stars member of $\sim$\,150 Myr old OCs and having the same mass distribution as the planet-hosting stars analysed in this work. The vertical solid red line marks the median value of the distribution shown by the blue line, while the two vertical dashed red lines indicate the 15.87$^{\mathrm{th}}$ and 84.14$^{\mathrm{th}}$ percentiles. Bottom: Same as the top panel, but for the $P_{\mathrm{rot,}600}$ PDF.}
    \label{fig:Prot}
\end{figure}

We combined all the $\ProtJ$ distributions we obtained from each analysed system finally obtaining the distribution shown by the blue histogram in Fig.~\ref{fig:Prot}. This distribution has a median value of $\bar{P}_{\mathrm{rot,150}}=7.9_{-5.3}^{+9.8}$ days, where the error bars are 1$\sigma$ uncertainties. We further used the results of \citet{johnstone15Prot150} to construct a $P_{\mathrm{rot\,J,150}}$ distribution considering stars member of $\sim$\,150 Myr old OCs and having the same mass distribution as that of the planet hosts analysed here. This distribution is shown by a black line in Fig.~\ref{fig:Prot} and has a median value of $\bar{P}_{\mathrm{rot\,J,150}}=2.2_{-1.4}^{+1.5}$ days, which is $\sim$\,1\,$\sigma$ smaller than the median value derived for our sample of planet-hosting stars. 

Figure \ref{fig:Prot} suggests that the rotation rate at an age of 150 Myr of the sample of planet-hosting stars analysed in this work is on average slower than what observed for the majority of stars. Because of the small number statistics it is not possible to conclude whether this is the result of SPI or not, but the most likely explanation for this difference is a selection bias. By construction, \textsc{Pasta}'s capability of constraining $\ProtJ$ of a given star hinges on it hosting at least one planet with a hydrogen-dominated atmosphere that has been and/or is still affected by atmospheric escape. On average, it is easier to find such planets orbiting stars that when young were slow rotators, because if they had been fast rotators the planets would have more likely lost their hydrogen envelope. Similar considerations hold for $\ProtJJ$, where $\bar{P}_{\mathrm{rotJ,}600}=6.3_{-1.5}^{+1.6}$ days is smaller than $\bar{P}_{\mathrm{rot,}600}=11.5_{-5.6}^{+10.9}$ by $\sim$\,$1\sigma$.

\subsection{Constraining planetary atmospheric accretion models}\label{sec:accretion}

One of the main aims of this work, and more in general of \textsc{Pasta}, is to enable one to look for correlations between the initial atmospheric mass fraction of close-in planets and system parameters to constrain planetary formation and atmospheric accretion models. In particular, we focus here on analysing the possible correlations present between initial atmospheric mass fraction and semi-major axis, planetary mass, and stellar mass.

\begin{figure}
    \centering
    \includegraphics[width=\columnwidth]{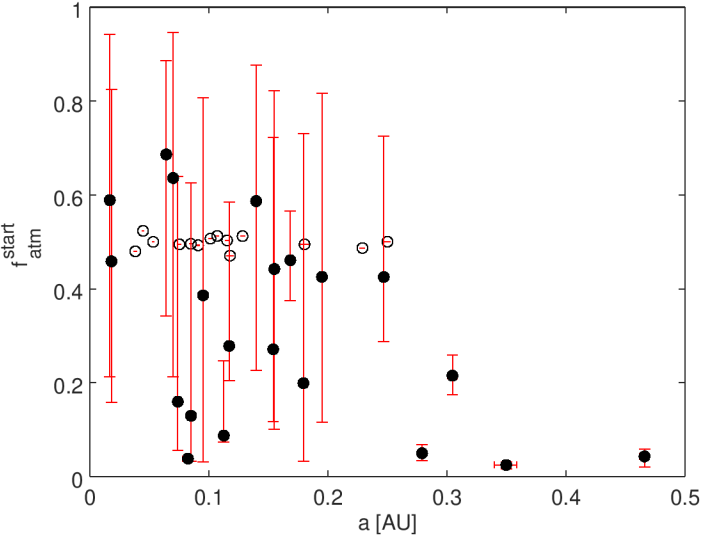}
    \includegraphics[width=\columnwidth]{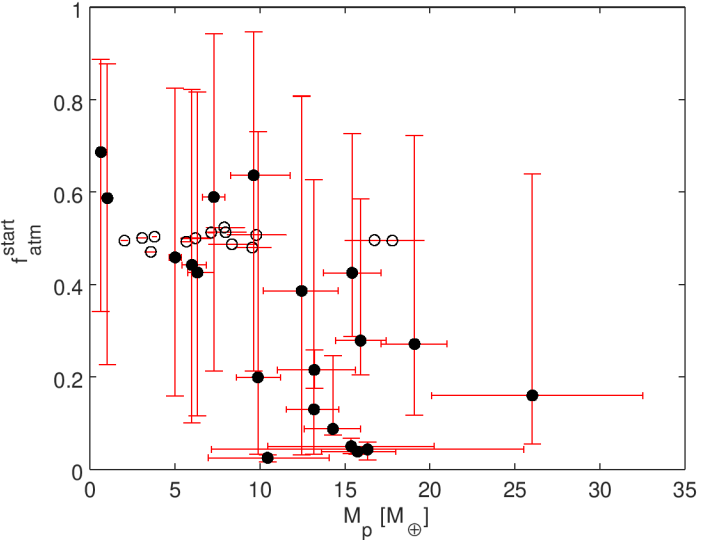}
    \includegraphics[width=\columnwidth]{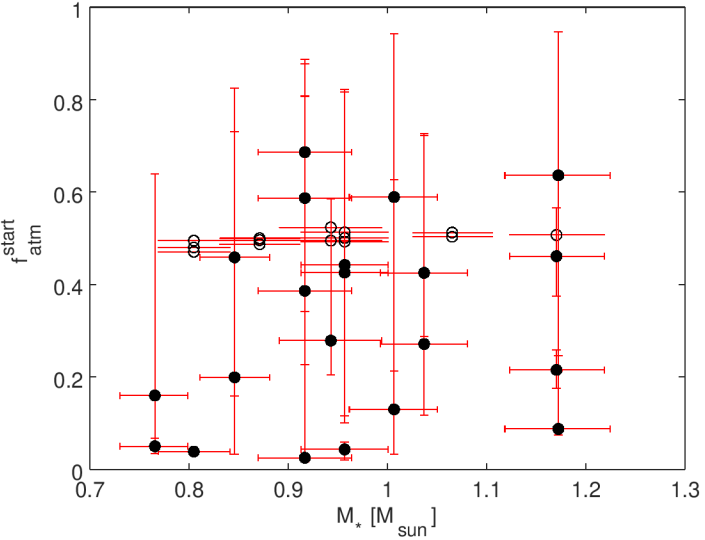}
    \caption{Initial atmospheric mass fractions as a function of planetary semi-major axis (top), planetary mass (middle), and stellar mass (bottom) for the sample of planetary systems considered here. For flat $\fas$ PDFs, the uncertainty would span the entire axis and is therefore not displayed for better visualisation; those data points are indicated by empty markers.}
    \label{fig:fatVsA}
\end{figure}

Figure~\ref{fig:fatVsA} shows the median $\fas$ values provided by \textsc{Pasta} as a function of semi-major axis, of planetary mass, and of stellar mass for the analysed systems. There seems to be a trend between $\fas$ and $a$, in which planets with $a\lesssim0.25$ AU have a larger $\fas$ compared to planets at larger $a$, a possible link between $\fas$ and $M_{\rm p}$, in which $\fas$ decreases with increasing $M_{\rm p}$, and a possible link between $\fas$ and $M_{\star}$, in which $\fas$ increases with increasing $M_{\star}$, but none of these correlations is statistically significant. 
In detail, because of small number statistics, particularly at large orbital separations, the trend between $\fas$ and $a$ has a coefficient of determination $R^2=0.25$. However, analyses of additional systems detected and/or measured for example by TESS \citep{ricker2015}, CHEOPS \citep{benz21}, or PLATO\citep{rauer14} may shed light on the possible existence of such a correlation. This trend is weak also because of the large uncertainties of the $\fas$ values obtained for the planets at short orbital separation. Indeed, for most of the planets with $a\lesssim0.25$ AU we obtained (almost) flat PDFs on $\fas$. This may be a result of planetary atmospheric evolution as planets farther away from their host star are less subject to atmospheric escape, and thus have a slower and more unique evolution compared to closer-in planets. The link between $\fas$ and $M_{\rm p}$ is not significant ($R^2=0.12$), but it shows the potential capability of \textsc{Pasta} to constrain atmospheric accretion processes. This plot further highlights that \textsc{Pasta} allows one to better constrain $\fas$ for larger mass planets, as the vast majority of flat $\fas$ posteriors is obtained for lower-mass planets. This is again possibly the result of planetary atmospheric evolution, as for larger mass planets the atmospheric escape is weaker and the evolution is slower. Even excluding the $\fas$ data points characterised by a flat posterior, also the link between $\fas$ and $M_{\star}$ is not significant ($R^2=0.068$). If statistically significant, such a correlation would go in the direction suggested by \citet{lozovsky2021} in which atmospheric accretion of hydrogen-dominated atmospheres might be more efficient for planets orbiting more massive stars \citep[see also][]{kennedy2008a,kennedy2008b,kennedy2009}. The analysis of additional systems may lead one to constrain this correlation.

\begin{figure}
    \centering
    \includegraphics[width=1.0\columnwidth]{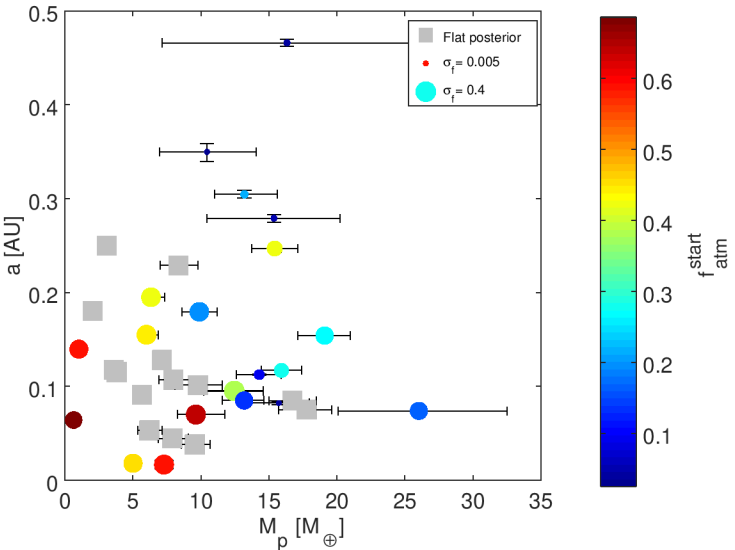}
    \includegraphics[width=1.0\columnwidth]{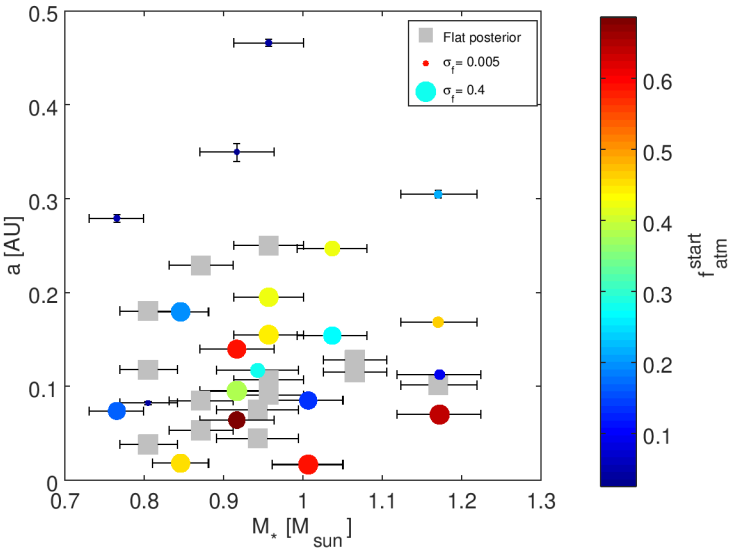}
    \includegraphics[width=1.0\columnwidth]{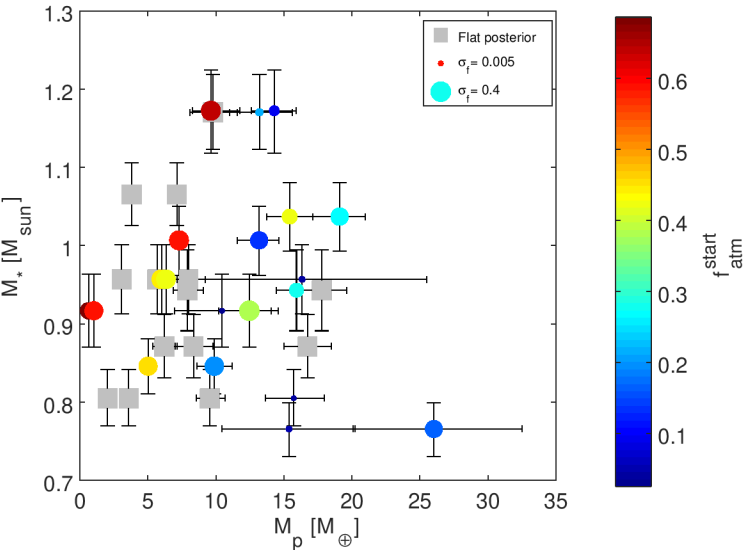}
    \caption{Planetary semi-major axis as a function of planetary mass (top) and stellar mass (middle), and stellar mass as a function of planetary mass (bottom) for the planetary systems considered here. In each panel, the initial atmospheric mass fraction ($\fas$) is colour-coded and the size of the data points is proportional to the uncertainty (at the $\sigma$ level) inferred from the $\fas$ posterior PDFs. Squares are for flat $\fas$ posterior PDFs.}
    \label{fig:aVsMs}
\end{figure}

In Fig.~\ref{fig:aVsMs}, we show the possible presence of multidimensional links among the parameters considered above. In particular, the top panel of Fig.~\ref{fig:aVsMs}, which shows how $\fas$ varies as a function of $a$ and $M_{\rm p}$, highlights that \textsc{Pasta} is capable of better constraining $\fas$ for higher-mass planets with a larger orbital separation. In fact, atmospheric loss is controlled by $\dot{M}_{\mathrm{atm}}$, which is positively correlated with $R_p$. Because of their low gravitational potential, low-mass planets can be subject of extremely large mass-loss rates and thus there is a degeneracy between atmospheric evolutionary tracks characterised by large $\fas$ and high $\dot{M}_{\mathrm{atm}}$ or small $\fas$ and low $\dot{M}_{\mathrm{atm}}$. For higher-mass planets, instead, the high gravitational potential limits the mass-loss rates and thus there is less degeneracy among the possible atmospheric evolutionary tracks, which leads to a more defined $\fas$. At equal planetary mass, the orbital separation plays the same role as planetary mass in controlling the escape and thus the degeneracy among evolutionary tracks. Therefore, on average, planets orbiting farther away from their host star lead to a more defined $\fas$. However, in general, Fig.~\ref{fig:aVsMs} leads to the same conclusions drawn by Fig.~\ref{fig:fatVsA}, without further particular additions.

\section{Conclusions}\label{sec:conclusions}

We have characterised the atmospheric evolution of planets composing a dozen exoplanetary systems. In particular, we constrained the initial atmospheric mass fraction (i.e. fraction of planetary atmospheric mass at the dispersal of the protoplanetary nebula) of the planets and the evolution of the rotation rate of the host stars. To this end, we employed a custom algorithm called \textsc{Pasta}, which is a major update to the code described by \citet{kubyshkina19ApJ}, and \citet{kubyshkina19AAKepler11}. Compared to the previous version of the algorithm, the major upgrade is the refinement of the gyrochronological relation and the treatment of the coefficients entering the empirical relations between stellar EUV emission, $L_X$ emission, and rotation period as jump parameters. The excellent prior-posterior agreement of all observational, empirically driven, and theoretically driven parameters suggests that the results on the unconstrained parameters are robust, particularly given the heterogeneity of the components making up our framework.

We found that the average PDF of the stellar rotation period at an age of 150\,Myr has a median value of $\ProtJ=7.9_{-5.3}^{+9.8}$, which is $\sim$\,$1\sigma$ higher than what was found by \citet{johnstone15Prot150} for coeval OC stars. This result, together with the spin-down rate distribution at early stages skewed towards low values, may be due to a selection bias. In fact, \textsc{Pasta} requires systems that host at least one close-in planet with a hydrogen-dominated atmosphere. Therefore, the presence of such a planet is favoured around stars that have evolved as slow rotators since the intense XUV emission of fast rotators would have more likely totally removed the planetary primary atmosphere.

We also found that the rotation rate distribution at an age older than 2\,Gyr for the planet hosts considered in this work is compatible with that found in the literature, though our distribution tends to be skewed towards lower spin-down rates. This is consistent with the conclusions independently drawn by \citet{kovacs15} and \citet{vanSaders16}, who noticed that field stars exhibit lower spin-down rates than their cluster counterparts. Similar conclusions were previously reached by \citet{brown14} and \citet{maxted15} when specifically comparing cluster stars with non-cluster planet-hosting stars. As suggested by \citet{kovacs15}, the differences arising between field stars and OC members may be due to environmental effects (e.g. a different strength of the magnetic field, density of the interstellar medium, or SPI).

We employed \textsc{Pasta}'s results to look for correlations between the planetary initial atmospheric mass fraction and semi-major axis, planetary mass, and stellar mass. We did not find any statistically significant correlation, partially due to the small number of analysed systems but mostly due to the large uncertainties on the values obtained for the initial atmospheric mass fractions. However, the results indicate that \textsc{Pasta} is better at constraining the initial atmospheric mass fraction of higher-mass planets orbiting farther away from the host star, which is likely due to the slower atmospheric evolution of these planets compared to that of lower-mass and closer-in planets that are subject to more intense stellar irradiation. Although not statistically significant, our results hint at the possible presence of a negative correlation of the initial atmospheric mass fraction with planetary mass and of a positive correlation with stellar mass. The latter, in particular, would agree with predictions by planetary atmospheric accretion models \citep{kennedy2008a,kennedy2008b,kennedy2009,lozovsky2021}.

This work represents a first step towards extracting information about the initial stages of the evolution of single planetary systems on the basis of the currently observed system parameters, further appropriately accounting for the observational uncertainties. Primarily, the algorithm requires precise and accurate parameters for a large number of systems. This is the goal of a number of ground- and space-based facilities. In particular, the TESS and CHEOPS missions, in conjunction with ground-based high-resolution spectrographs, are providing measurements of the required quality for a number of systems that will enable us in the near future to significantly enlarge the sample size. In the future, the PLATO mission will enable us to significantly increase the sample size, still providing the required accuracy on the system parameters. In parallel, we will aim at further improving the models behind \textsc{Pasta} by adding physics (e.g. the inclusion of elements in addition to hydrogen, a self-consistent calculation of the heating efficiency in the escape models, non-solar composition stellar evolutionary tracks, and the evolution of the orbital separation) and thus reliability to the results.

\begin{acknowledgements}
DK received funding from the European Research Council (ERC) under the European Union’s Horizon 2020 research and innovation programme (grant agreement No 817540, ASTROFLOW). We thank contributors to the Python Programming Language and the free
and open-source community, including: \textsc{MC3} \citep{cubillos17}, \textsc{Numpy} \citep{vanderWaltEtal2011numpy}, \textsc{SciPy} \citep{JonesEtal2001scipy}, \textsc{Matplotlib} \citep{Hunter2007ieeeMatplotlib}, \textsc{Corner} \citep{foremanMackey16corner}.
\end{acknowledgements}


\bibliographystyle{aa} % style aa.bst
\bibliography{biblio} % your references Yourfile.bib

\begin{appendix}
\section{Plots of the set priors and obtained posteriors for each considered system}\label{app:plots}
After the detailed discussion commenting the results of the K2-285 system in Sect.~\ref{sec:results}, we present here plots showing the outputs of \textsc{Pasta} referring to all the other considered systems.
In general, the posteriors of the stellar parameters agree well with their priors for all analysed systems. In a few cases, namely for KOI-94 and Kepler-48, the posterior and prior for the $q_{\mathrm{SF}}$ and $m_{\mathrm{SF}}$ coefficients of Eq.~(\ref{eq:L_EUV}) are in tension (last row of Figs.~\ref{fig:KOI-94star} and \ref{fig:Kepler-48star}). However, the posterior estimates are $(\hat{q}_{\mathrm{SF}}; \hat{m}_{\mathrm{SF}})=(3.71_{-1.58}^{+1.57}; 0.820\pm0.057)$ and $(6.53_{-1.56}^{+1.61}; 0.926_{-0.057}^{+0.058})$ for KOI-94 and Kepler-48, respectively, which differ less than 1$\sigma$ from their respective reference values. We remark that, on the one hand, combining together the results coming from several systems leads to a good agreement between the posterior and prior of the empirically derived conversion coefficients, which confirms the global consistency of \textsc{Pasta}'s framework (see discussion in Sect.~\ref{sec:gyroExp}). On the other hand, on a star-by-star basis $\hat{q}_{\mathrm{SF}}$ and $\hat{m}_{\mathrm{SF}}$ may occasionally differ from their empirical counterpart, which stresses the importance of giving enough degrees of freedom to the framework.

For all systems but Kepler-11, Kepler-36, and Kepler-48, there is always one planet (two in the case of Kepler-411 and KOI-94) whose present-day atmospheric and core mass impose a strong constraint on the atmospheric evolution so that the $\fas$-PDF is well defined. As already seen with K2-285, the atmospheric modelling had the positive side effect of constraining the masses of some of the other planets (e.g. $M_{b,\mathrm{K-18}}$ of Kepler-18, Fig.~\ref{fig:Kepler-18planets}, leftmost panel of the second row; $M_{e,\mathrm{K-20}}$ and $M_{f,\mathrm{K-20}}$ of Kepler-20, Fig.~\ref{fig:Kepler-20planets}, two rightmost panels of the second row). We theoretically estimated $\hat{M}_{b,\mathrm{K-18}}=7.9_{-1.1}^{+1.2}\,M_{\oplus}$ (reducing the uncertainties by a factor of $\sim$\,3), and $\hat{M}_{e,\mathrm{K-20}}=0.650_{-0.061}^{+0.062}\,M_{\oplus}$ and $\hat{M}_{f,\mathrm{K-20}}=1.03_{-0.20}^{+0.22}$ (improving from the upper limits given as priors).

As emphasised in Sect.~\ref{sec:rotperiods}, the host stars considered in this work were generally slow rotators when they were young, if compared to their OC counterparts, except for Kepler-411 whose median $\bar{P}_{\mathrm{rot,150}}=4.0_{-2.7}^{+3.5}$ days is slightly lower than the comparison sample, which has $\bar{P}_{\mathrm{rotJ,150}}=5.2_{-4.7}^{+3.2}$. As a consequence of being a moderately fast rotator at 150 Myr, when young Kepler-411 was likely in the $L_X$ saturation regime, hence the posterior exhibits a peak at high $L_X$ values ($\log{L_X}$\,$\sim$\,29.5, Fig.~ \ref{fig:Kepler-411star}, leftmost panel of the second row) that is much more pronounced than the corresponding peaks obtained for the other stars.

Finally, Kepler-11 deserves a special, separated comment. The star hosts six planets (from b to g) for which there is no general consensus about their masses in the literature. Therefore, we merged different probability distributions according to the results published by several authors to produce the reference priors for the planetary masses (see Sect.~\ref{sec:sample}). 
For each planet with a non-flat mass prior (b to f), our posterior theoretical estimates $\hat{M}_p$ differ less than 1.5$\sigma$ from our adopted prior modes (Fig.~\ref{fig:Kepler-11planets}, second row), mostly as a result of the large prior uncertainties especially on $M_b$ and $M_c$. We also note that the $\hat{M}_p$ values differ less than 1$\sigma$ from the corresponding values proposed by \citet{hadden14} or \citet{lissauer11}.
The loose constraints we imposed on $M_p$ do not enable \textsc{Pasta} to derive sharply peaked $\fas$-PDFs, except partly for $f_{\mathrm{atm,}g}^{\mathrm{start}}=0.043_{-0.023}^{+0.016}$ (see Fig.~\ref{fig:Kepler-11planets}, last row), which is so distant from its host ($\bar{a}_g=0.4660\pm0.0040$ AU) that it has basically retained its original atmospheric content \citep[see][for a discussion about this]{kubyshkina19AAKepler11}. The different evolutionary scenarios that may have characterised the Kepler-11 system are also reflected by the lack of constraints on the mass of planet g, whose posterior follows the flat prior, thus adding no information. We remark that all the $M_p$ values present in the literature for this system are based on transit time variations, which suffer of degeneracies when used to constrain planetary masses and eccentricities \citep{hadden14}. As this is a challenging measurement, it is not surprising to see some kind of tension between \textsc{Pasta}'s outputs and the adopted priors, and a more accurate study of the evolution of the Kepler-11 system will need to wait until more precise planetary masses become available.

\begin{figure*}
    \resizebox{\hsize}{!}{ \includegraphics{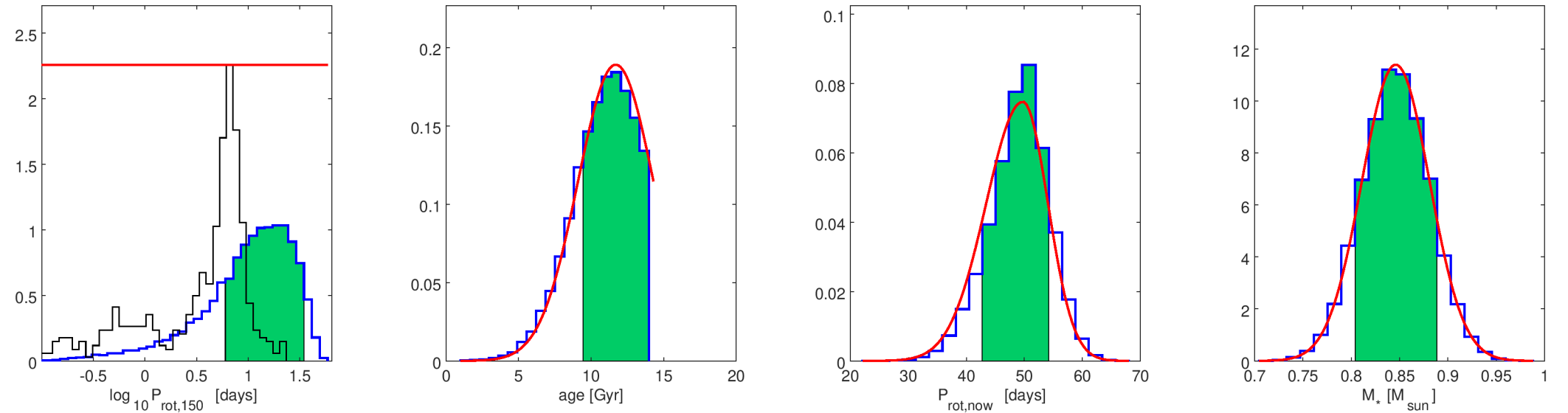}} \\
    \resizebox{\hsize}{!}{ \includegraphics{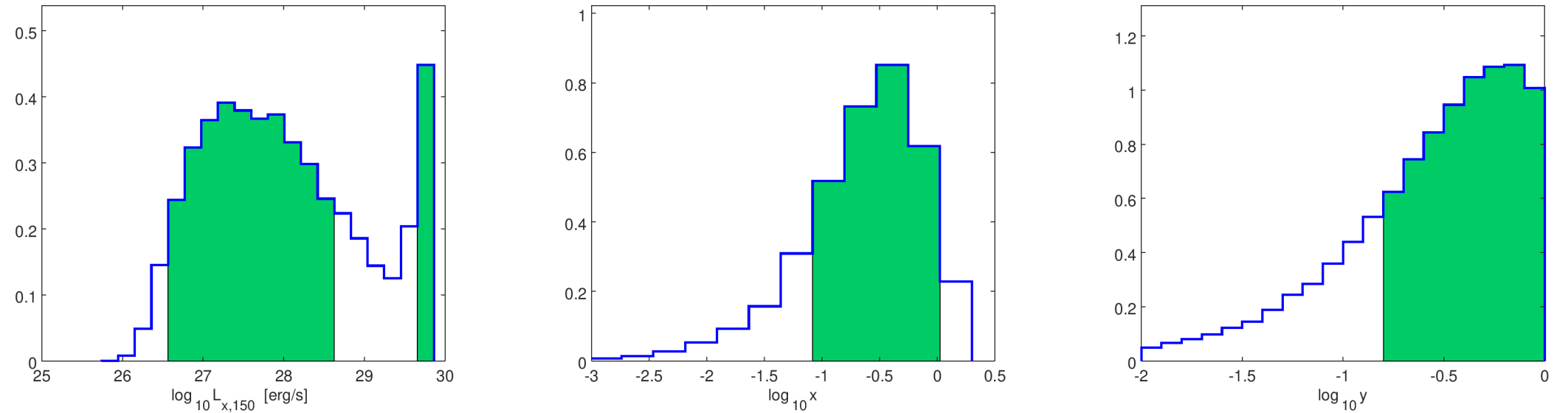}} \\
    \resizebox{\hsize}{!}{ \includegraphics{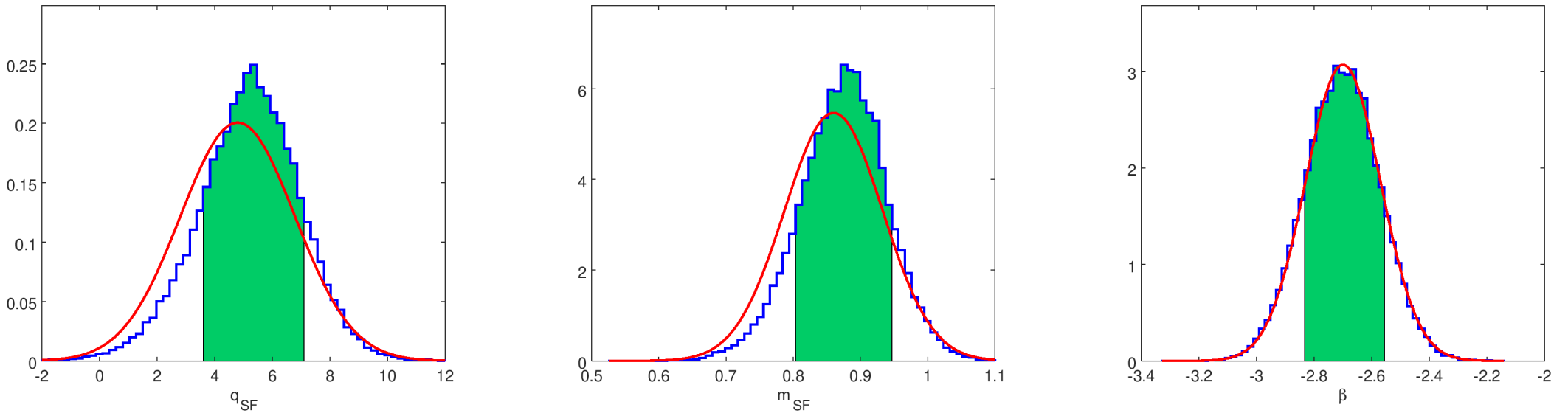}}
    \caption{Same as Fig.~\ref{fig:K2-285star}, but for the star-related properties of HD\,3167.}
    \label{fig:HD3167star}
\end{figure*}
\begin{figure*}
    \resizebox{\hsize}{!}{ \includegraphics{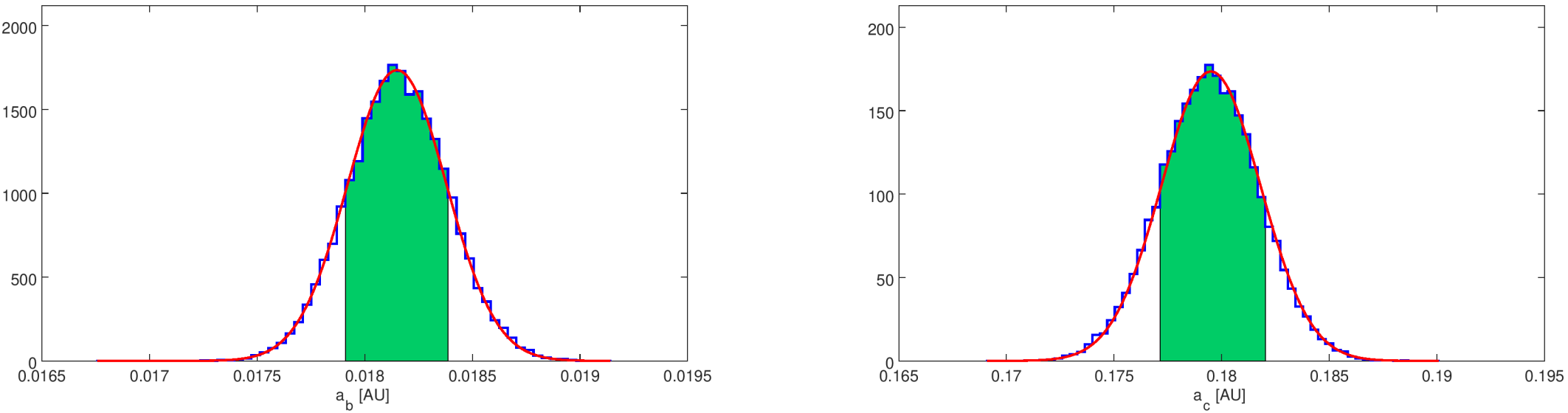}} \\
    \resizebox{\hsize}{!}{ \includegraphics{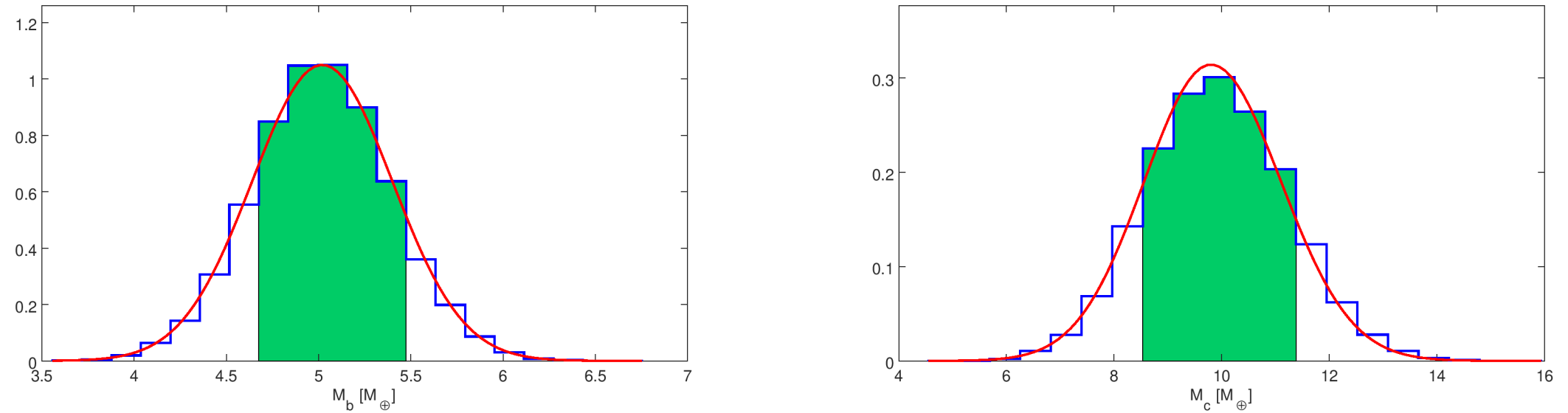}} \\
    \resizebox{\hsize}{!}{ \includegraphics{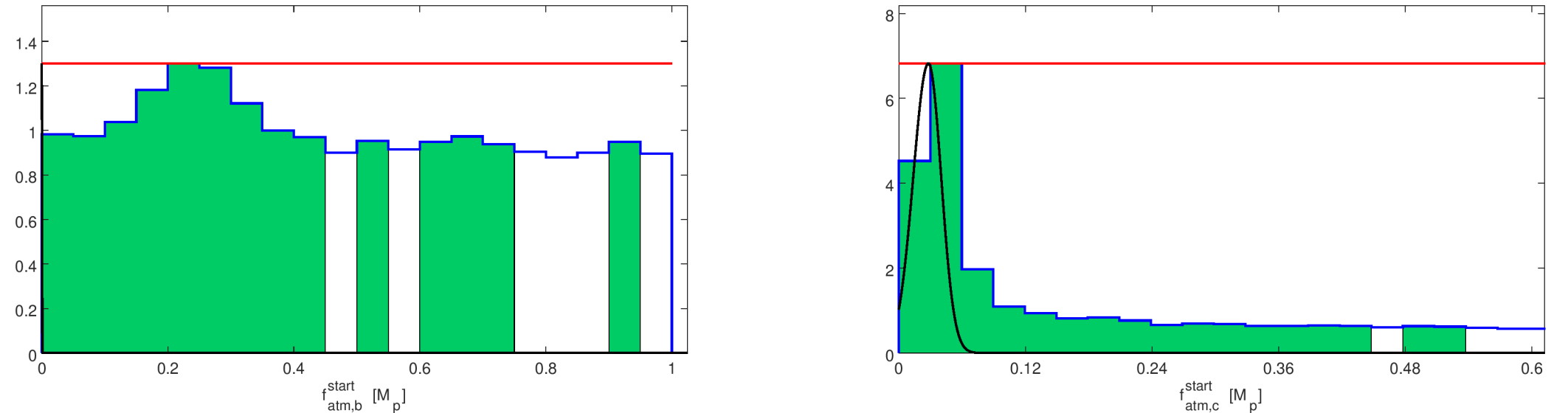}}
    \caption{Same as Fig.~\ref{fig:K2-285planets}, but for the planetary parameters of the HD\,3167 system.}
    \label{fig:HD3167planets}
\end{figure*}

\begin{figure*}
    \resizebox{\hsize}{!}{ \includegraphics{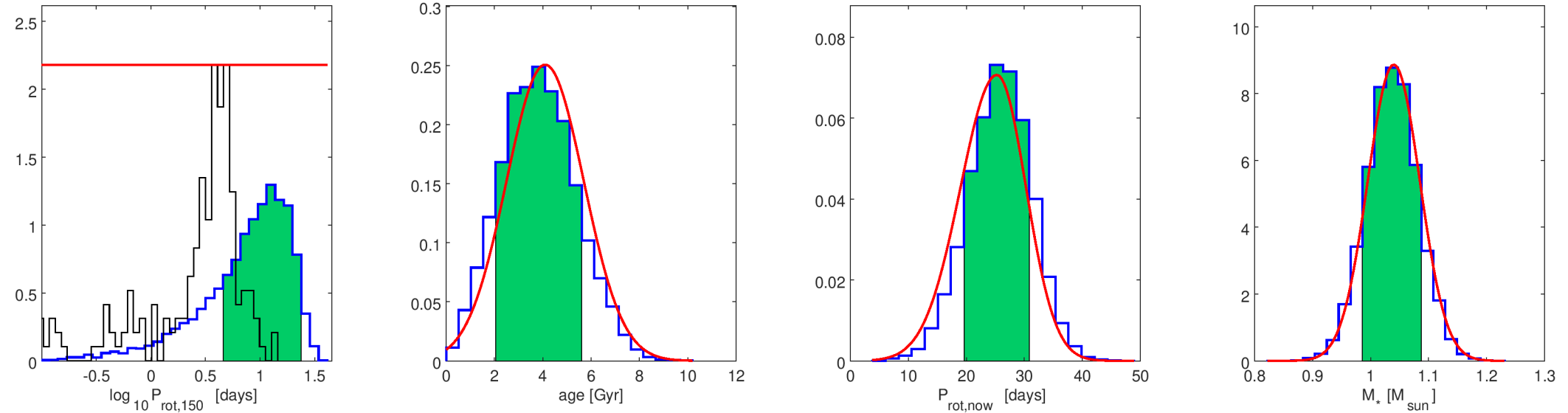}} \\
    \resizebox{\hsize}{!}{ \includegraphics{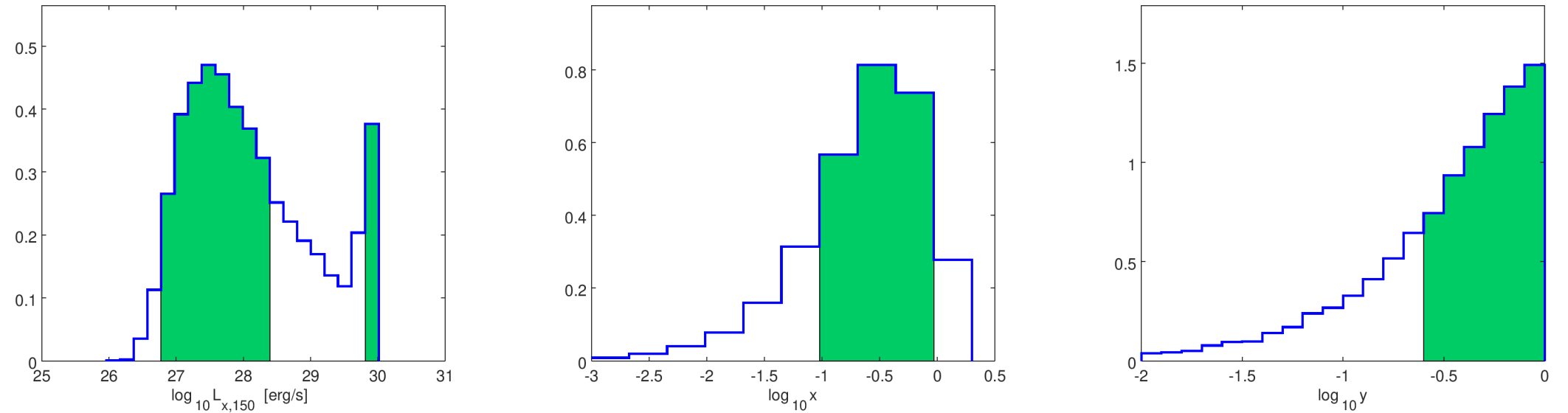}} \\
    \resizebox{\hsize}{!}{ \includegraphics{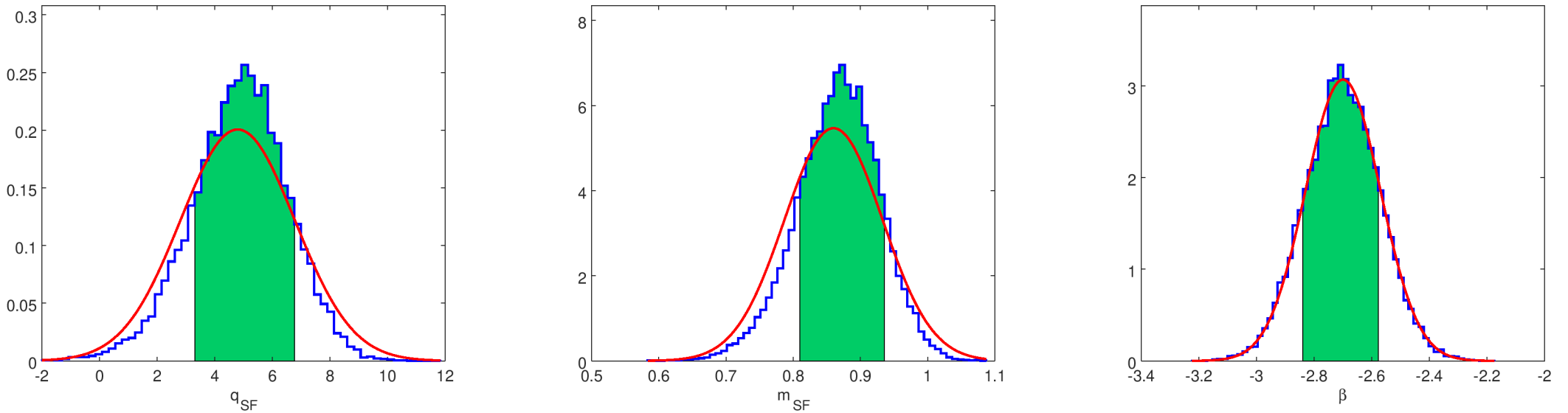}}
    \caption{Same as Fig.~\ref{fig:K2-285star}, but for the star-related properties of K2-24.}
    \label{fig:K2-24star}
\end{figure*}
\begin{figure*}
    \resizebox{\hsize}{!}{ \includegraphics{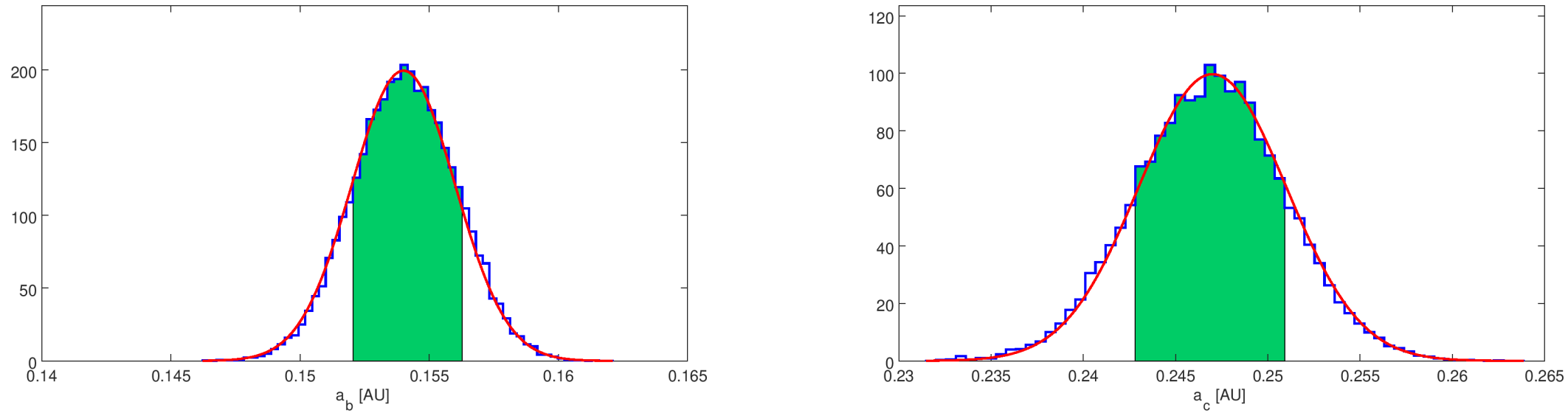}} \\
    \resizebox{\hsize}{!}{ \includegraphics{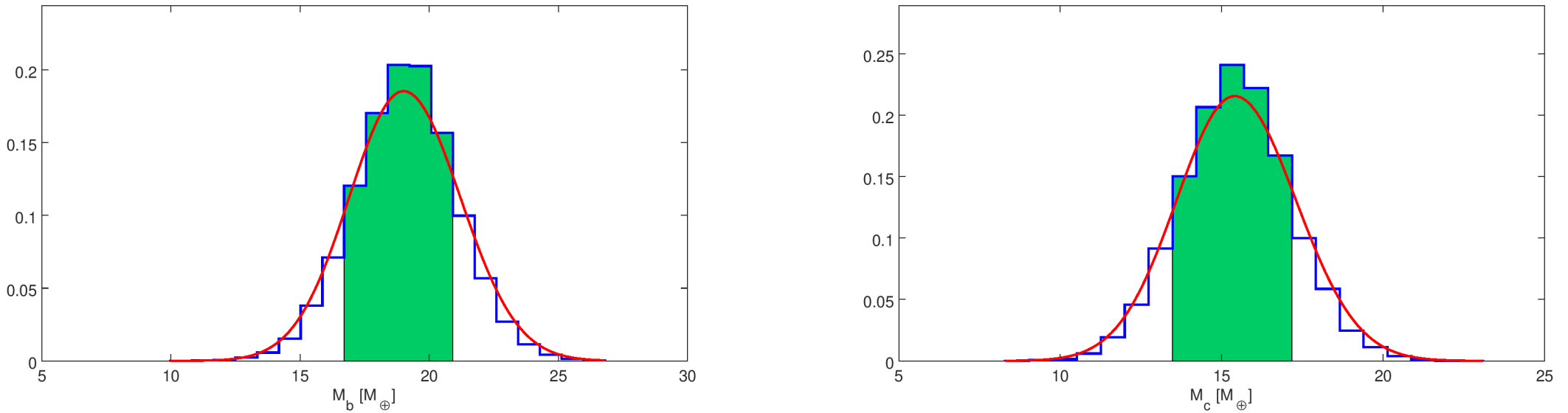}} \\
    \resizebox{\hsize}{!}{ \includegraphics{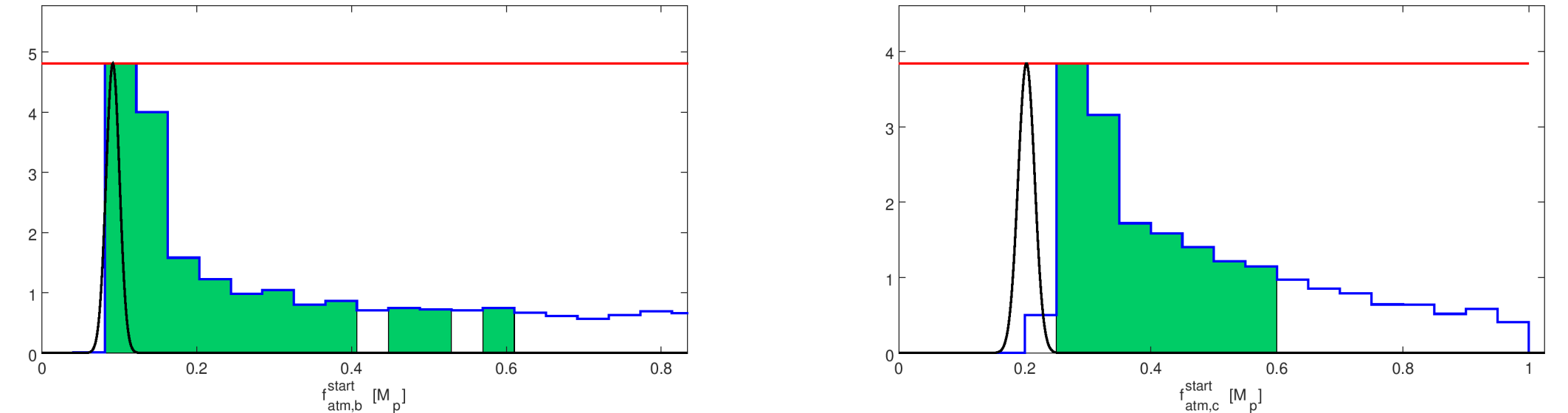}}
    \caption{Same as Fig.~\ref{fig:K2-285planets}, but for the planetary parameters of the K2-24 system.}
    \label{fig:K2-24planets}
\end{figure*}

\begin{figure*}
    \resizebox{\hsize}{!}{ \includegraphics{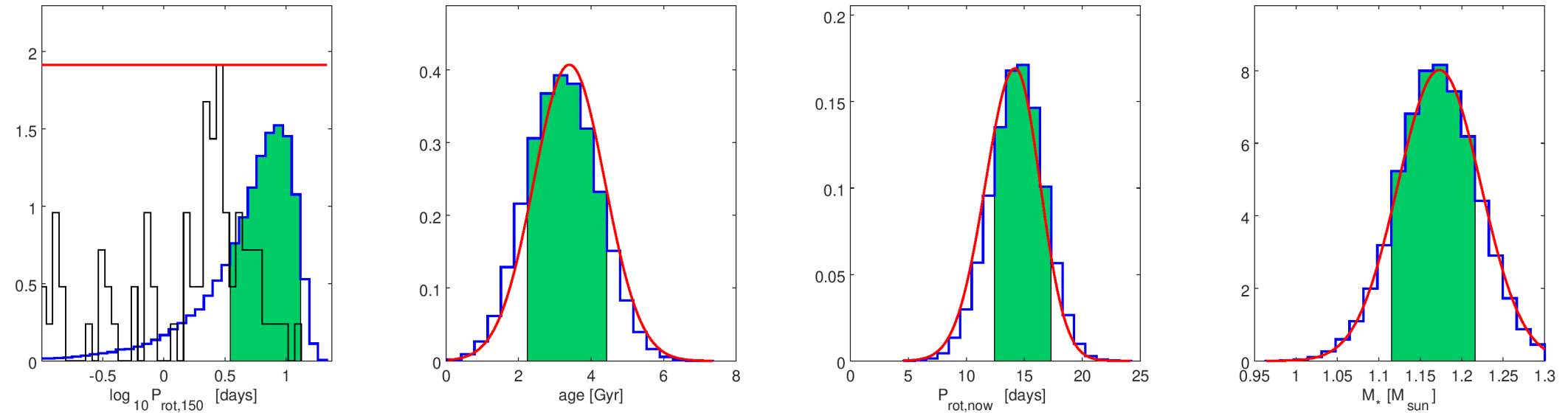}} \\
    \resizebox{\hsize}{!}{ \includegraphics{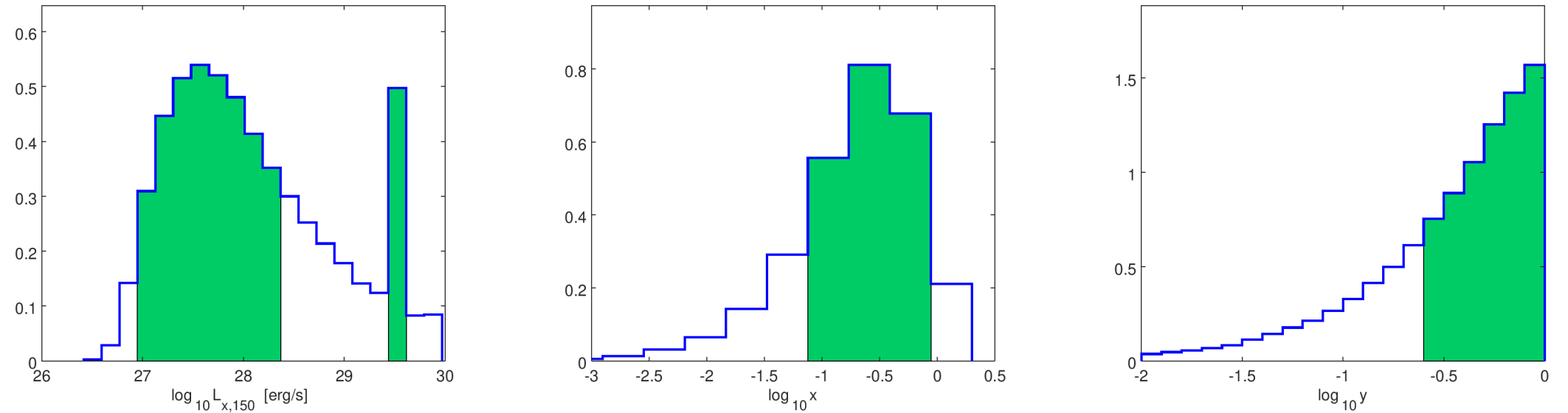}} \\
    \resizebox{\hsize}{!}{ \includegraphics{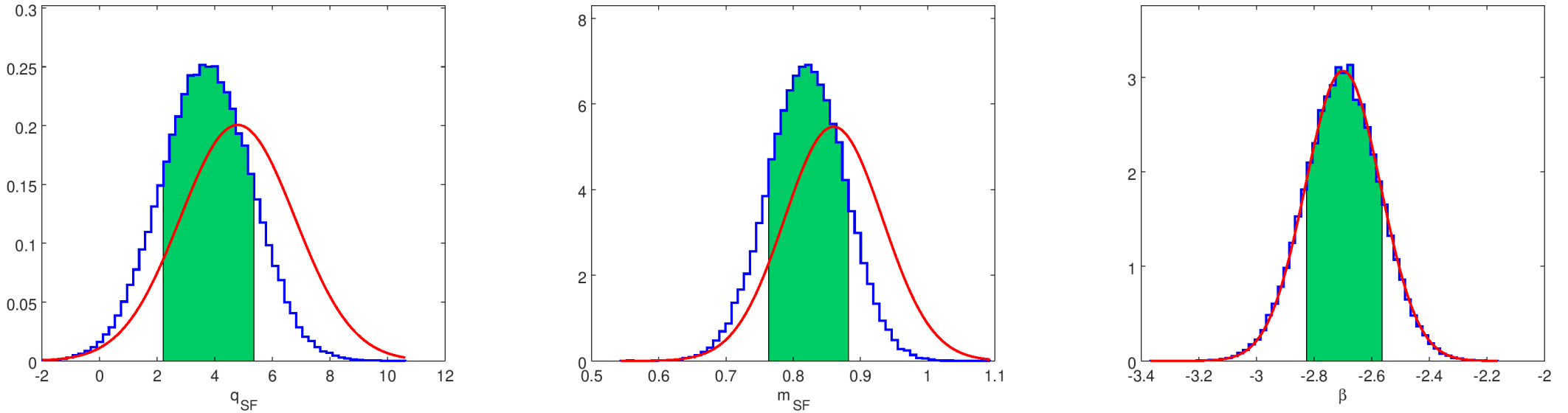}}
    \caption{Same as Fig.~\ref{fig:K2-285star}, but for the star-related properties of KOI-94.}
    \label{fig:KOI-94star}
\end{figure*}
\begin{figure*}
    \resizebox{\hsize}{!}{ \includegraphics{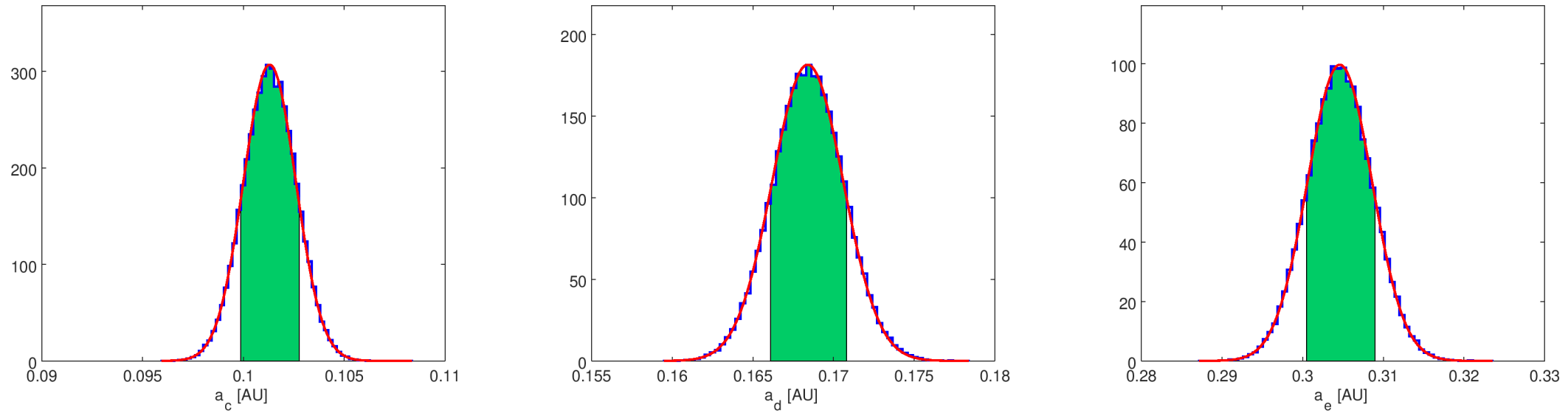}} \\
    \resizebox{\hsize}{!}{ \includegraphics{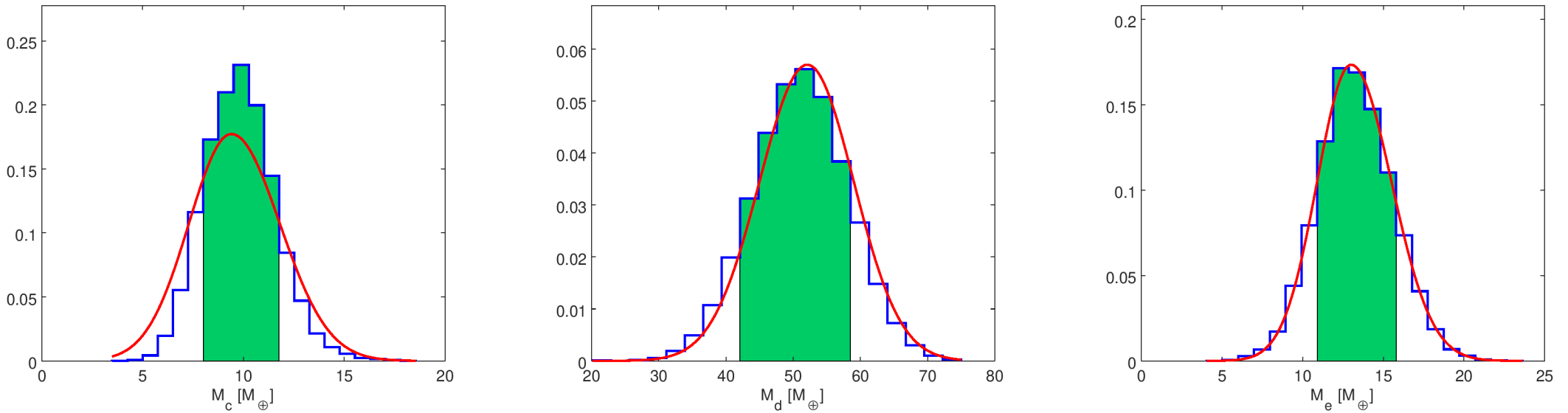}} \\
    \resizebox{\hsize}{!}{ \includegraphics{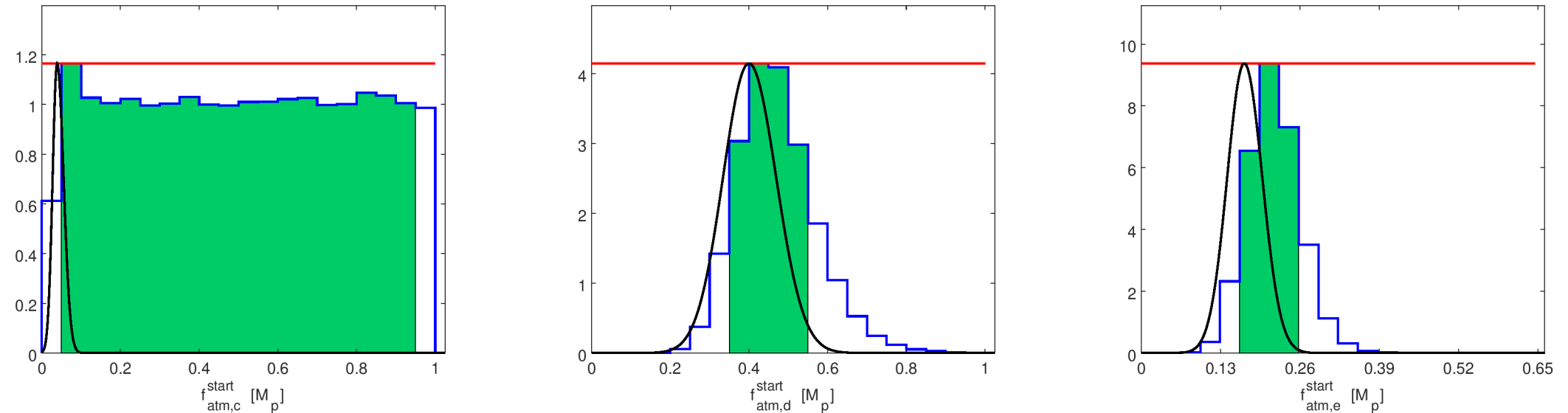}}
    \caption{Same as Fig.~\ref{fig:K2-285planets}, but for the planetary parameters of the KOI-94 system.}
    \label{fig:KOI-94planets}
\end{figure*}

\begin{figure*}
    \resizebox{\hsize}{!}{ \includegraphics{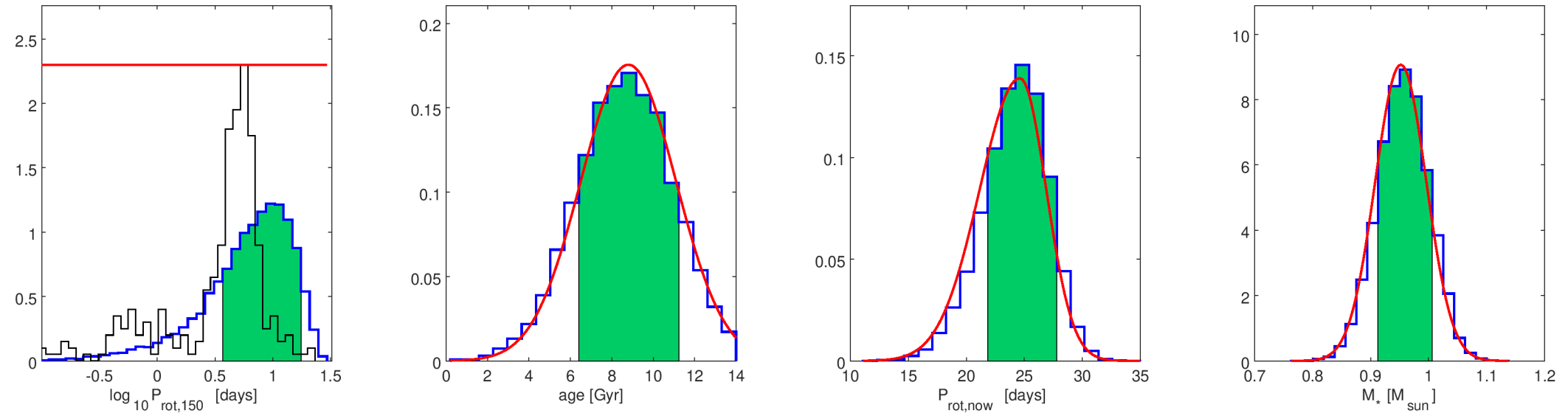}} \\
    \resizebox{\hsize}{!}{ \includegraphics{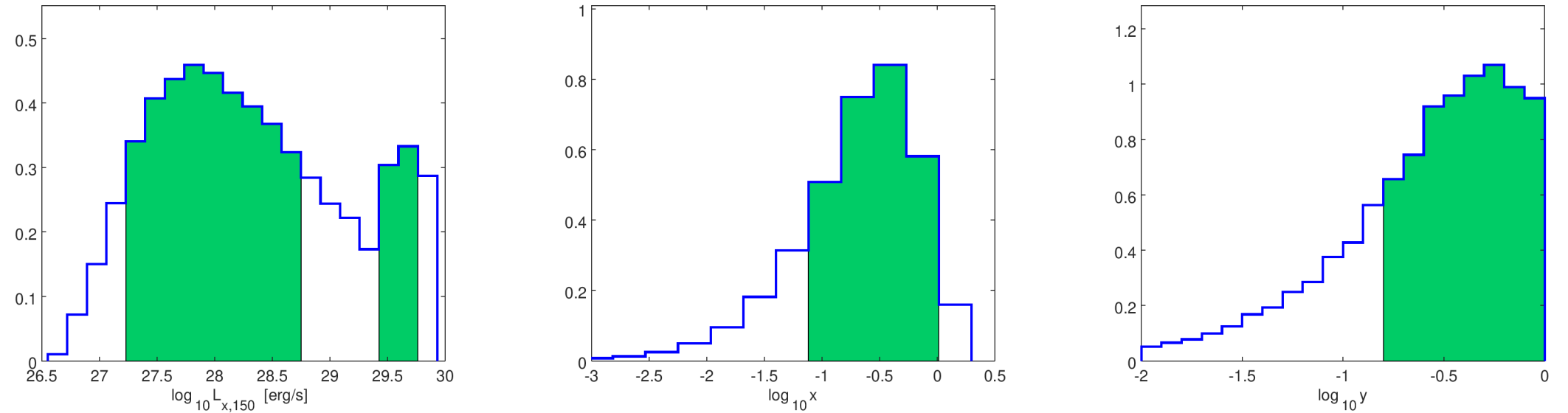}} \\
    \resizebox{\hsize}{!}{ \includegraphics{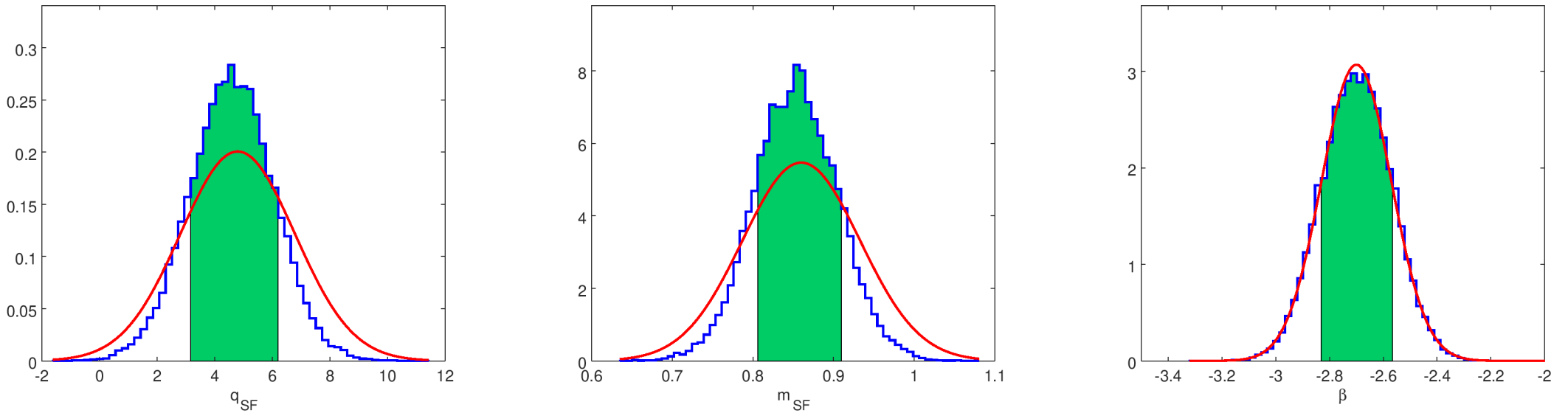}}
    \caption{Same as Fig.~\ref{fig:K2-285star}, but for the star-related properties of Kepler-11.}
    \label{fig:Kepler-11star}
\end{figure*}
\begin{figure*}
    \resizebox{\hsize}{!}{ \includegraphics{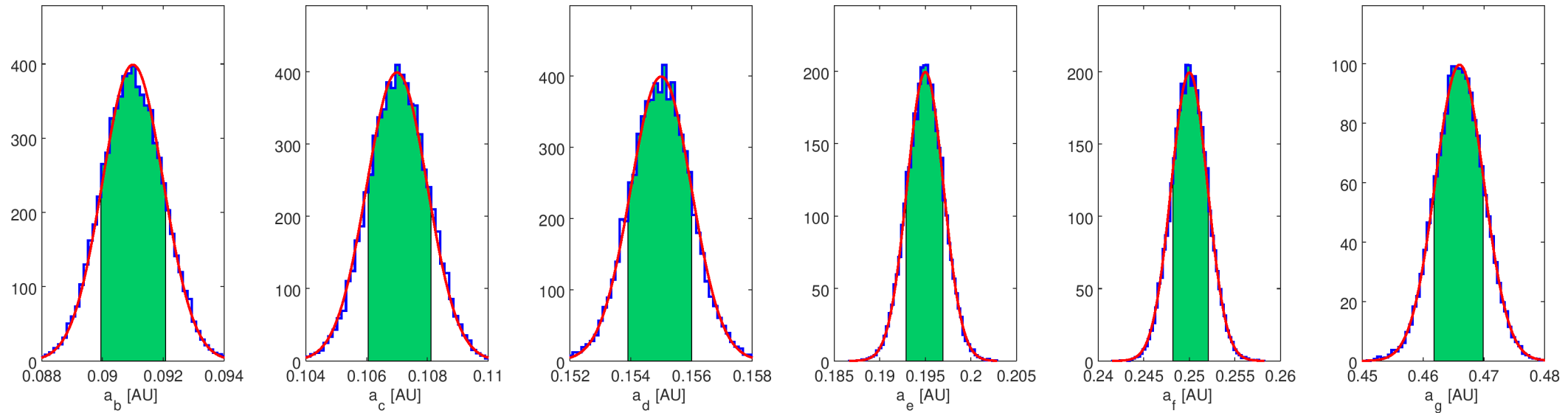}} \\
    \resizebox{\hsize}{!}{ \includegraphics{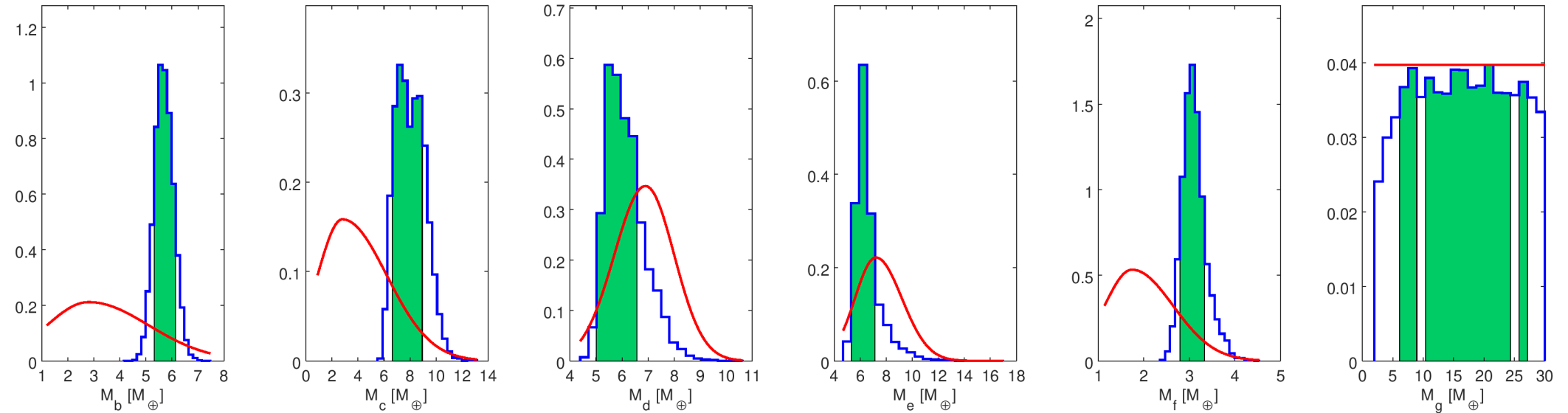}} \\
    \resizebox{\hsize}{!}{ \includegraphics{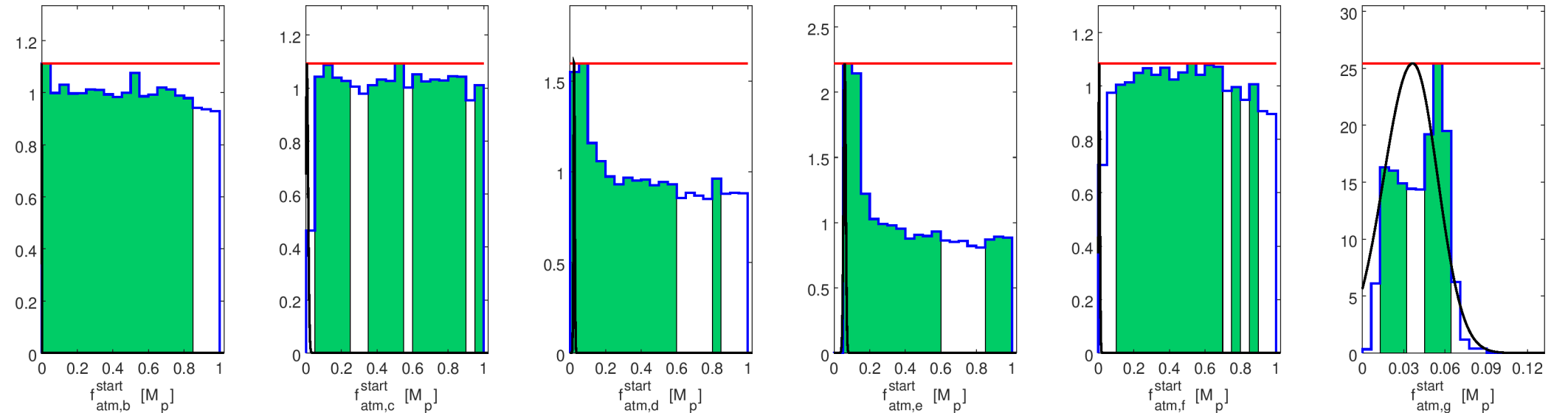}}
    \caption{Same as Fig.~\ref{fig:K2-285planets}, but for the planetary parameters of the Kepler-11 system.}
    \label{fig:Kepler-11planets}
\end{figure*}

\begin{figure*}
    \resizebox{\hsize}{!}{ \includegraphics{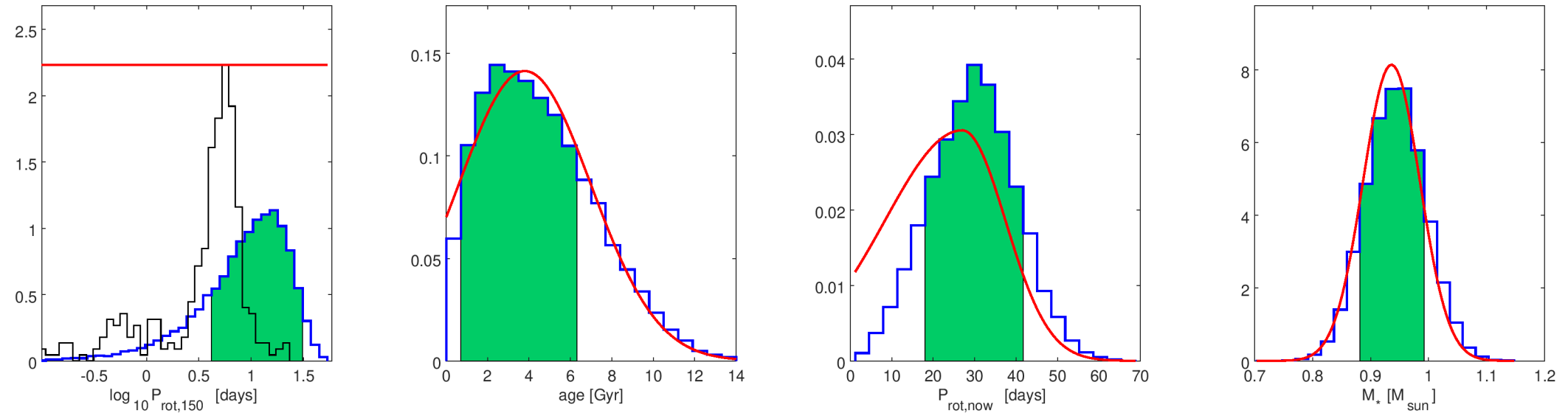}} \\
    \resizebox{\hsize}{!}{ \includegraphics{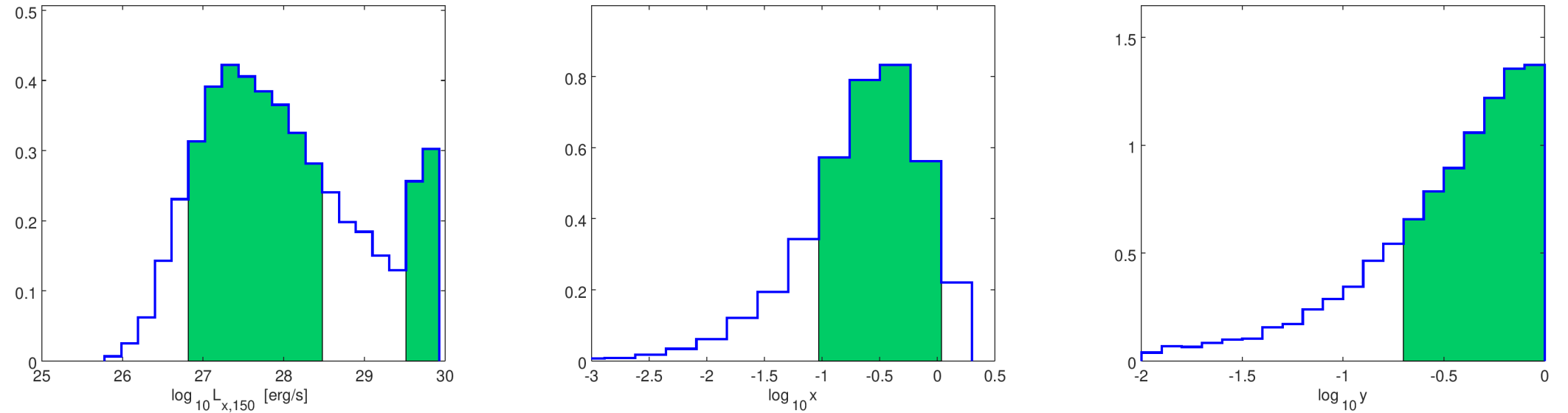}} \\
    \resizebox{\hsize}{!}{ \includegraphics{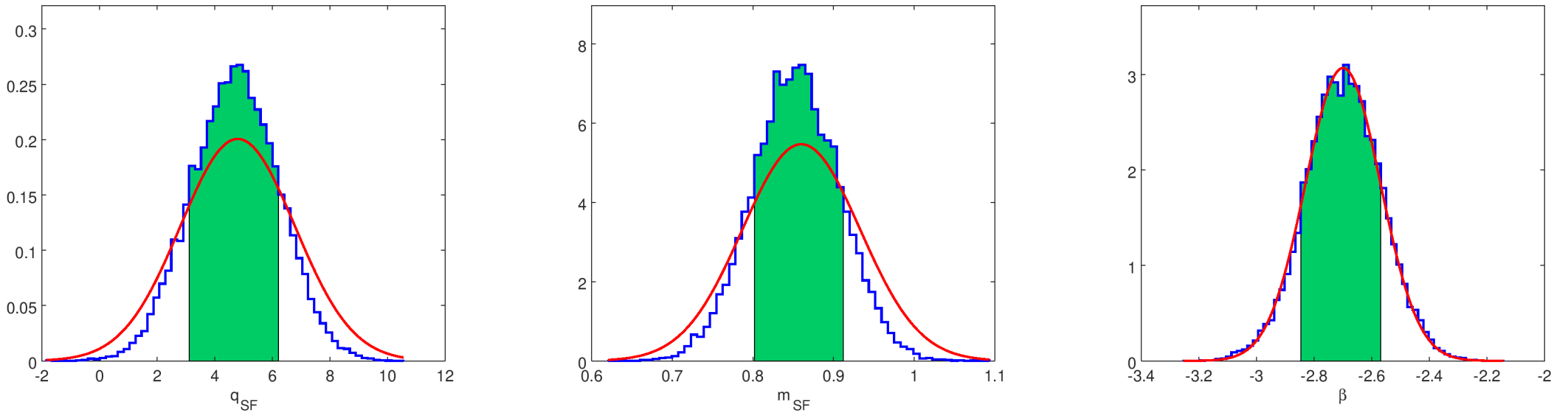}}
    \caption{Same as Fig.~\ref{fig:K2-285star}, but for the star-related properties of Kepler-18.}
    \label{fig:Kepler-18star}
\end{figure*}
\begin{figure*}
    \resizebox{\hsize}{!}{ \includegraphics{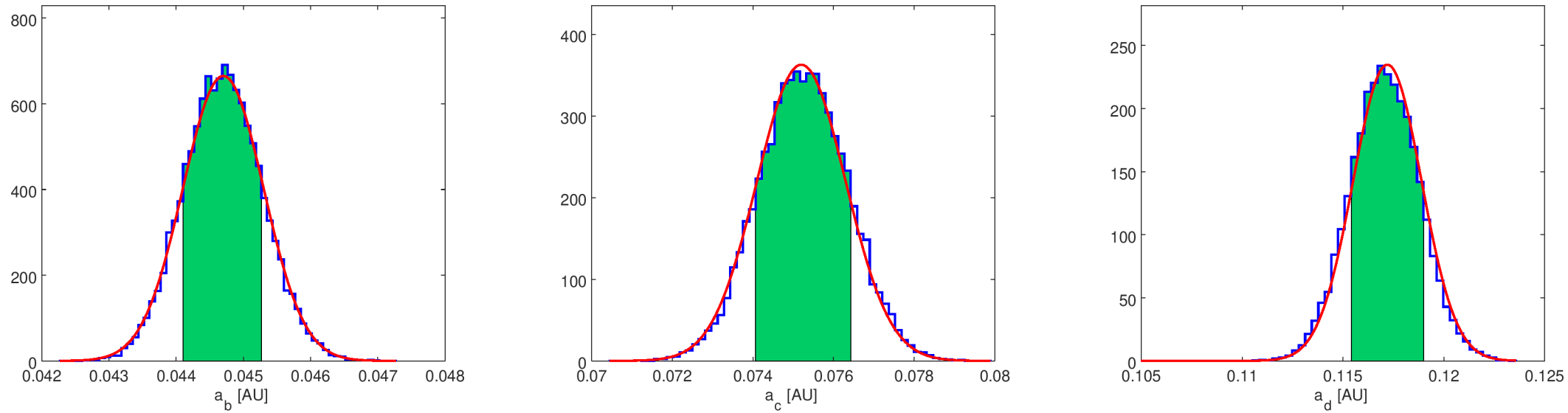}} \\
    \resizebox{\hsize}{!}{ \includegraphics{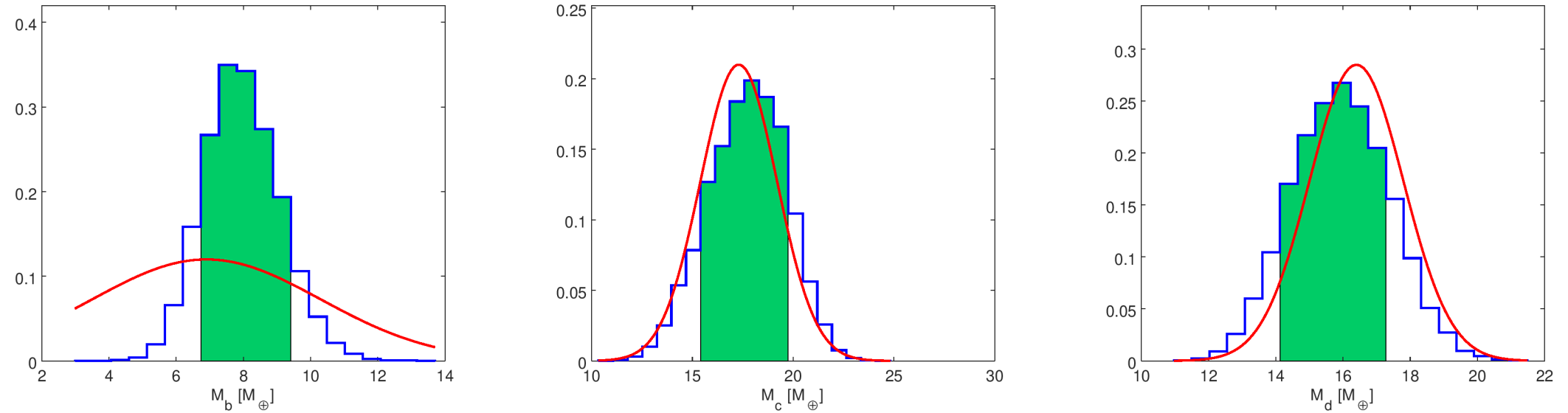}} \\
    \resizebox{\hsize}{!}{ \includegraphics{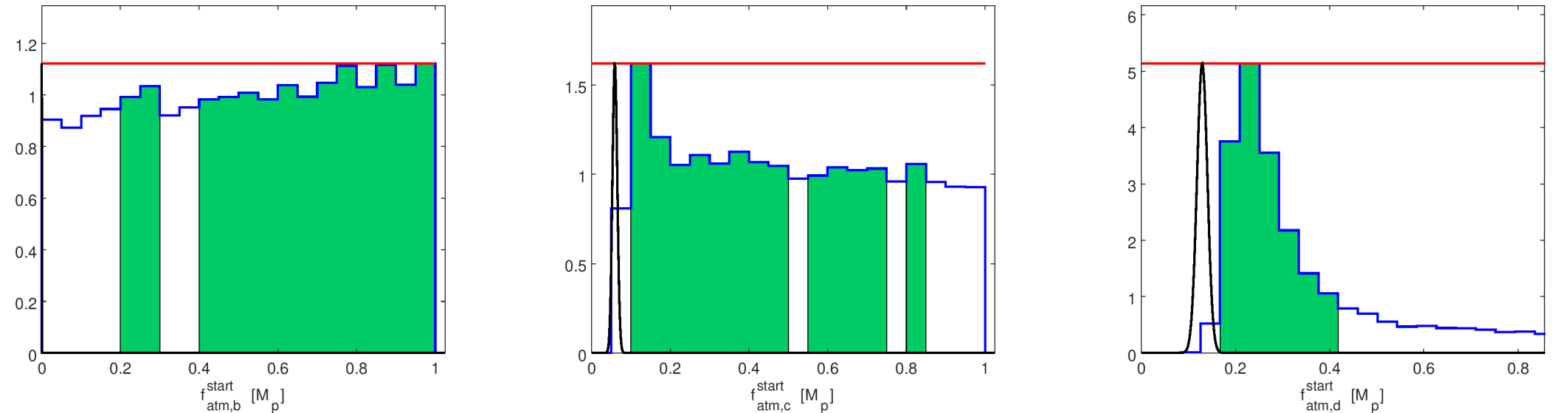}}
    \caption{Same as Fig.~\ref{fig:K2-285planets}, but for the planetary parameters of the Kepler-18 system.}
    \label{fig:Kepler-18planets}
\end{figure*}

\begin{figure*}
    \resizebox{\hsize}{!}{ \includegraphics{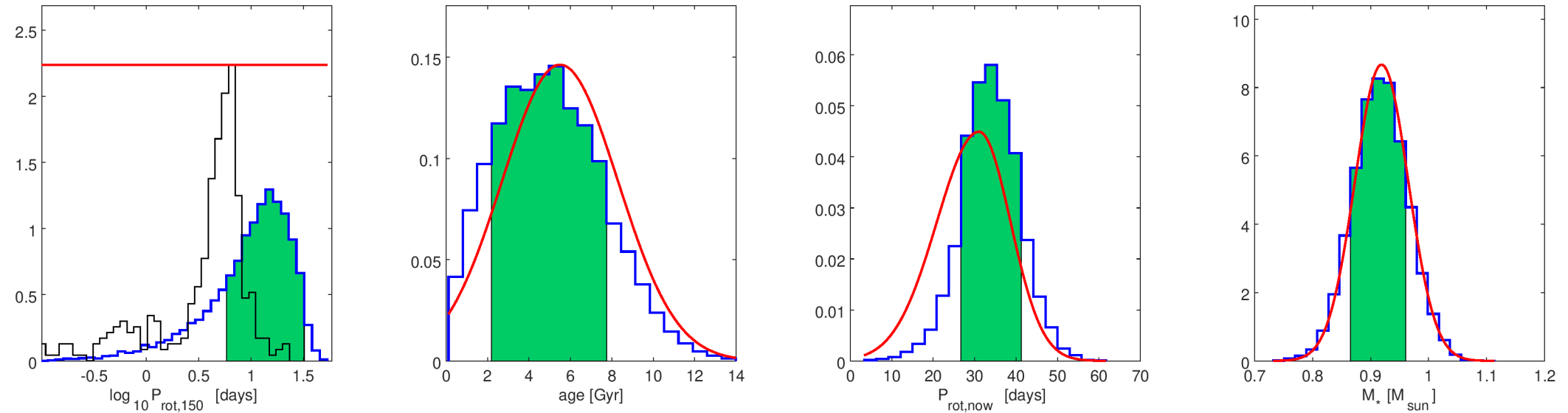}} \\
    \resizebox{\hsize}{!}{ \includegraphics{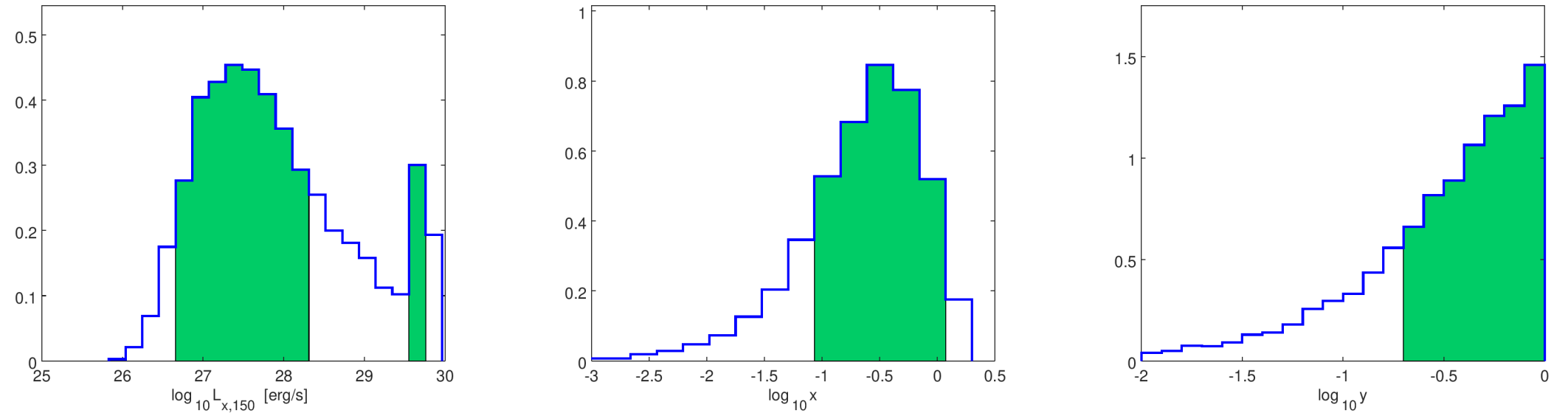}} \\
    \resizebox{\hsize}{!}{ \includegraphics{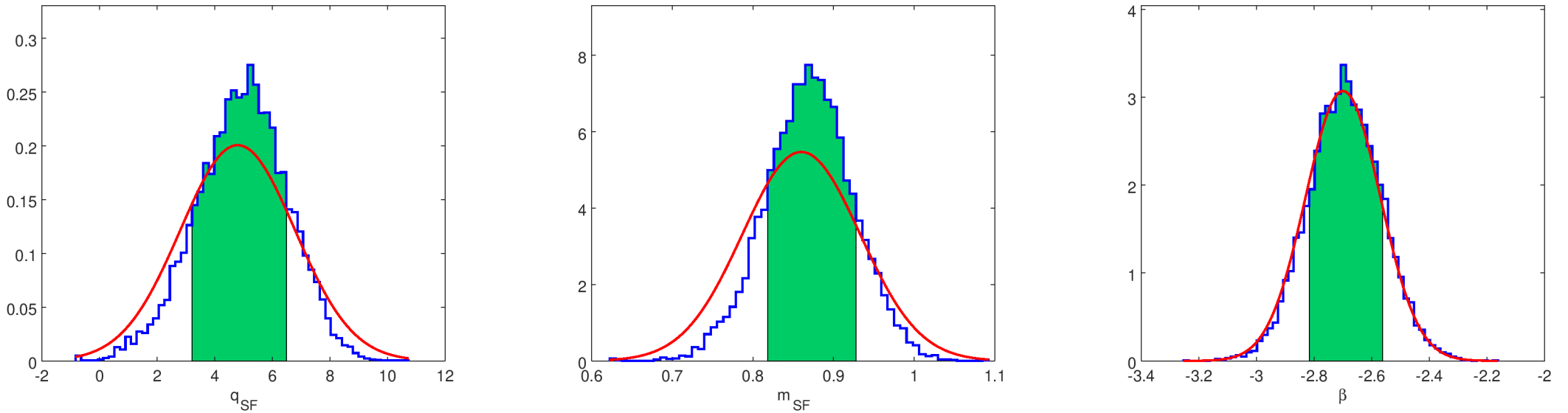}}
    \caption{Same as Fig.~\ref{fig:K2-285star}, but for the star-related properties of Kepler-20.}
    \label{fig:Kepler-20star}
\end{figure*}
\begin{figure*}
    \resizebox{\hsize}{!}{ \includegraphics{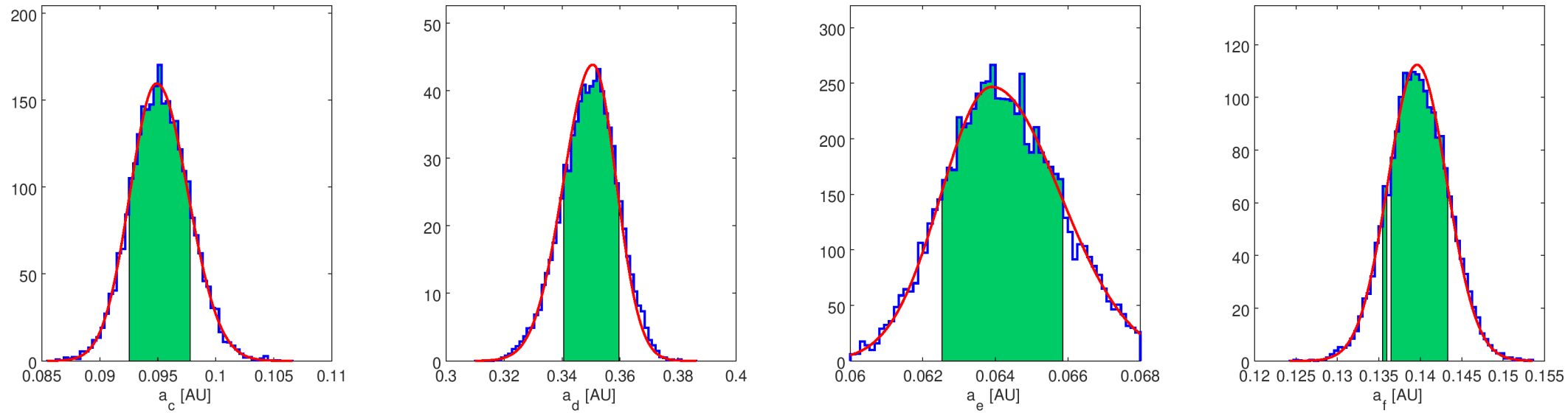}} \\
    \resizebox{\hsize}{!}{ \includegraphics{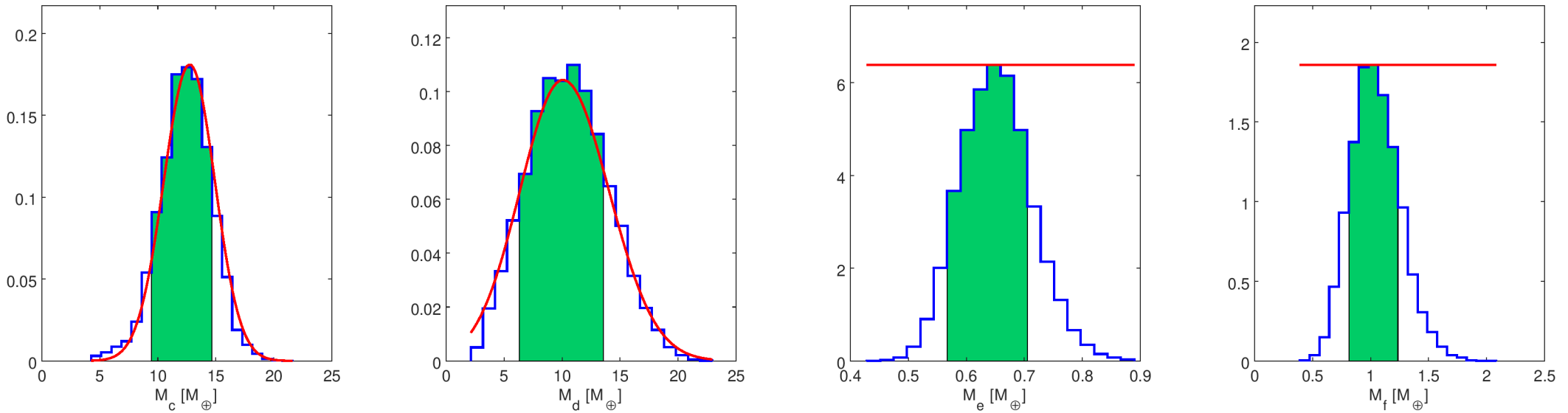}} \\
    \resizebox{\hsize}{!}{ \includegraphics{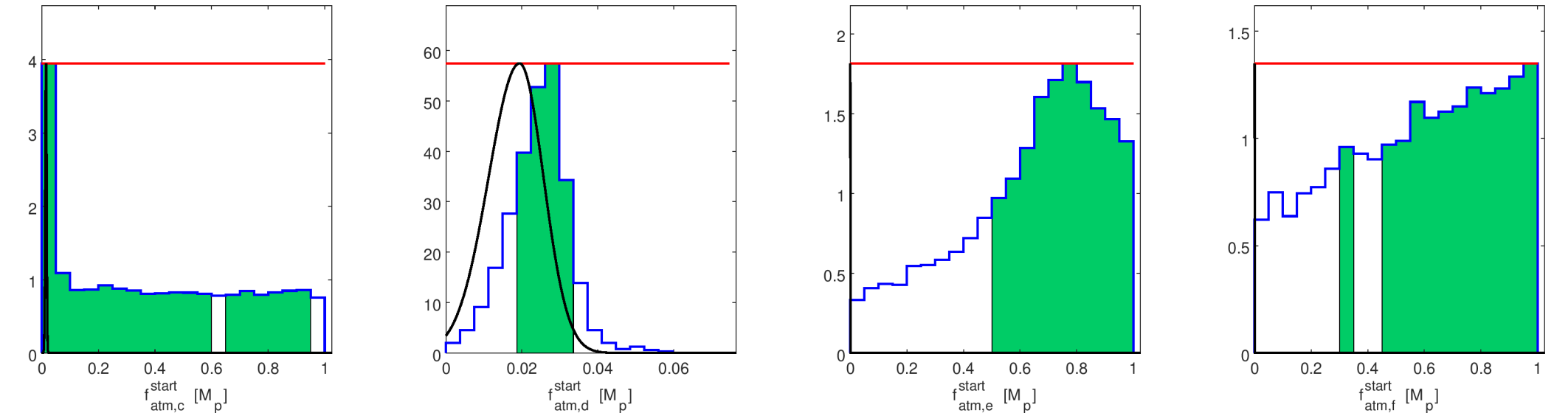}}
    \caption{Same as Fig.~\ref{fig:K2-285planets}, but for the planetary parameters of the Kepler-20 system.}
    \label{fig:Kepler-20planets}
\end{figure*}

\begin{figure*}
    \resizebox{\hsize}{!}{ \includegraphics{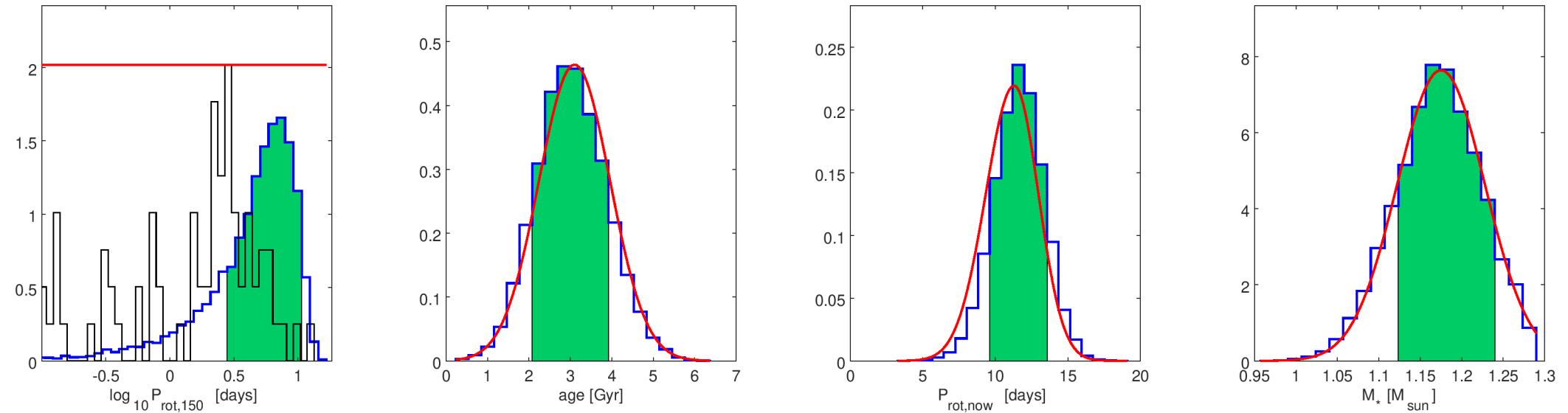}} \\
    \resizebox{\hsize}{!}{ \includegraphics{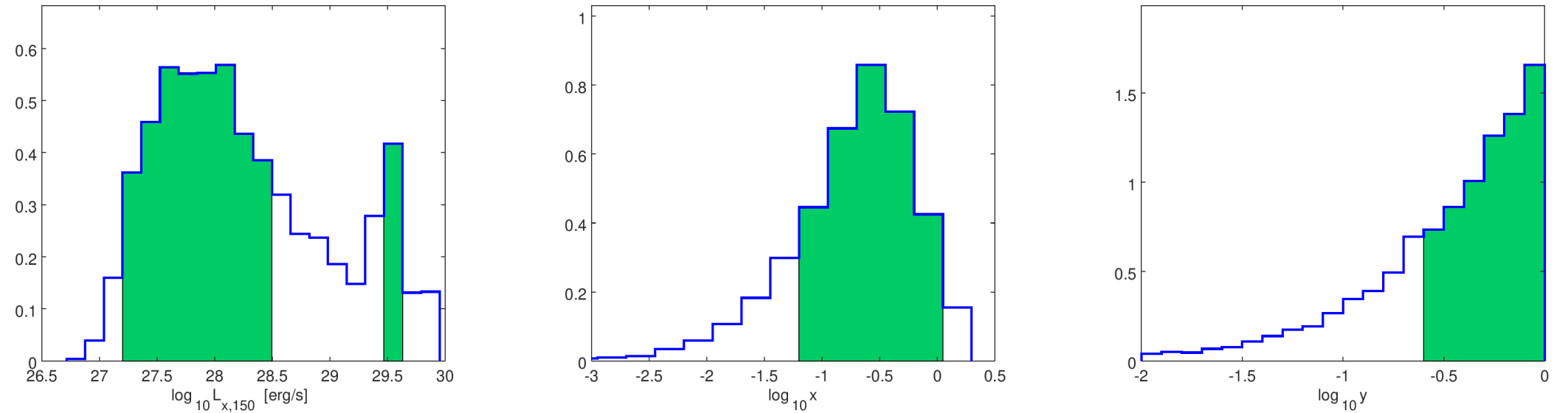}} \\
    \resizebox{\hsize}{!}{ \includegraphics{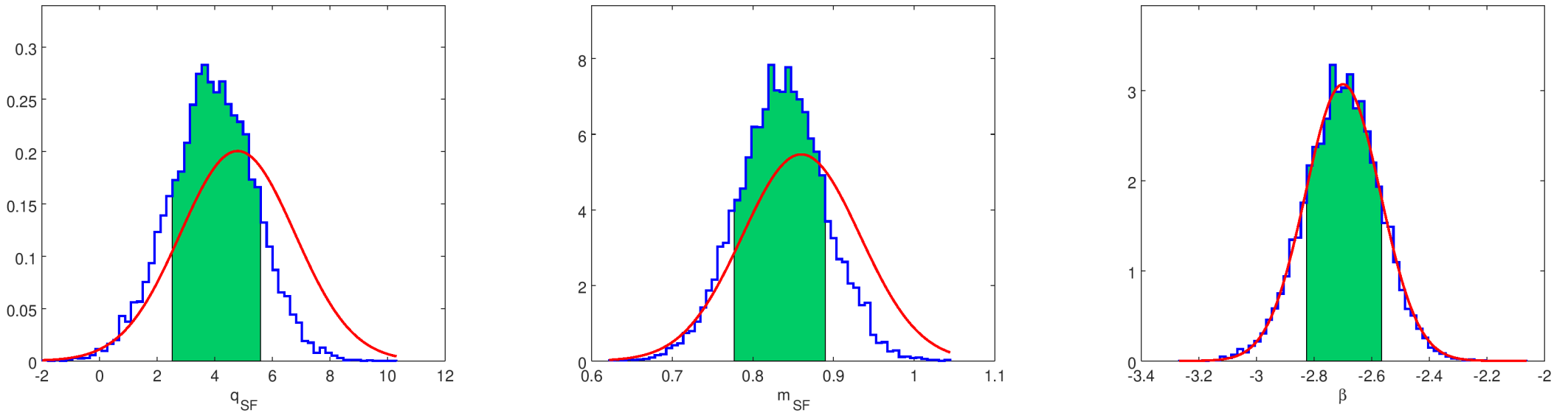}}
    \caption{Same as Fig.~\ref{fig:K2-285star}, but for the star-related properties of Kepler-25.}
    \label{fig:Kepler-25star}
\end{figure*}
\begin{figure*}
    \resizebox{\hsize}{!}{ \includegraphics{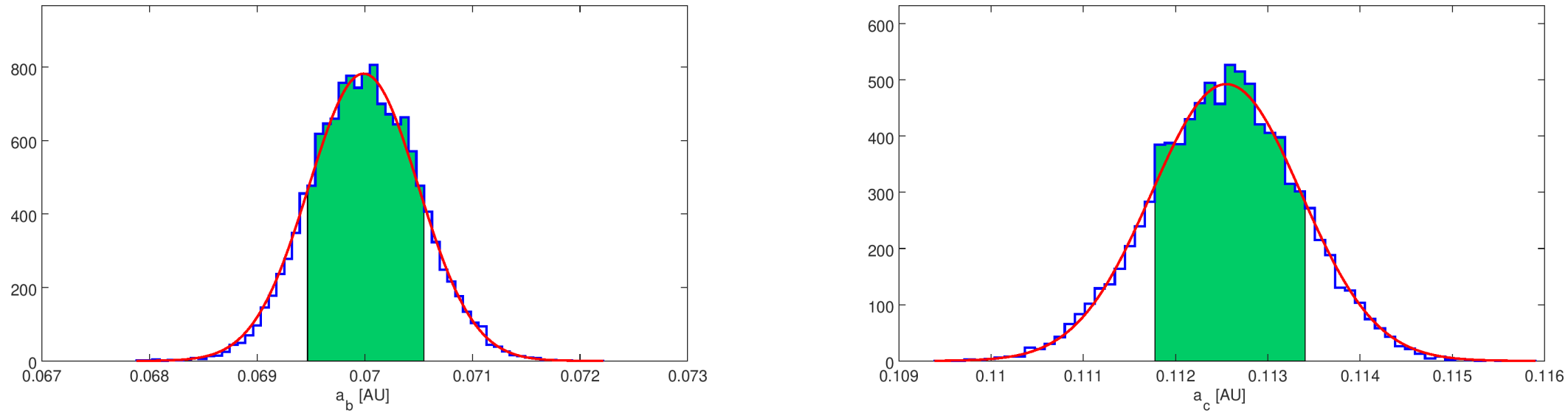}} \\
    \resizebox{\hsize}{!}{ \includegraphics{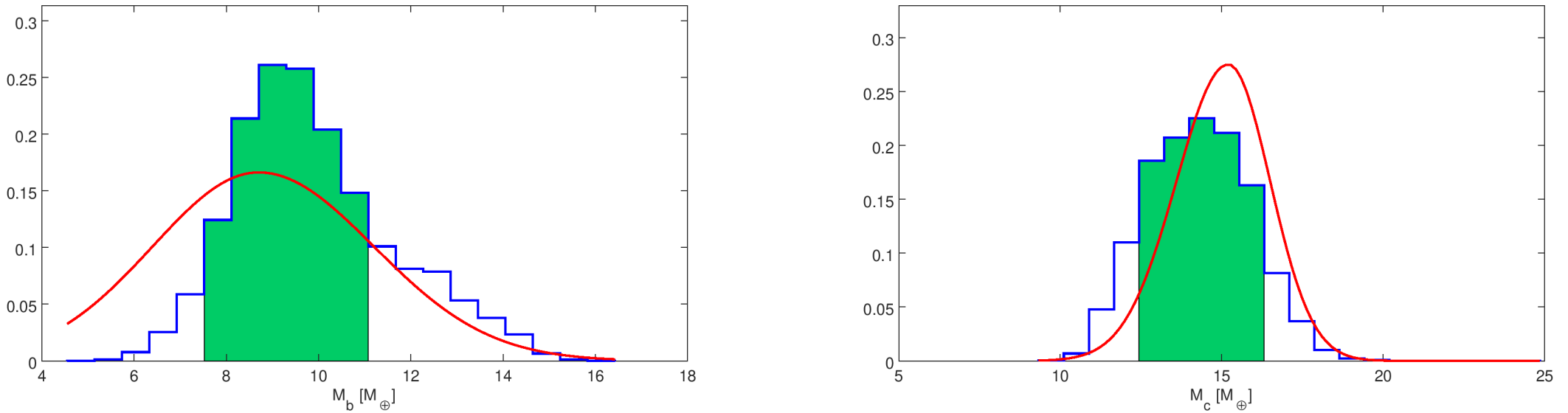}} \\
    \resizebox{\hsize}{!}{ \includegraphics{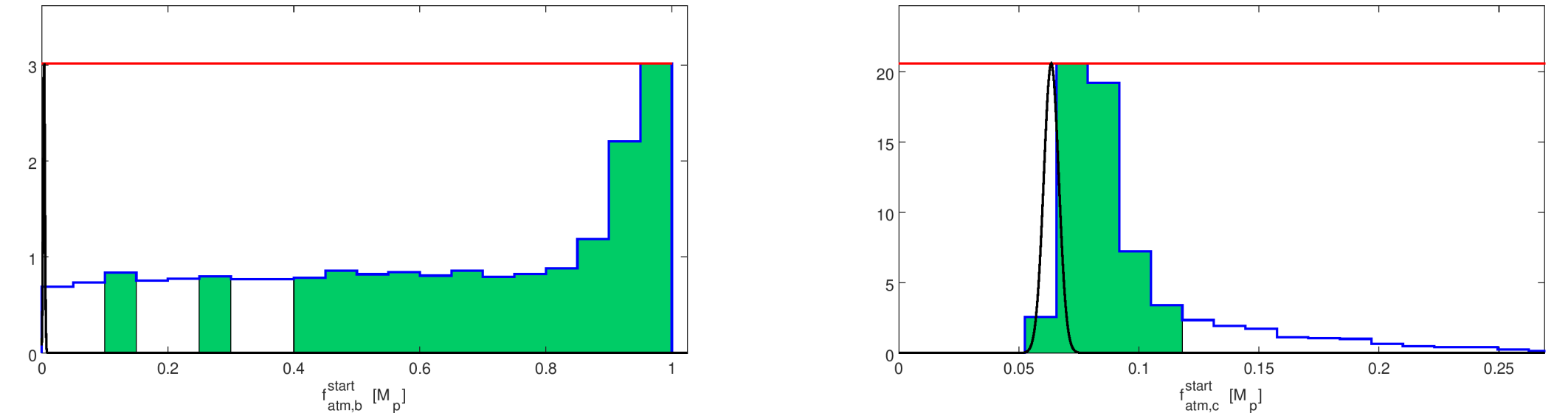}}
    \caption{Same as Fig.~\ref{fig:K2-285planets}, but for the planetary parameters of the Kepler-25 system.}
    \label{fig:Kepler-25planets}
\end{figure*}

\begin{figure*}
    \resizebox{\hsize}{!}{ \includegraphics{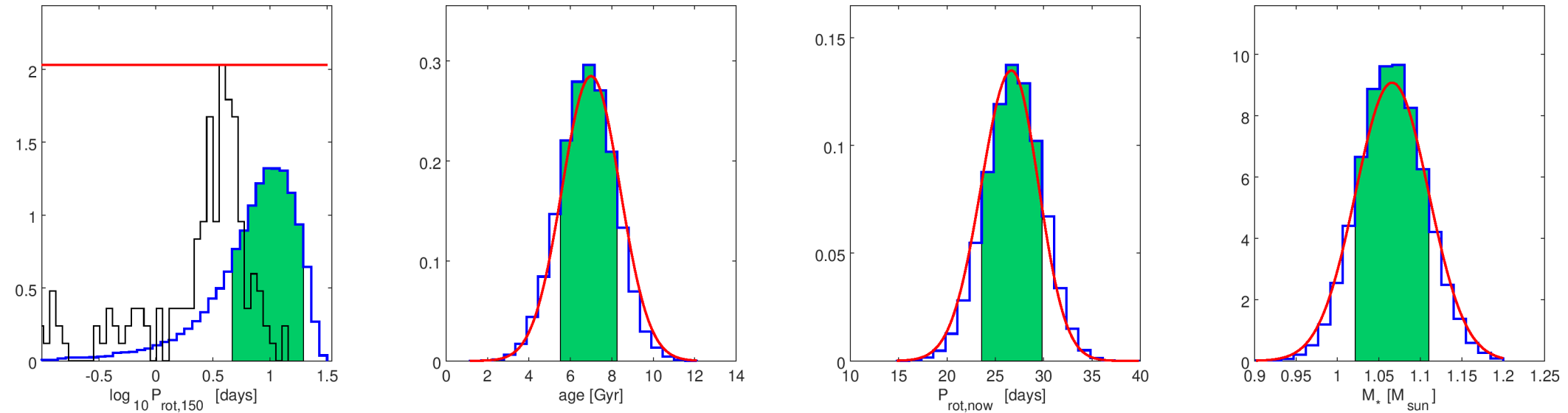}} \\
    \resizebox{\hsize}{!}{ \includegraphics{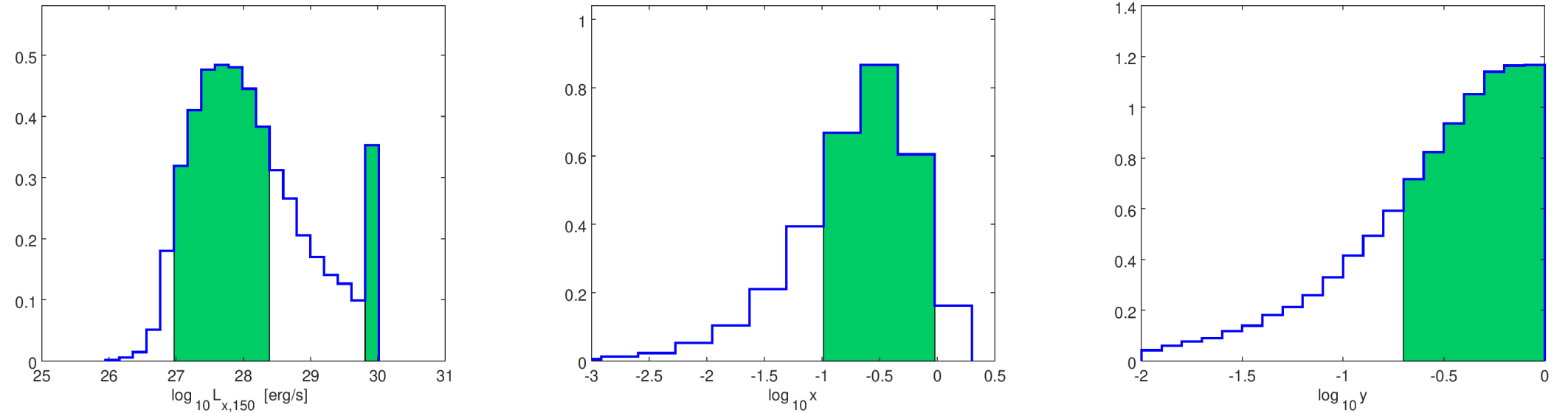}} \\
    \resizebox{\hsize}{!}{ \includegraphics{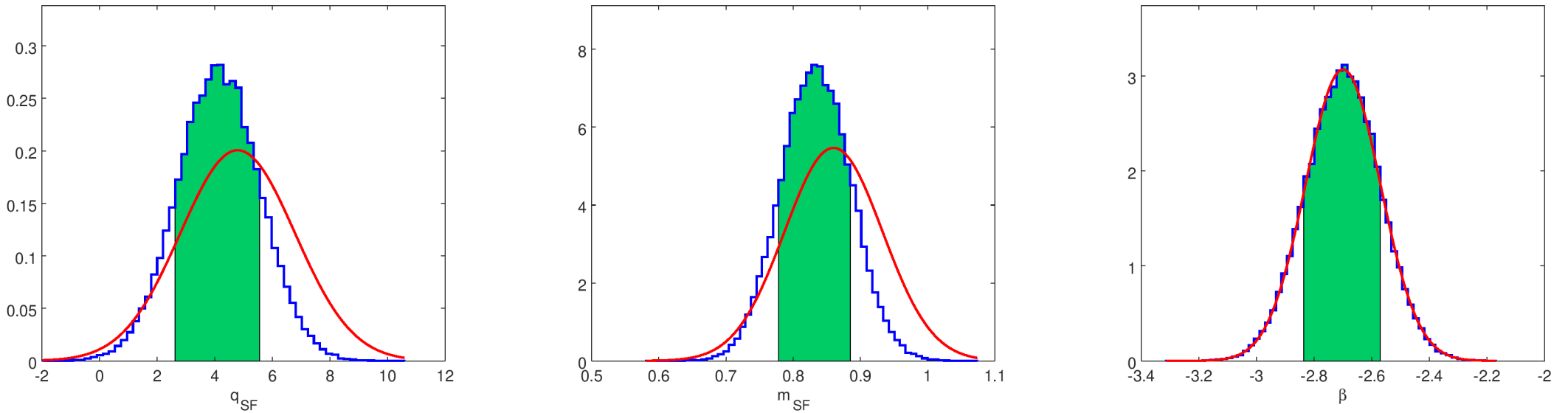}}
    \caption{Same as Fig.~\ref{fig:K2-285star}, but for the star-related properties of Kepler-36.}
    \label{fig:Kepler-36star}
\end{figure*}
\begin{figure*}
    \resizebox{\hsize}{!}{ \includegraphics{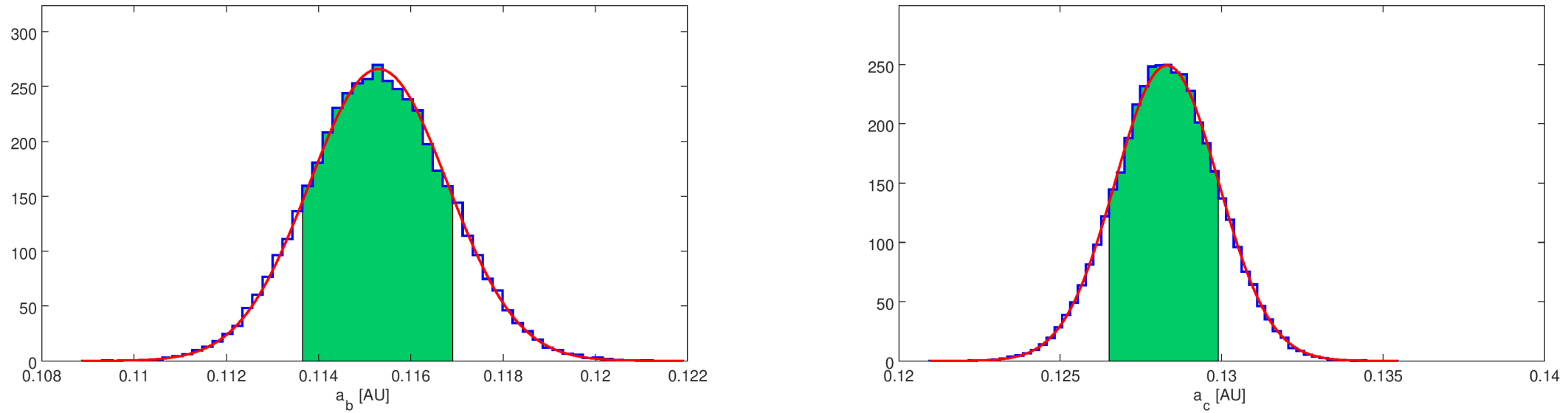}} \\
    \resizebox{\hsize}{!}{ \includegraphics{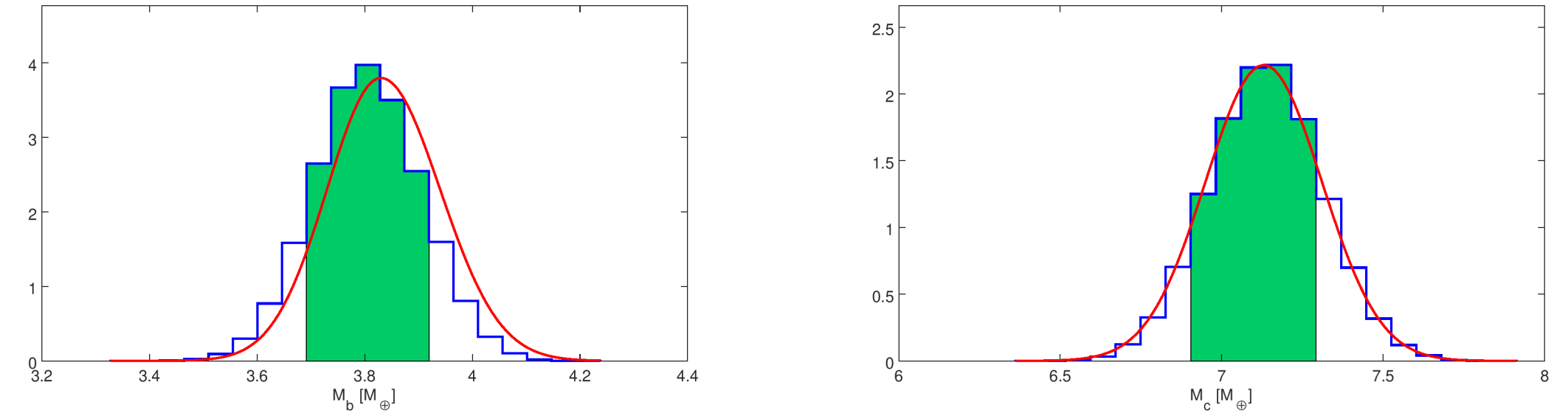}} \\
    \resizebox{\hsize}{!}{ \includegraphics{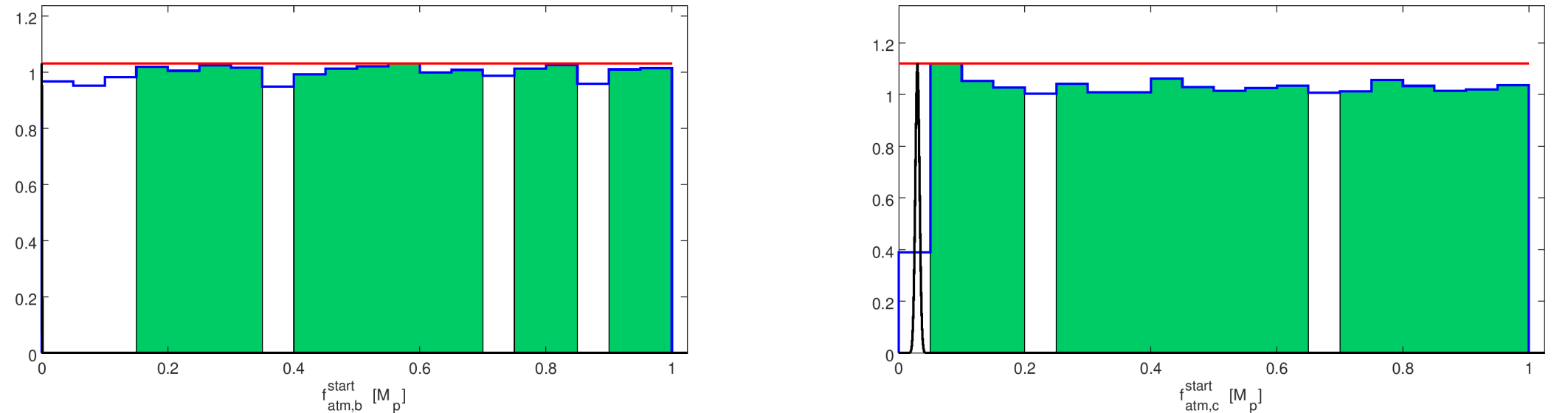}}
    \caption{Same as Fig.~\ref{fig:K2-285planets}, but for the planetary parameters of the Kepler-36 system.}
    \label{fig:Kepler-36planets}
\end{figure*}

\begin{figure*}
    \resizebox{\hsize}{!}{ \includegraphics{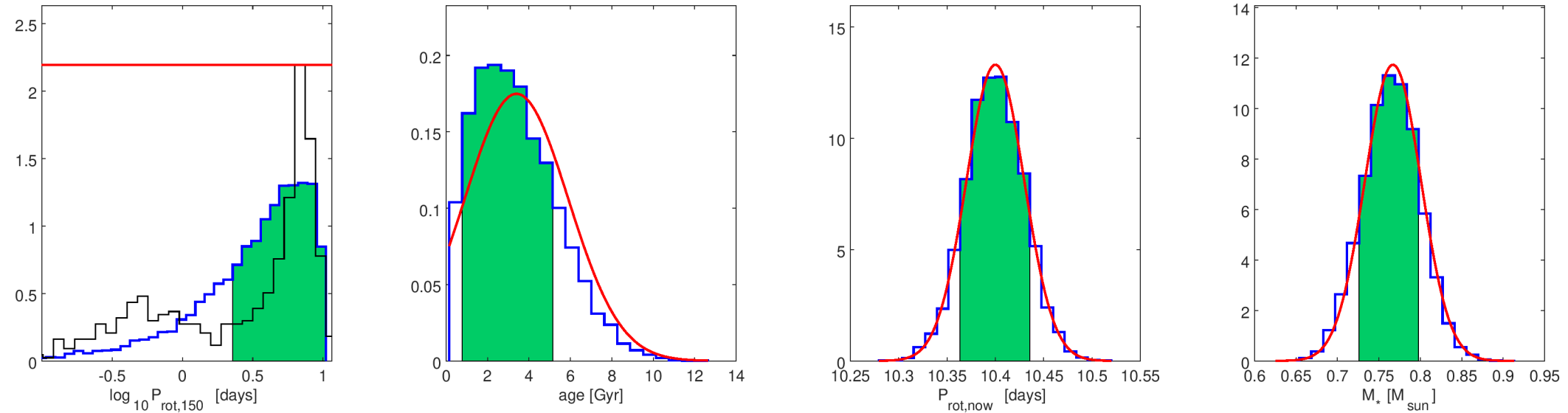}} \\
    \resizebox{\hsize}{!}{ \includegraphics{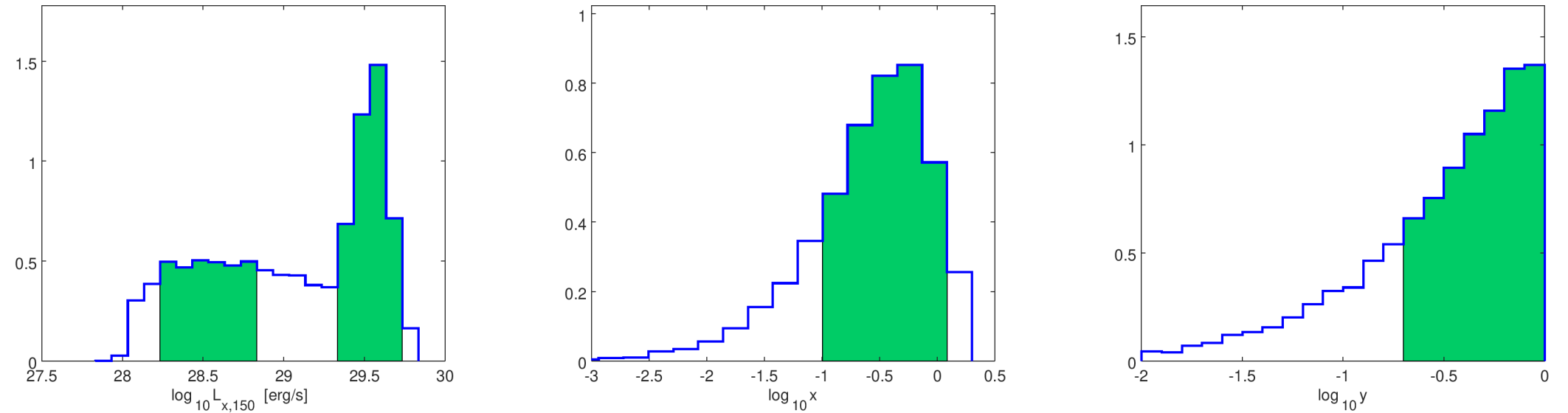}} \\
    \resizebox{\hsize}{!}{ \includegraphics{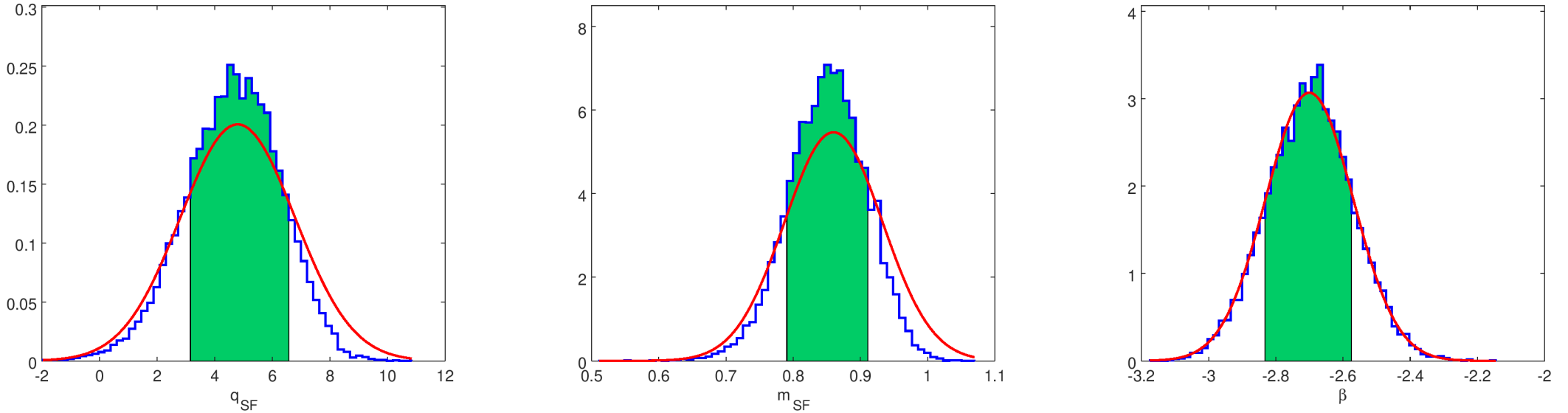}}
    \caption{Same as Fig.~\ref{fig:K2-285star}, but for the star-related properties of Kepler-411.}
    \label{fig:Kepler-411star}
\end{figure*}
\begin{figure*}
    \resizebox{\hsize}{!}{ \includegraphics{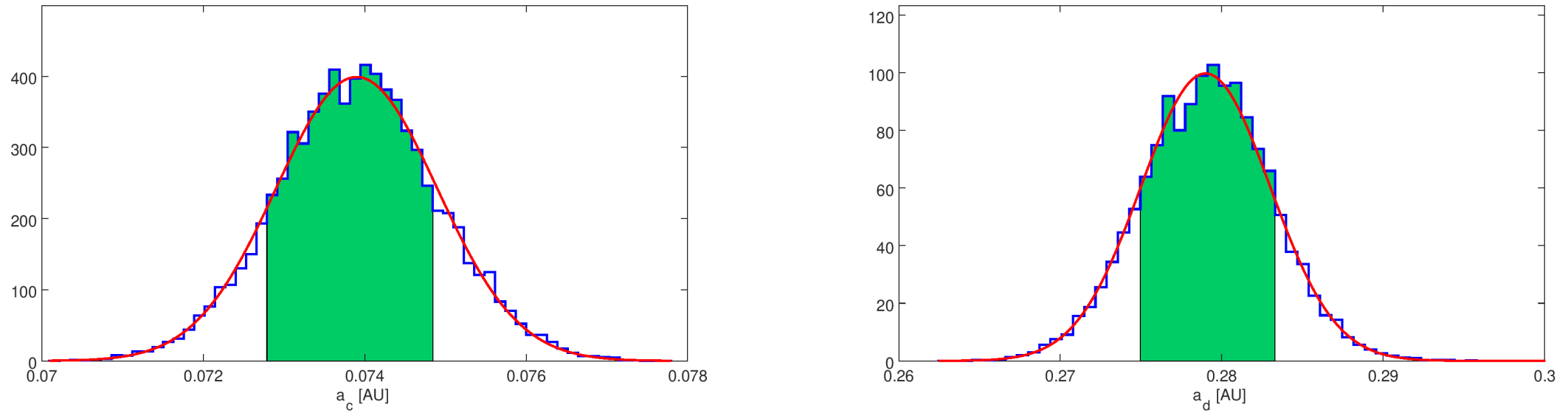}} \\
    \resizebox{\hsize}{!}{ \includegraphics{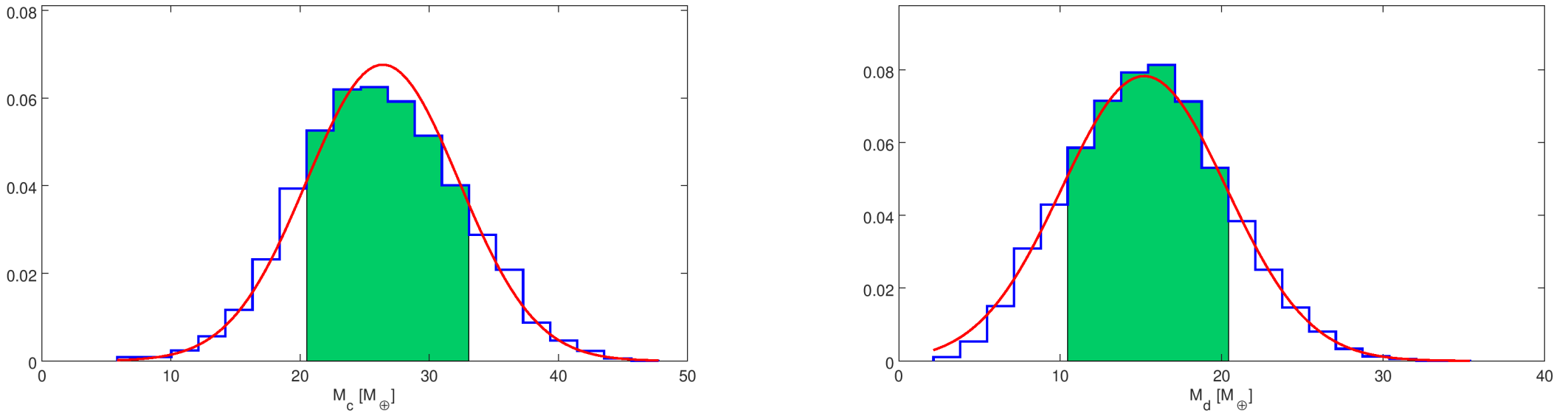}} \\
    \resizebox{\hsize}{!}{ \includegraphics{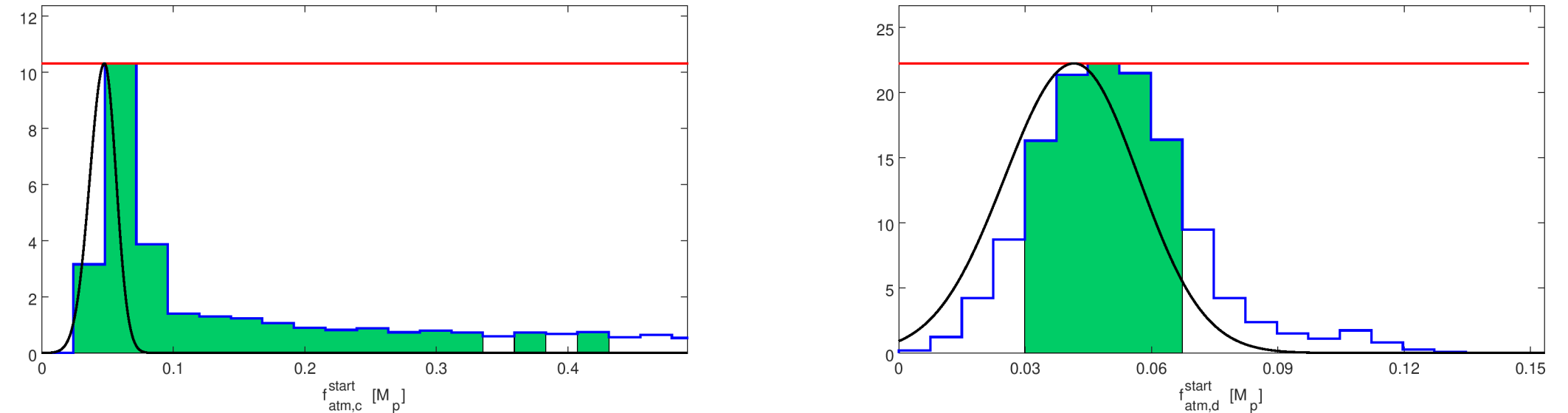}}
    \caption{Same as Fig.~\ref{fig:K2-285planets}, but for the planetary parameters of the Kepler-411 system.}
    \label{fig:Kepler-411planets}
\end{figure*}

\begin{figure*}
    \resizebox{\hsize}{!}{ \includegraphics{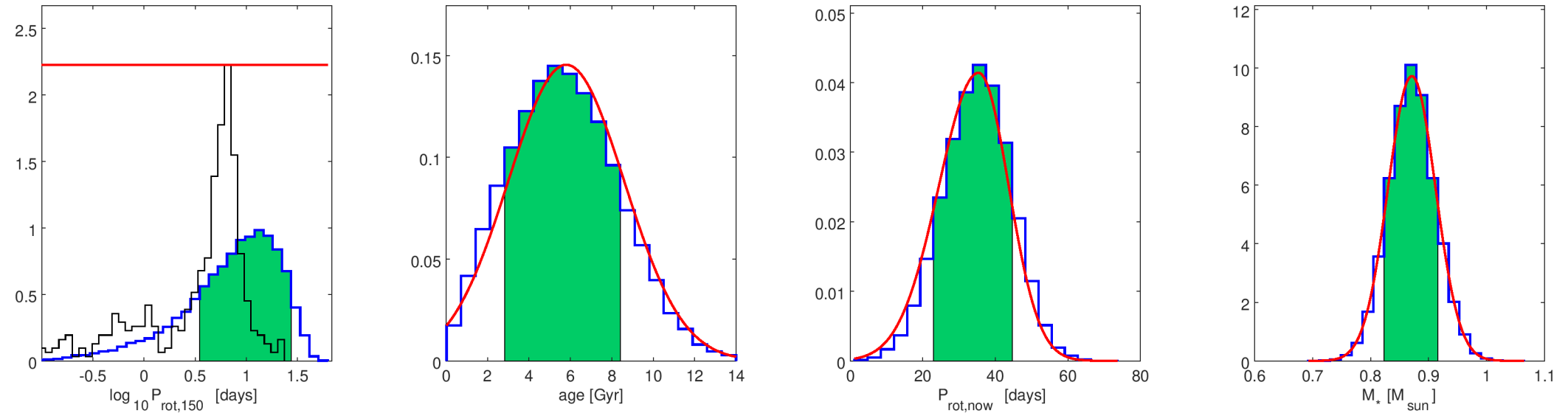}} \\
    \resizebox{\hsize}{!}{ \includegraphics{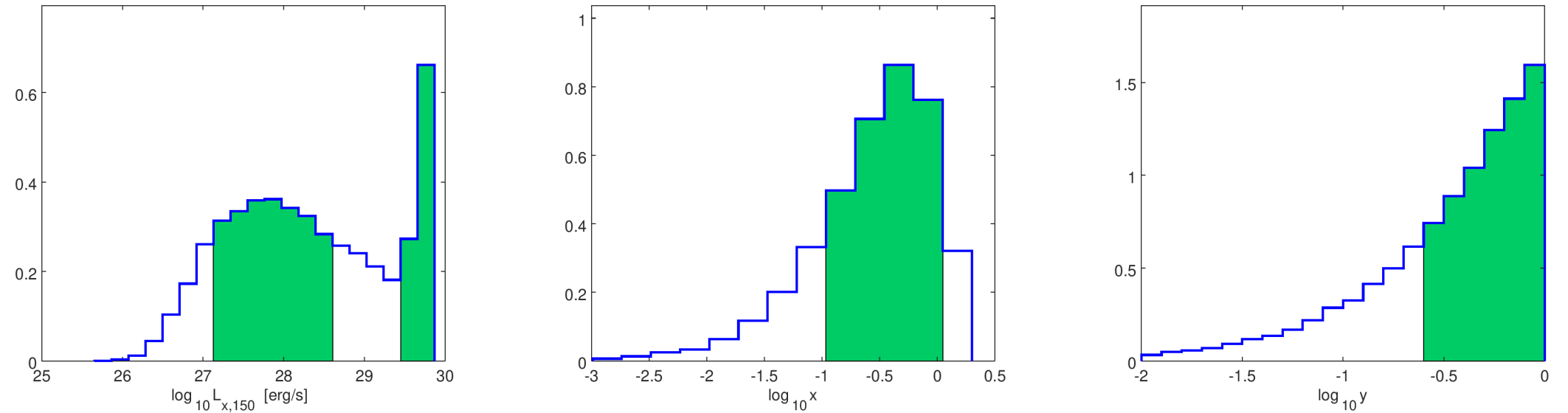}} \\
    \resizebox{\hsize}{!}{ \includegraphics{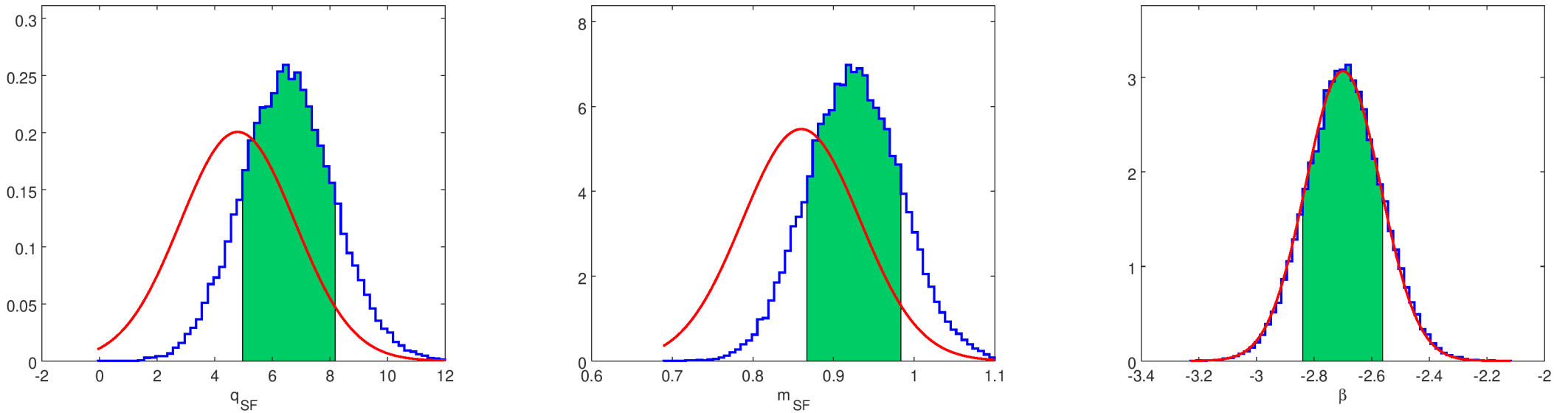}}
    \caption{Same as Fig.~\ref{fig:K2-285star}, but for the star-related properties of Kepler-48.}
    \label{fig:Kepler-48star}
\end{figure*}
\begin{figure*}
    \resizebox{\hsize}{!}{ \includegraphics{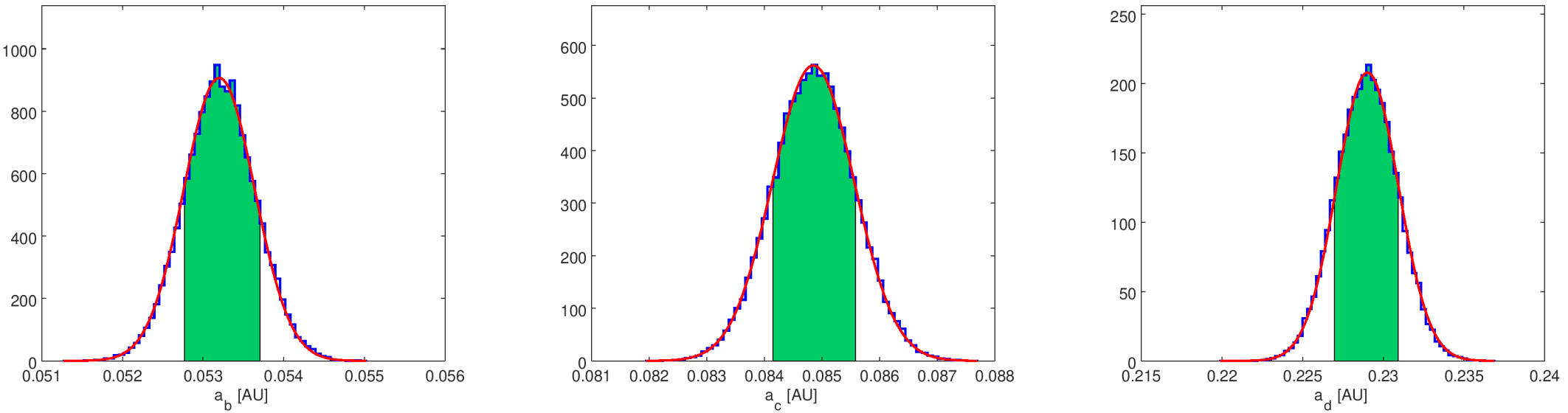}} \\
    \resizebox{\hsize}{!}{ \includegraphics{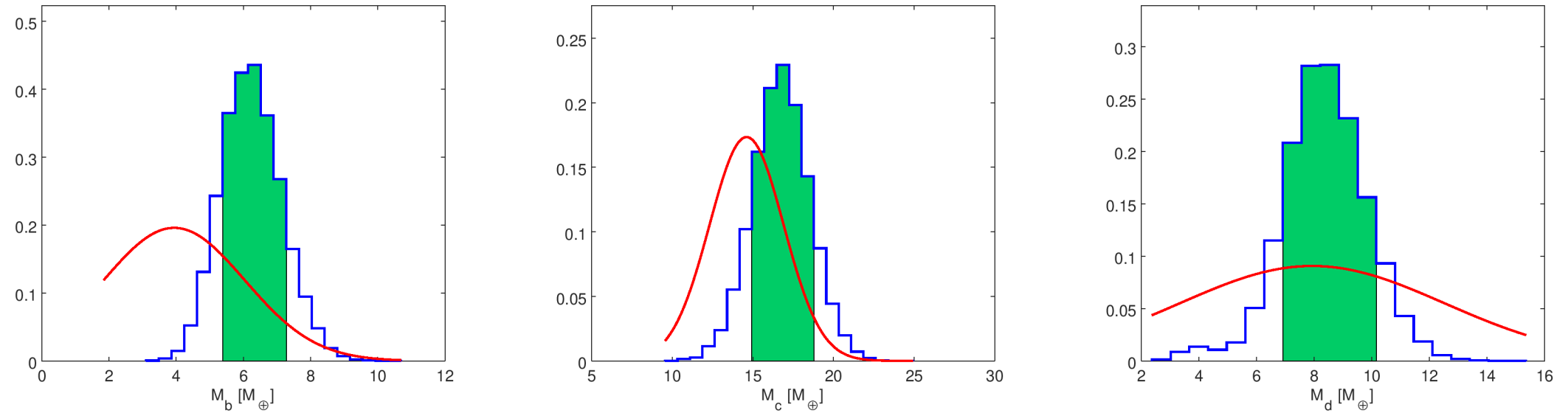}} \\
    \resizebox{\hsize}{!}{ \includegraphics{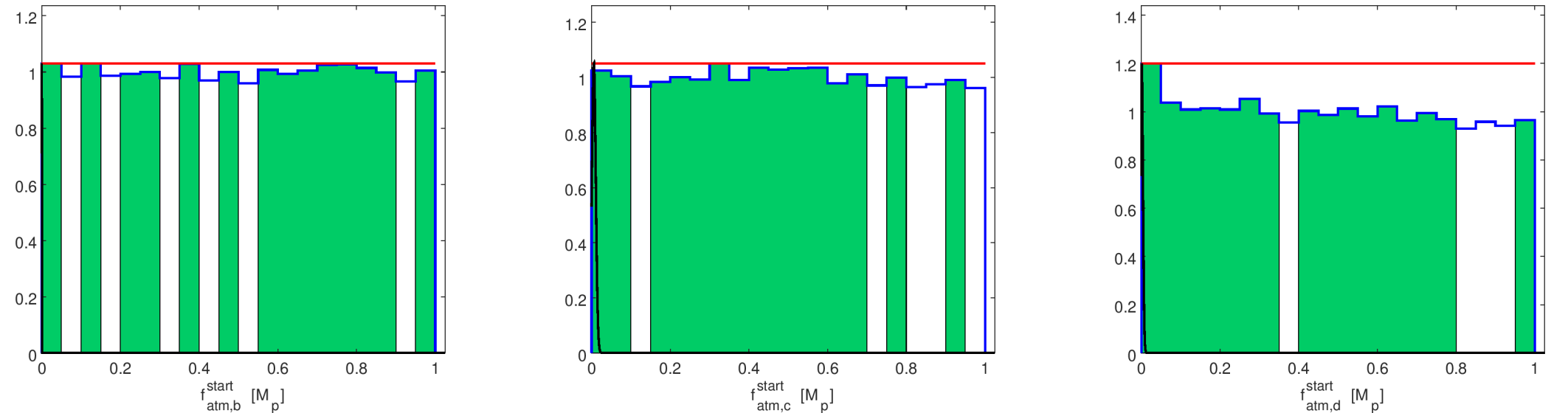}}
    \caption{Same as Fig.~\ref{fig:K2-285planets}, but for the planetary parameters of the Kepler-48 system.}
    \label{fig:Kepler-48planets}
\end{figure*}

\begin{figure*}
    \resizebox{\hsize}{!}{ \includegraphics{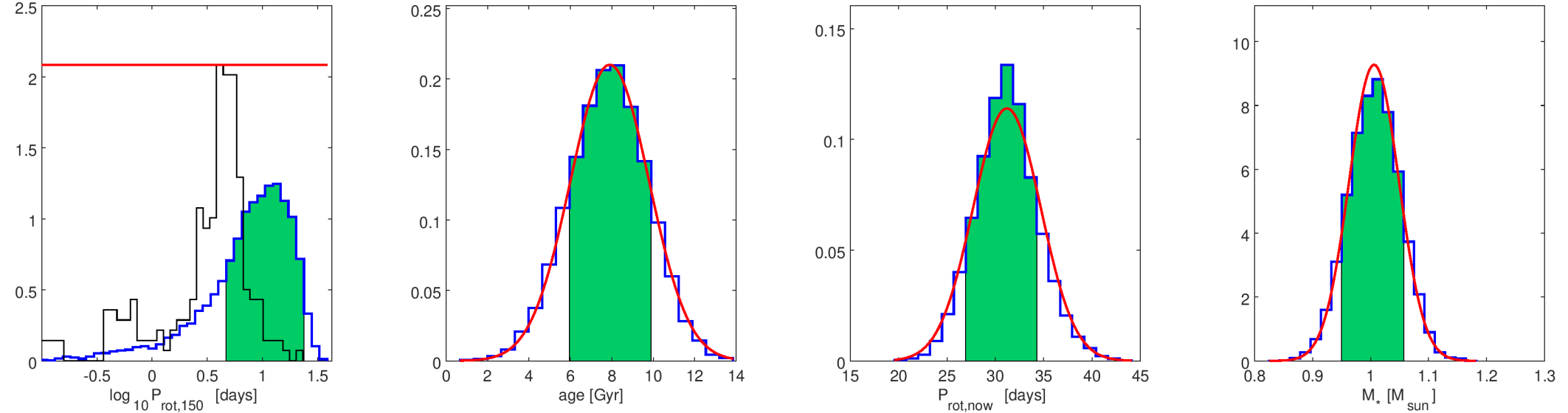}} \\
    \resizebox{\hsize}{!}{ \includegraphics{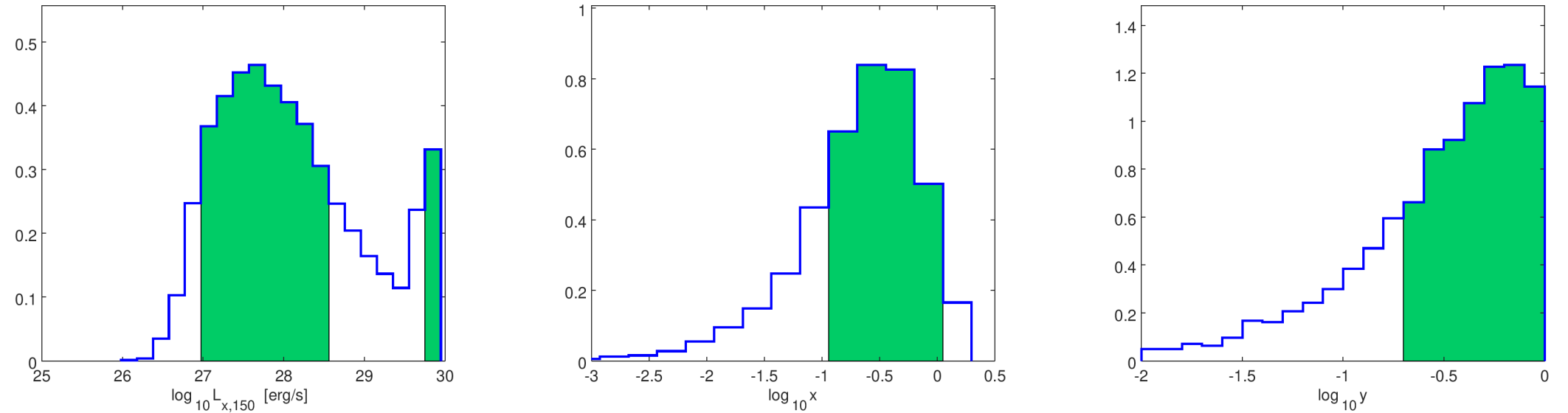}} \\
    \resizebox{\hsize}{!}{ \includegraphics{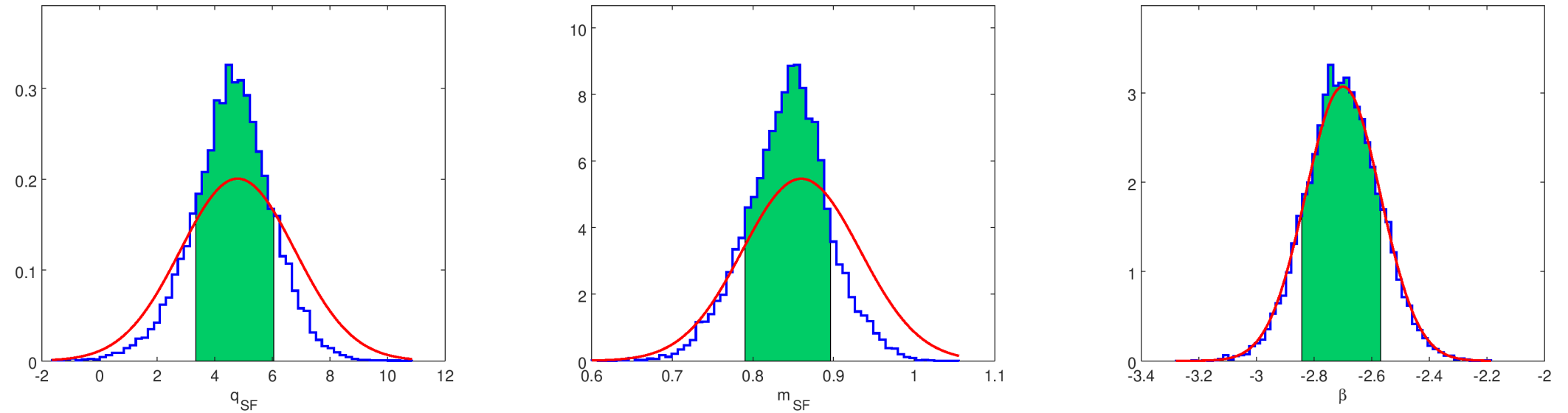}}
    \caption{Same as Fig.~\ref{fig:K2-285star}, but for the star-related properties of WASP-47.}
    \label{fig:WASP-47star}
\end{figure*}
\begin{figure*}
    \resizebox{\hsize}{!}{ \includegraphics{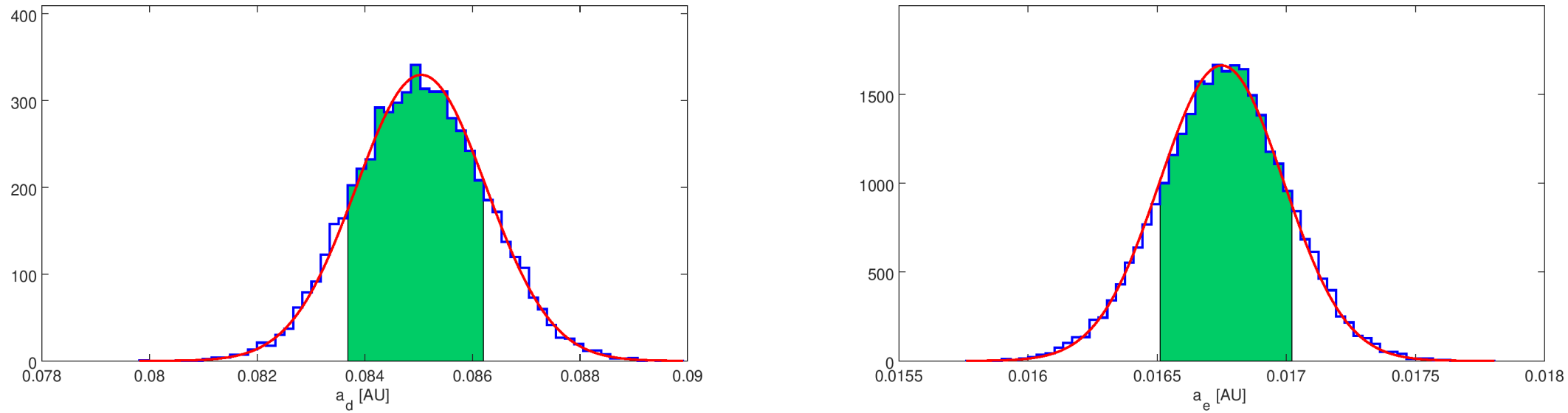}} \\
    \resizebox{\hsize}{!}{ \includegraphics{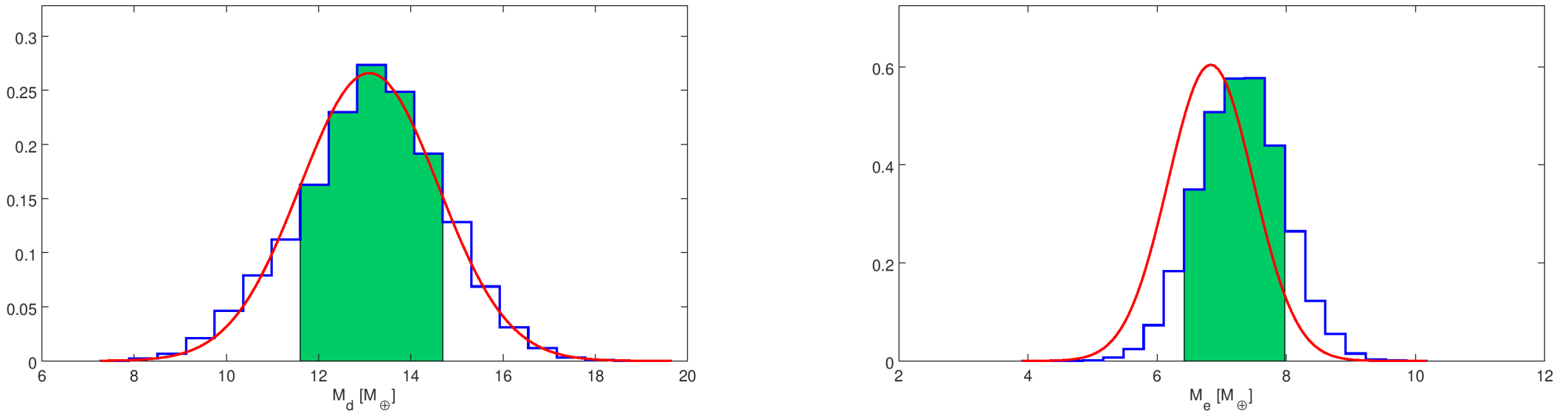}} \\
    \resizebox{\hsize}{!}{ \includegraphics{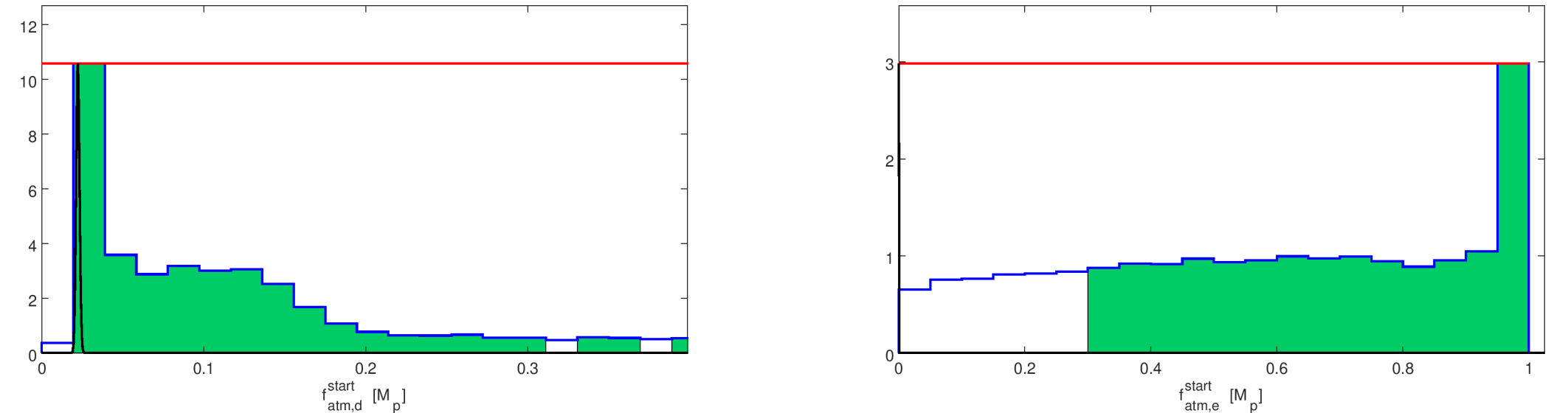}}
    \caption{Same as Fig.~\ref{fig:K2-285planets}, but for the planetary parameters of the WASP-47 system.}
    \label{fig:WASP-47planets}
\end{figure*}

\end{appendix}

\end{document}

%% file: planets.tex
\begin{table*}
\caption{Planetary radii $R_{\rm p}$, masses $M_{\rm p}$, and semi-major axes $a$ taken from the literature and adopted in this work, followed by the main results obtained with \textsc{Pasta} (median values and 1-$\sigma$ confidence interval), that is the stellar rotation period at 150 Myr ($\ProtJ$), the theoretically-estimated planetary mass ($\hat{M}_{\rm p}$), and the atmospheric mass fraction at $t=t_{\mathrm{disk}}=5$\,Myr ($\fas$).}
\label{tab:planets}
\centering
\begin{tabular}{lccclccc}
\hline\hline
Planet & $R_{\rm p}$ [$R_{\oplus}$] & $M_{\rm p}$ [$M_{\oplus}$] & $a$ [AU] & Source & $\ProtJ$ [d] & $\hat{M}_{\rm p}$ [$M_{\oplus}$] & $\fas$ \\
\hline
HD 3167 b & $1.7_{-0.15}^{+0.18}$ & $5.02\pm 0.38$ & $0.01815\pm 0.00023$ & Ch17 & 
            \multirow{2}*{$11.8_{-8.2}^{+13.7}$} & $5.02_{-0.36}^{+0.37}$ & $0.45_{-0.30}^{+0.37}$ \\ 
HD 3167 c & $3.01_{-0.28}^{+0.42}$ & $9.8_{-1.24}^{+1.3}$ & $0.1795\pm 0.0023$ & Ch17 &
                                                & $9.9\pm1.3$ & $0.20_{-0.17}^{+0.53}$ \\ 
K2-24 b & $5.4\pm 0.2$ & $19_{-2.1}^{+2.2}$ & $0.154\pm 0.002$\tablefootmark{(a)} & Pe18 &
            \multirow{2}*{$9.1_{-6.2}^{+8.4}$}  & $19.1_{-2.0}^{+1.9}$ & $0.27_{-0.15}^{+0.45}$ \\
K2-24 c & $7.5\pm 0.3$ & $15.4_{-1.8}^{+1.9}$ & $0.247\pm 0.004$\tablefootmark{(a)} & Pe18 &
                                                & $15.4\pm1.7$ & $0.43_{-0.14}^{+0.30}$ \\
K2-285 b & $2.59\pm 0.06$ & $9.68_{-1.37}^{+1.21}$ & $0.03817_{-0.00092}^{+0.00095}$ & Pa19 &
            \multirow{4}*{$15.3_{-9.4}^{+13.5}$}  & $9.5_{-1.0}^{+1.1}$ & $0.48\pm0.33$ \\ 
K2-285 c & $3.53\pm 0.08$ & $15.68_{-2.13}^{+2.28}$ & $0.0824\pm 0.0018$ & Pa19 &
                                            & $15.7_{-2.1}^{+2.3}$ & $0.0384_{-0.0053}^{+0.0052}$\\ 
K2-285 d & $2.48\pm 0.06$ & $<6.5$ & $0.1178\pm 0.0029$ & Pa19 &
                                                & $3.59_{-0.34}^{+0.33}$ & $0.47\pm0.36$ \\
K2-285 e & $1.95\pm 0.05$ & $<10.7$ & $0.1804_{-0.0043}^{+0.0042}$ & Pa19 &
                                                & $2.04_{-0.18}^{+0.20}$ & $0.49_{-0.35}^{+0.34}$ \\
KOI-94 c & $4.32\pm 0.41$ & $9.4_{-2.1}^{+2.4}$\tablefootmark{(b)} & $0.1013\pm 0.0013$ & We13 &
            \multirow{3}*{$6.2_{-3.9}^{+4.3}$}  & $9.8_{-1.7}^{+1.8}$ & $0.51\pm0.34$ \\
KOI-94 d & $11.27\pm 1.06$ & $52.1_{-7.1}^{+6.9}$\tablefootmark{(b)} & $0.1684\pm 0.0022$ & We13 &
                                                & $51.1_{-7.2}^{+7.0}$ & $0.46_{-0.09}^{+0.11}$ \\
KOI-94 e & $6.56\pm 0.62$ & $13_{-2.1}^{+2.5}$\tablefootmark{(b)} & $0.3046\pm 0.0040$ & We13 &
                                                & $13.2_{-2.2}^{+2.4}$ & $0.215_{-0.040}^{+0.044}$ \\
Kepler-11 b & $1.8_{-0.05}^{+0.03}$ & $2.8_{-1.6}^{+2.3}$\tablefootmark{(c)} & $0.091\pm 0.001$ & Li13 &
            \multirow{6}*{$7.1_{-4.5}^{+6.8}$}  & $5.68_{-0.35}^{+0.38}$ & $0.49\pm0.34$ \\ 
Kepler-11 c & $2.87_{-0.06}^{+0.05}$ & $2.8_{-1.9}^{+3.4}$\tablefootmark{(c)} & $0.107\pm 0.001$ & Li13 &                                          & $8.0_{-1.1}^{+1.2}$ & $0.51_{-0.34}^{+0.33}$ \\ 
Kepler-11 d & $3.12_{-0.07}^{+0.06}$ & $6.9_{-1.2}^{+1.1}$\tablefootmark{(d)} & $0.155\pm 0.001$ & Li13 &                                          & $6.00_{-0.59}^{+0.85}$ & $0.44_{-0.34}^{+0.38}$ \\ 
Kepler-11 e & $4.19_{-0.09}^{+0.07}$ & $7.2_{-1.6}^{+2.0}$\tablefootmark{(d)} & $0.195\pm 0.002$ & Li13&
                                                & $6.33_{-0.57}^{+0.99}$ & $0.43_{-0.31}^{+0.39}$ \\ 
Kepler-11 f & $2.49_{-0.07}^{+0.04}$ & $1.7_{-0.6}^{+0.9}$\tablefootmark{(d)} & $0.25\pm 0.002$ & Li13&
                                                & $3.08_{-0.22}^{+0.27}$ & $0.50\pm0.33$ \\ 
Kepler-11 g & $3.33_{-0.08}^{+0.06}$ & $<25$ & $0.466\pm 0.004$ & Li13 &
                                                & $16.3\pm9.2$ & $0.043_{-0.023}^{+0.016}$ \\
Kepler-18 b & $2\pm 0.1$ & $6.9\pm 3.4$ & $0.0447\pm 0.0006$ & Co11 &
            \multirow{3}*{$10.1_{-6.8}^{+10.7}$}  & $7.9_{-1.1}^{+1.2}$ & $0.52_{-0.35}^{+0.33}$ \\
Kepler-18 c & $5.49\pm 0.26$ & $17.3\pm 1.9$ & $0.0752\pm 0.0011$ & Co11 &
                                                & $17.8_{-2.1}^{+1.8}$ & $0.49_{-0.31}^{+0.34}$ \\
Kepler-18 d & $6.98\pm 0.33$ & $16.4\pm 1.4$ & $0.1172\pm 0.0017$ & Co11 &
                                                & $15.9\pm1.5$ & $0.28_{-0.07}^{+0.31}$ \\
Kepler-20 c & $3.047_{-0.056}^{+0.064}$ & $12.75_{-2.24}^{+2.17}$ & $0.0949_{-0.0023}^{+0.0027}$ & Bu16 &       \multirow{4}*{$11.6_{-7.8}^{+10.5}$} & $12.5_{-2.2}^{+2.1}$ & $0.39_{-0.36}^{+0.42}$ \\
Kepler-20 d & $2.744_{-0.055}^{+0.073}$ & $10.1_{-3.7}^{+4.0}$ & $0.3506_{-0.0101}^{+0.0081}$ & Bu16&
                                              & $10.5_{-3.5}^{+3.6}$ & $0.0249_{-0.0086}^{+0.0063}$ \\
Kepler-20 e & $0.865_{-0.028}^{+0.026}$ & $<3.08$\tablefootmark{(e)} & $0.0639_{-0.0014}^{+0.0019}$ & Bu16 &                                         & $0.650_{-0.061}^{+0.062}$ & $0.69_{-0.34}^{+0.20}$ \\
Kepler-20 f & $1.003_{-0.089}^{+0.05}$ & $<14.3$\tablefootmark{(e)} & $0.1396_{-0.0035}^{+0.0036}$ & Bu16 &                                          & $1.03_{-0.20}^{+0.22}$ & $0.59_{-0.36}^{+0.29}$ \\
Kepler-25 b & $2.748_{-0.035}^{+0.038}$ & $8.7_{-2.3}^{+2.5}$ & $0.06999\pm 0.00051$\tablefootmark{(f)} & Mi19 & \multirow{2}*{$5.3_{-3.2}^{+3.4}$} & $9.6_{-1.4}^{+2.1}$ & $0.63_{-0.42}^{+0.31}$ \\
Kepler-25 c & $5.217_{-0.065}^{+0.07}$ & $15.2_{-1.6}^{+1.3}$ & $0.11255\pm 0.00081$\tablefootmark{(f)} & Mi19 &            & $14.3_{-1.7}^{+1.6}$ & $0.09_{-0.01}^{+0.16}$ \\
Kepler-36 b & $1.498_{-0.049}^{+0.061}$ & $3.83_{-0.10}^{+0.11}$ & $0.1153\pm 0.0015$\tablefootmark{(g)} & Vi20 & \multirow{2}*{$8.6_{-5.3}^{+7.2}$} & $3.80\pm0.10$ & $0.50\pm0.34$ \\
Kepler-36 c & $3.679_{-0.091}^{+0.096}$ & $7.13\pm 0.18$ & $0.1283\pm 0.0016$\tablefootmark{(g)} & Vi20 &                                          & $7.14_{-0.18}^{+0.17}$ & $0.51\pm0.33$ \\
Kepler-411 c & $4.421\pm 0.062$ & $26.4\pm 5.9$ & $0.0739\pm 0.001$ & Su19 &
            \multirow{2}*{$4.0_{-2.7}^{+3.5}$} & $26.0_{-5.9}^{+6.5}$ & $0.16_{-0.10}^{+0.48}$ \\
Kepler-411 d & $3.319\pm 0.104$ & $15.2\pm 5.1$ & $0.279\pm 0.004$ & Su19 &
                                                & $15.4\pm4.9$ & $0.050_{-0.016}^{+0.018}$ \\
Kepler-48 b & $1.88\pm 0.10$ & $3.9\pm 2.1$ & $0.05302\pm 0.00044$\tablefootmark{(f)} & Ma14 &
            \multirow{3}*{$8.6_{-6.2}^{+11.2}$} & $6.21_{-0.86}^{+0.94}$ & $0.50\pm0.34$ \\
Kepler-48 c & $2.71\pm 0.14$ & $14.6\pm 2.3$ & $0.08485\pm 0.00071$\tablefootmark{(f)} & Ma14 &
                                                & $16.7_{-1.7}^{+1.8}$ & $0.50\pm0.34$ \\
Kepler-48 d & $2.04\pm 0.11$ & $7.9\pm 4.6$ & $0.22902\pm 0.00192$\tablefootmark{(f)} & Ma14 &
                                                & $8.4_{-1.3}^{+1.4}$ & $0.49_{-0.34}^{+0.35}$ \\
WASP-47 d & $3.576\pm 0.046$ & $13.1\pm 1.5$ & $0.08505_{-0.00121}^{+0.019}$\tablefootmark{(f)} & Va17 &           \multirow{2}*{$8.8_{-5.7}^{+8.2}$}  & $13.2_{-1.6}^{+1.5}$ & $0.13_{-0.10}^{+0.50}$ \\
WASP-47 e & $6.35\pm 0.64$ & $6.83\pm 0.66$ & $0.01675_{-0.00024}^{+0.0038}$\tablefootmark{(f)} & Va17 &                                               & $7.29_{-0.66}^{+0.65}$ & $0.59_{-0.38}^{+0.35}$ \\
\hline
\end{tabular}
\tablefoot{Ch17: \citet{christiansen17}; Pe18: \citet{petigura18}; Pa19: \citet{palle19}; We13: \citet{weiss13}; Li13: \citet{lissauer13}; Co11: \citet{cochran11}; Bu16: \citet{buchhave16};
Mi19: \citet{mills19}; Vi20: \citet{vissapragada20}; Su19: \citet{sun19}; Ma14: \citet{marcy14}; Va17: \citet{vanderburg17} \\
\tablefoottext{a}{From \citet{petigura16}}
\tablefoottext{b}{From \citet{masuda13}}
\tablefoottext{c}{Combination of estimates from \citet{lissauer11}, \citet{lissauer13}, \citet{borsato14}, and \citet{hadden14}}
\tablefoottext{d}{Same as \textit{(c)} + \citet{haddenLithwick17}}
\tablefoottext{e}{From \citet{fressin12}}
\tablefoottext{f}{Computed from Kepler III law}
\tablefoottext{g}{From \citet{carter12}}
}
\end{table*}

%% file: stars.tex
%-------------------------------------------------------------
%                      A rotated Two column Table in landscape  
%-------------------------------------------------------------
\begin{sidewaystable*}
\caption{Stellar parameters.}
\label{tab:stars}
\centering
\begin{tabular}{lcccccccccc} 
\hline\hline             %$\sigma_{\mathrm{low}}\Prot$ & $\sigma_{\mathrm{up}}\Prot$ &
\multirow{2}*{Star} & $T_{\mathrm{eff,}\star}$ & $\log{g_{\star}}$ & [Fe/H]$_{\star}$ & $G_{\star}$ & $\pi_{\star}$ & $\log{R'_{\mathrm{HK}}}$ & $v\sin{i_{\star}}$ & $P_{\mathrm{rot,}\star}$ & $M_{\star}$ & $t_{\star}$ \\
 & [K] & [cgs] & [dex] & [mag] & [mas] & [dex] & [km/s] & [d] & [$M_{\odot}$] & [Gyr] \\
\hline
HD 3167 & $5261\pm60$ & $4.47\pm0.05$ & $0.04\pm0.05$ & $8.7622\pm0.0003$ & $21.118\pm0.066$ & $-5.04$ & 1.7 & $49.7_{-6.3}^{+4.4}$ & $0.846\pm0.035$ & $11.7\pm2.6$ \\
K2-24 & $5625\pm60$ & $4.29\pm0.05$ & $0.34\pm0.04$ & $10.9072\pm0.0003$ & $5.840\pm0.050$ & $-5.24$ & 2 & $25.3_{-6.3}^{+5.0}$ & $1.040\pm0.045$ & $4.1\pm1.6$ \\
K2-285 & $4975\pm95$ & $4.4\pm0.1$ & $0\pm0.05$ & $11.804\pm0.0003$ & $6.425\pm0.048$ &  & 3.9 & $47.1_{-11.6}^{+9.3}$ & $0.807\pm0.038$ & $8.9\pm3.6$ \\
KOI-94 & $6182\pm58$ & $4.181\pm0.066$ & $0.0228\pm0.002$ & $12.1882\pm0.0002$ & $2.068\pm0.022$ &  & 7.3 & $14.2_{-2.6}^{+2.2}$ & $1.173\pm0.050$ & $3.4\pm1.0$ \\
Kepler-11 & $5666\pm60$ & $4.279\pm0.071$ & $0.002\pm0.04$ & $13.7062\pm0.0003$ & $1.518\pm0.015$ & $-4.82$ & 2.2 & $24.62_{-2.27}^{+3.47}$ & $0.952\pm0.044$ & $8.8\pm2.3$ \\
Kepler-18 & $5345\pm100$ & $4.31\pm0.12$ & $0.19\pm0.06$ & $13.5393\pm0.0002$ & $2.280\pm0.017$ &  & 4 & $27.0_{-18.7}^{+10.5}$ & $0.936\pm0.049$ & $3.8\pm3.2$ \\
Kepler-20 & $5495\pm50$ & $4.446\pm0.010$ & $0.07\pm0.08$ & $12.4535\pm0.0001$ & $3.510\pm0.022$ &  & 2 & $31.1_{-10.2}^{+7.6}$ & $0.919\pm0.046$ & $5.5\pm2.8$ \\
Kepler-25 & $6270\pm79$ & $4.278\pm0.030$ & $-0.04\pm0.1$ & $10.6319\pm0.0003$ & $4.082\pm0.024$ &  & 9.5 & $11.3_{-2.0}^{+1.7}$ & $1.175\pm0.053$ & $3.1\pm0.9$ \\
Kepler-36 & $5979\pm60$ & $4.1\pm0.1$ & $-0.18\pm0.04$ & $12.0767\pm0.0002$ & $1.865\pm0.022$ &  & 4.9 & $26.7_{-3.1}^{+2.8}$ & $1.066\pm0.044$ & $7.0\pm1.4$ \\
Kepler-411 & $4906\pm50$ & $4.60\pm0.01$ & $0.05\pm0.11$ & $12.2327\pm0.0006$ & $6.478\pm0.019$ &  &  & $10.4\pm 0.03$ & $0.767\pm0.034$ & $3.4\pm2.5$ \\
Kepler-48 & $5194\pm73$ & $4.487\pm0.050$ & $0.17\pm0.07$ & $13.0772\pm0.0003$ & $3.231\pm0.019$ &  & 0.5 & $35.4_{-11.0}^{+8.3}$ & $0.871\pm0.041$ & $5.8\pm2.8$ \\
WASP-47 & $5552\pm75$ & $4.3437\pm0.0063$ & $0.38\pm0.05$ & $11.7889\pm0.0003$ & $3.750\pm0.068$ &  & 1.8 & $31.2\pm 3.5$ & $1.006\pm0.043$ & $7.9\pm1.9$ \\
\hline
\end{tabular}
\end{sidewaystable*}
%

%% file: evolution.bbl
\begin{thebibliography}{98}
\expandafter\ifx\csname natexlab\endcsname\relax\def\natexlab#1{#1}\fi

\bibitem[{{Alexander} {et~al.}(2014){Alexander}, {Pascucci}, {Andrews},
  {Armitage}, \& {Cieza}}]{alexander14}
{Alexander}, R., {Pascucci}, I., {Andrews}, S., {Armitage}, P., \& {Cieza}, L.
  2014, in Protostars and Planets VI, ed. H.~{Beuther}, R.~S. {Klessen}, C.~P.
  {Dullemond}, \& T.~{Henning}, 475

\bibitem[{{Barnes}(2003)}]{barnes03}
{Barnes}, S.~A. 2003, \apj, 586, 464

\bibitem[{{Barnes}(2007)}]{barnes07}
{Barnes}, S.~A. 2007, \apj, 669, 1167

\bibitem[{{Barnes}(2009)}]{barnes09}
{Barnes}, S.~A. 2009, in IAU Symposium, Vol. 258, IAU Symposium, ed. E.~E.
  {Mamajek}, D.~R. {Soderblom}, \& R.~F.~G. {Wyse}, 345--356

\bibitem[{{Barnes}(2010)}]{barnes10}
{Barnes}, S.~A. 2010, \apj, 722, 222

\bibitem[{{Benz} {et~al.}(2021){Benz}, {Broeg}, {Fortier}, {Rando}, {Beck},
  {Beck}, {Queloz}, {Ehrenreich}, {Maxted}, {Isaak}, {Billot}, {Alibert},
  {Alonso}, {Ant{\'o}nio}, {Asquier}, {Bandy}, {B{\'a}rczy}, {Barrado},
  {Barros}, {Baumjohann}, {Bekkelien}, {Bergomi}, {Biondi}, {Bonfils},
  {Borsato}, {Brandeker}, {Busch}, {Cabrera}, {Cessa}, {Charnoz}, {Chazelas},
  {Collier Cameron}, {Corral Van Damme}, {Cortes}, {Davies}, {Deleuil},
  {Deline}, {Delrez}, {Demangeon}, {Demory}, {Erikson}, {Farinato}, {Fossati},
  {Fridlund}, {Futyan}, {Gandolfi}, {Garcia Munoz}, {Gillon}, {Guterman},
  {Gutierrez}, {Hasiba}, {Heng}, {Hernandez}, {Hoyer}, {Kiss}, {Kovacs},
  {Kuntzer}, {Laskar}, {Lecavelier des Etangs}, {Lendl}, {L{\'o}pez}, {Lora},
  {Lovis}, {L{\"u}ftinger}, {Magrin}, {Malvasio}, {Marafatto}, {Michaelis}, {de
  Miguel}, {Modrego}, {Munari}, {Nascimbeni}, {Olofsson}, {Ottacher},
  {Ottensamer}, {Pagano}, {Palacios}, {Pall{\'e}}, {Peter}, {Piazza}, {Piotto},
  {Pizarro}, {Pollaco}, {Ragazzoni}, {Ratti}, {Rauer}, {Ribas}, {Rieder},
  {Rohlfs}, {Safa}, {Salatti}, {Santos}, {Scandariato}, {S{\'e}gransan},
  {Simon}, {Smith}, {Sordet}, {Sousa}, {Steller}, {Szab{\'o}}, {Szoke},
  {Thomas}, {Tschentscher}, {Udry}, {Van Grootel}, {Viotto}, {Walter},
  {Walton}, {Wildi}, \& {Wolter}}]{benz21}
{Benz}, W., {Broeg}, C., {Fortier}, A., {et~al.} 2021, Experimental Astronomy,
  51, 109

\bibitem[{{Bonfanti} {et~al.}(2021){Bonfanti}, {Delrez}, {Hooton}, {Wilson},
  {Fossati}, {Alibert}, {Hoyer}, {Mustill}, {Osborn}, {Adibekyan}, {Gandolfi},
  {Salmon}, {Sousa}, {Tuson}, {Van Grootel}, {Cabrera}, {Nascimbeni}, {Maxted},
  {Barros}, {Billot}, {Bonfils}, {Borsato}, {Broeg}, {Davies}, {Deleuil},
  {Demangeon}, {Fridlund}, {Lacedelli}, {Lendl}, {Persson}, {Santos},
  {Scandariato}, {Szab{\'o}}, {Collier Cameron}, {Udry}, {Benz}, {Beck},
  {Ehrenreich}, {Fortier}, {Isaak}, {Queloz}, {Alonso}, {Asquier}, {Bandy},
  {B{\'a}rczy}, {Barrado}, {Barrag{\'a}n}, {Baumjohann}, {Beck}, {Bekkelien},
  {Bergomi}, {Brandeker}, {Busch}, {Cessa}, {Charnoz}, {Chazelas}, {Corral Van
  Damme}, {Demory}, {Erikson}, {Farinato}, {Futyan}, {Garcia Mu{\~n}oz},
  {Gillon}, {Guedel}, {Guterman}, {Hasiba}, {Heng}, {Hernandez}, {Kiss},
  {Kuntzer}, {Laskar}, {Lecavelier des Etangs}, {Lovis}, {Magrin}, {Malvasio},
  {Marafatto}, {Michaelis}, {Munari}, {Olofsson}, {Ottacher}, {Ottensamer},
  {Pagano}, {Pall{\'e}}, {Peter}, {Piazza}, {Piotto}, {Pollacco}, {Ragazzoni},
  {Rando}, {Ratti}, {Rauer}, {Ribas}, {Rieder}, {Rohlfs}, {Safa}, {Salatti},
  {S{\'e}gransan}, {Simon}, {Smith}, {Sordet}, {Steller}, {Thomas},
  {Tschentscher}, {Van Eylen}, {Viotto}, {Walter}, {Walton}, {Wildi}, \&
  {Wolter}}]{bonfanti21}
{Bonfanti}, A., {Delrez}, L., {Hooton}, M.~J., {et~al.} 2021, \aap, 646, A157

\bibitem[{{Bonfanti} {et~al.}(2016){Bonfanti}, {Ortolani}, \&
  {Nascimbeni}}]{bonfanti16}
{Bonfanti}, A., {Ortolani}, S., \& {Nascimbeni}, V. 2016, \aap, 585, A5

\bibitem[{{Bonfanti} {et~al.}(2015){Bonfanti}, {Ortolani}, {Piotto}, \&
  {Nascimbeni}}]{bonfanti15}
{Bonfanti}, A., {Ortolani}, S., {Piotto}, G., \& {Nascimbeni}, V. 2015, \aap,
  575, A18

\bibitem[{{Borsato} {et~al.}(2014){Borsato}, {Marzari}, {Nascimbeni}, {Piotto},
  {Granata}, {Bedin}, \& {Malavolta}}]{borsato14}
{Borsato}, L., {Marzari}, F., {Nascimbeni}, V., {et~al.} 2014, \aap, 571, A38

\bibitem[{{Brown}(2014)}]{brown14}
{Brown}, D.~J.~A. 2014, \mnras, 442, 1844

\bibitem[{{Buchhave} {et~al.}(2016){Buchhave}, {Dressing}, {Dumusque}, {Rice},
  {Vanderburg}, {Mortier}, {Lopez-Morales}, {Lopez}, {Lundkvist}, {Kjeldsen},
  {Affer}, {Bonomo}, {Charbonneau}, {Collier Cameron}, {Cosentino}, {Figueira},
  {Fiorenzano}, {Harutyunyan}, {Haywood}, {Johnson}, {Latham}, {Lovis},
  {Malavolta}, {Mayor}, {Micela}, {Molinari}, {Motalebi}, {Nascimbeni}, {Pepe},
  {Phillips}, {Piotto}, {Pollacco}, {Queloz}, {Sasselov}, {S{\'e}gransan},
  {Sozzetti}, {Udry}, \& {Watson}}]{buchhave16}
{Buchhave}, L.~A., {Dressing}, C.~D., {Dumusque}, X., {et~al.} 2016, \aj, 152,
  160

\bibitem[{{Carter} {et~al.}(2012){Carter}, {Agol}, {Chaplin}, {Basu},
  {Bedding}, {Buchhave}, {Christensen-Dalsgaard}, {Deck}, {Elsworth},
  {Fabrycky}, {Ford}, {Fortney}, {Hale}, {Handberg}, {Hekker}, {Holman},
  {Huber}, {Karoff}, {Kawaler}, {Kjeldsen}, {Lissauer}, {Lopez}, {Lund},
  {Lundkvist}, {Metcalfe}, {Miglio}, {Rogers}, {Stello}, {Borucki}, {Bryson},
  {Christiansen}, {Cochran}, {Geary}, {Gilliland}, {Haas}, {Hall}, {Howard},
  {Jenkins}, {Klaus}, {Koch}, {Latham}, {MacQueen}, {Sasselov}, {Steffen},
  {Twicken}, \& {Winn}}]{carter12}
{Carter}, J.~A., {Agol}, E., {Chaplin}, W.~J., {et~al.} 2012, Science, 337, 556

\bibitem[{{Choi} {et~al.}(2016){Choi}, {Dotter}, {Conroy}, {Cantiello},
  {Paxton}, \& {Johnson}}]{choi16}
{Choi}, J., {Dotter}, A., {Conroy}, C., {et~al.} 2016, \apj, 823, 102

\bibitem[{{Christiansen} {et~al.}(2017){Christiansen}, {Vanderburg}, {Burt},
  {Fulton}, {Batygin}, {Benneke}, {Brewer}, {Charbonneau}, {Ciardi}, {Collier
  Cameron}, {Coughlin}, {Crossfield}, {Dressing}, {Greene}, {Howard}, {Latham},
  {Molinari}, {Mortier}, {Mullally}, {Pepe}, {Rice}, {Sinukoff}, {Sozzetti},
  {Thompson}, {Udry}, {Vogt}, {Barman}, {Batalha}, {Bouchy}, {Buchhave},
  {Butler}, {Cosentino}, {Dupuy}, {Ehrenreich}, {Fiorenzano}, {Hansen},
  {Henning}, {Hirsch}, {Holden}, {Isaacson}, {Johnson}, {Knutson}, {Kosiarek},
  {L{\'o}pez-Morales}, {Lovis}, {Malavolta}, {Mayor}, {Micela}, {Motalebi},
  {Petigura}, {Phillips}, {Piotto}, {Rogers}, {Sasselov}, {Schlieder},
  {S{\'e}gransan}, {Watson}, \& {Weiss}}]{christiansen17}
{Christiansen}, J.~L., {Vanderburg}, A., {Burt}, J., {et~al.} 2017, \aj, 154,
  122

\bibitem[{{Cochran} {et~al.}(2011){Cochran}, {Fabrycky}, {Torres}, {Fressin},
  {D{\'e}sert}, {Ragozzine}, {Sasselov}, {Fortney}, {Rowe}, {Brugamyer},
  {Bryson}, {Carter}, {Ciardi}, {Howell}, {Steffen}, {Borucki}, {Koch}, {Winn},
  {Welsh}, {Uddin}, {Tenenbaum}, {Still}, {Seager}, {Quinn}, {Mullally},
  {Miller}, {Marcy}, {MacQueen}, {Lucas}, {Lissauer}, {Latham}, {Knutson},
  {Kinemuchi}, {Johnson}, {Jenkins}, {Isaacson}, {Howard}, {Horch}, {Holman},
  {Henze}, {Haas}, {Gilliland}, {Gautier}, {Ford}, {Fischer}, {Everett},
  {Endl}, {Demory}, {Deming}, {Charbonneau}, {Caldwell}, {Buchhave}, {Brown},
  \& {Batalha}}]{cochran11}
{Cochran}, W.~D., {Fabrycky}, D.~C., {Torres}, G., {et~al.} 2011, \apjs, 197, 7

\bibitem[{{Collier Cameron} {et~al.}(2009){Collier Cameron}, {Davidson},
  {Hebb}, {Skinner}, {Anderson}, {Christian}, {Clarkson}, {Enoch}, {Irwin},
  {Joshi}, {Haswell}, {Hellier}, {Horne}, {Kane}, {Lister}, {Maxted}, {Norton},
  {Parley}, {Pollacco}, {Ryans}, {Scholz}, {Skillen}, {Smalley}, {Street},
  {West}, {Wilson}, \& {Wheatley}}]{collierCameron09}
{Collier Cameron}, A., {Davidson}, V.~A., {Hebb}, L., {et~al.} 2009, \mnras,
  400, 451

\bibitem[{{Cubillos} {et~al.}(2017){Cubillos}, {Harrington}, {Loredo}, {Lust},
  {Blecic}, \& {Stemm}}]{cubillos17}
{Cubillos}, P., {Harrington}, J., {Loredo}, T.~J., {et~al.} 2017, \aj, 153, 3

\bibitem[{{Davis} \& {Wheatley}(2009)}]{davis09}
{Davis}, T.~A. \& {Wheatley}, P.~J. 2009, \mnras, 396, 1012

\bibitem[{{Delrez} {et~al.}(2021){Delrez}, {Ehrenreich}, {Alibert}, {Bonfanti},
  {Borsato}, {Fossati}, {Hooton}, {Hoyer}, {Pozuelos}, {Salmon}, {Sulis},
  {Wilson}, {Adibekyan}, {Bourrier}, {Brandeker}, {Charnoz}, {Deline},
  {Guterman}, {Haldemann}, {Hara}, {Oshagh}, {Sousa}, {Van Grootel}, {Alonso},
  {Anglada-Escud{\'e}}, {B{\'a}rczy}, {Barrado}, {Barros}, {Baumjohann},
  {Beck}, {Bekkelien}, {Benz}, {Billot}, {Bonfils}, {Broeg}, {Cabrera},
  {Collier Cameron}, {Davies}, {Deleuil}, {Delisle}, {Demangeon}, {Demory},
  {Erikson}, {Fortier}, {Fridlund}, {Futyan}, {Gandolfi}, {Garcia Mu{\~n}oz},
  {Gillon}, {Guedel}, {Heng}, {Kiss}, {Laskar}, {Lecavelier des Etangs},
  {Lendl}, {Lovis}, {Maxted}, {Nascimbeni}, {Olofsson}, {Osborn}, {Pagano},
  {Pall{\'e}}, {Piotto}, {Pollacco}, {Queloz}, {Rauer}, {Ragazzoni}, {Ribas},
  {Santos}, {Scandariato}, {S{\'e}gransan}, {Simon}, {Smith}, {Steller},
  {Szab{\'o}}, {Thomas}, {Udry}, \& {Walton}}]{delrez2021}
{Delrez}, L., {Ehrenreich}, D., {Alibert}, Y., {et~al.} 2021, Nature Astronomy,
  5, 775

\bibitem[{{Denissenkov} {et~al.}(2010){Denissenkov}, {Pinsonneault},
  {Terndrup}, \& {Newsham}}]{denissenkov10}
{Denissenkov}, P.~A., {Pinsonneault}, M., {Terndrup}, D.~M., \& {Newsham}, G.
  2010, \apj, 716, 1269

\bibitem[{{Dorn} {et~al.}(2017){Dorn}, {Venturini}, {Khan}, {Heng}, {Alibert},
  {Helled}, {Rivoldini}, \& {Benz}}]{dorn2017}
{Dorn}, C., {Venturini}, J., {Khan}, A., {et~al.} 2017, \aap, 597, A37

\bibitem[{{Erkaev} {et~al.}(2007){Erkaev}, {Kulikov}, {Lammer}, {Selsis},
  {Langmayr}, {Jaritz}, \& {Biernat}}]{erkaev07}
{Erkaev}, N.~V., {Kulikov}, Y.~N., {Lammer}, H., {et~al.} 2007, \aap, 472, 329

\bibitem[{Foreman-Mackey(2016)}]{foremanMackey16corner}
Foreman-Mackey, D. 2016, The Journal of Open Source Software, 1, 24

\bibitem[{{Fossati} {et~al.}(2017){Fossati}, {Erkaev}, {Lammer}, {Cubillos},
  {Odert}, {Juvan}, {Kislyakova}, {Lendl}, {Kubyshkina}, \&
  {Bauer}}]{fossati17}
{Fossati}, L., {Erkaev}, N.~V., {Lammer}, H., {et~al.} 2017, \aap, 598, A90

\bibitem[{{Fressin} {et~al.}(2012){Fressin}, {Torres}, {Rowe}, {Charbonneau},
  {Rogers}, {Ballard}, {Batalha}, {Borucki}, {Bryson}, {Buchhave}, {Ciardi},
  {D{\'e}sert}, {Dressing}, {Fabrycky}, {Ford}, {Gautier}, {Henze}, {Holman},
  {Howard}, {Howell}, {Jenkins}, {Koch}, {Latham}, {Lissauer}, {Marcy},
  {Quinn}, {Ragozzine}, {Sasselov}, {Seager}, {Barclay}, {Mullally}, {Seader},
  {Still}, {Twicken}, {Thompson}, \& {Uddin}}]{fressin12}
{Fressin}, F., {Torres}, G., {Rowe}, J.~F., {et~al.} 2012, \nat, 482, 195

\bibitem[{{Fulton} \& {Petigura}(2018)}]{fulton18}
{Fulton}, B.~J. \& {Petigura}, E.~A. 2018, \aj, 156, 264

\bibitem[{{Fulton} {et~al.}(2017){Fulton}, {Petigura}, {Howard}, {Isaacson},
  {Marcy}, {Cargile}, {Hebb}, {Weiss}, {Johnson}, {Morton}, {Sinukoff},
  {Crossfield}, \& {Hirsch}}]{fulton17}
{Fulton}, B.~J., {Petigura}, E.~A., {Howard}, A.~W., {et~al.} 2017, \aj, 154,
  109

\bibitem[{{Ginzburg} {et~al.}(2016){Ginzburg}, {Schlichting}, \&
  {Sari}}]{ginzburg2016}
{Ginzburg}, S., {Schlichting}, H.~E., \& {Sari}, R. 2016, \apj, 825, 29

\bibitem[{{Ginzburg} {et~al.}(2018){Ginzburg}, {Schlichting}, \&
  {Sari}}]{ginzburg2018}
{Ginzburg}, S., {Schlichting}, H.~E., \& {Sari}, R. 2018, \mnras, 476, 759

\bibitem[{{Gorti} {et~al.}(2016){Gorti}, {Liseau}, {S{\'a}ndor}, \&
  {Clarke}}]{gorti16}
{Gorti}, U., {Liseau}, R., {S{\'a}ndor}, Z., \& {Clarke}, C. 2016, \ssr, 205,
  125

\bibitem[{{Gupta} \& {Schlichting}(2019)}]{gupta2019}
{Gupta}, A. \& {Schlichting}, H.~E. 2019, \mnras, 487, 24

\bibitem[{{Gupta} \& {Schlichting}(2020)}]{gupta2020}
{Gupta}, A. \& {Schlichting}, H.~E. 2020, \mnras, 493, 792

\bibitem[{{Hadden} \& {Lithwick}(2014)}]{hadden14}
{Hadden}, S. \& {Lithwick}, Y. 2014, \apj, 787, 80

\bibitem[{{Hadden} \& {Lithwick}(2017)}]{haddenLithwick17}
{Hadden}, S. \& {Lithwick}, Y. 2017, \aj, 154, 5

\bibitem[{{Hunter}(2007)}]{Hunter2007ieeeMatplotlib}
{Hunter}, J.~D. 2007, Computing In Science \& Engineering, 9, 90

\bibitem[{{Jeans}(1925)}]{jeans25}
{Jeans}, J.~H. 1925, The dynamical theory of gases (Cambridge: At the
  University Press)

\bibitem[{{Jin} \& {Mordasini}(2018)}]{jin18}
{Jin}, S. \& {Mordasini}, C. 2018, \apj, 853, 163

\bibitem[{{Jin} {et~al.}(2014){Jin}, {Mordasini}, {Parmentier}, {van Boekel},
  {Henning}, \& {Ji}}]{jin14}
{Jin}, S., {Mordasini}, C., {Parmentier}, V., {et~al.} 2014, \apj, 795, 65

\bibitem[{{Johnstone} {et~al.}(2021){Johnstone}, {Bartel}, \&
  {G{\"u}del}}]{johnstone21}
{Johnstone}, C.~P., {Bartel}, M., \& {G{\"u}del}, M. 2021, \aap, 649, A96

\bibitem[{{Johnstone} {et~al.}(2015{\natexlab{a}}){Johnstone}, {G{\"u}del},
  {Brott}, \& {L{\"u}ftinger}}]{johnstone15Prot150}
{Johnstone}, C.~P., {G{\"u}del}, M., {Brott}, I., \& {L{\"u}ftinger}, T.
  2015{\natexlab{a}}, \aap, 577, A28

\bibitem[{{Johnstone} {et~al.}(2015{\natexlab{b}}){Johnstone}, {G{\"u}del},
  {St{\"o}kl}, {Lammer}, {Tu}, {Kislyakova}, {L{\"u}ftinger}, {Odert},
  {Erkaev}, \& {Dorfi}}]{johnstone15models}
{Johnstone}, C.~P., {G{\"u}del}, M., {St{\"o}kl}, A., {et~al.}
  2015{\natexlab{b}}, \apjl, 815, L12

\bibitem[{Jones {et~al.}(2001)Jones, Oliphant, Peterson,
  {et~al.}}]{JonesEtal2001scipy}
Jones, E., Oliphant, T., Peterson, P., {et~al.} 2001, {SciPy}: Open source
  scientific tools for {Python}

\bibitem[{{Kennedy} \& {Kenyon}(2008{\natexlab{a}})}]{kennedy2008b}
{Kennedy}, G.~M. \& {Kenyon}, S.~J. 2008{\natexlab{a}}, \apj, 682, 1264

\bibitem[{{Kennedy} \& {Kenyon}(2008{\natexlab{b}})}]{kennedy2008a}
{Kennedy}, G.~M. \& {Kenyon}, S.~J. 2008{\natexlab{b}}, \apj, 673, 502

\bibitem[{{Kennedy} \& {Kenyon}(2009)}]{kennedy2009}
{Kennedy}, G.~M. \& {Kenyon}, S.~J. 2009, \apj, 695, 1210

\bibitem[{{Kimura} {et~al.}(2016){Kimura}, {Kunitomo}, \&
  {Takahashi}}]{kimura16}
{Kimura}, S.~S., {Kunitomo}, M., \& {Takahashi}, S.~Z. 2016, \mnras, 461, 2257

\bibitem[{{Kov{\'a}cs}(2015)}]{kovacs15}
{Kov{\'a}cs}, G. 2015, \aap, 581, A2

\bibitem[{{Krenn} {et~al.}(2021){Krenn}, {Fossati}, {Kubyshkina}, \&
  {Lammer}}]{krenn2021}
{Krenn}, A.~F., {Fossati}, L., {Kubyshkina}, D., \& {Lammer}, H. 2021, \aap,
  650, A94

\bibitem[{{Kubyshkina} {et~al.}(2019{\natexlab{a}}){Kubyshkina}, {Cubillos},
  {Fossati}, {Erkaev}, {Johnstone}, {Kislyakova}, {Lammer}, {Lendl}, {Odert},
  \& {G{\"u}del}}]{kubyshkina19ApJ}
{Kubyshkina}, D., {Cubillos}, P.~E., {Fossati}, L., {et~al.}
  2019{\natexlab{a}}, \apj, 879, 26

\bibitem[{{Kubyshkina} {et~al.}(2018{\natexlab{a}}){Kubyshkina}, {Fossati},
  {Erkaev}, {Cubillos}, {Johnstone}, {Kislyakova}, {Lammer}, {Lendl}, \&
  {Odert}}]{kubyshkina18ApJHBA}
{Kubyshkina}, D., {Fossati}, L., {Erkaev}, N.~V., {et~al.} 2018{\natexlab{a}},
  \apjl, 866, L18

\bibitem[{{Kubyshkina} {et~al.}(2018{\natexlab{b}}){Kubyshkina}, {Fossati},
  {Erkaev}, {Johnstone}, {Cubillos}, {Kislyakova}, {Lammer}, {Lendl}, \&
  {Odert}}]{kubyshkina18AAgrids}
{Kubyshkina}, D., {Fossati}, L., {Erkaev}, N.~V., {et~al.} 2018{\natexlab{b}},
  \aap, 619, A151

\bibitem[{{Kubyshkina} {et~al.}(2019{\natexlab{b}}){Kubyshkina}, {Fossati},
  {Mustill}, {Cubillos}, {Davies}, {Erkaev}, {Johnstone}, {Kislyakova},
  {Lammer}, {Lendl}, \& {Odert}}]{kubyshkina19AAKepler11}
{Kubyshkina}, D., {Fossati}, L., {Mustill}, A.~J., {et~al.} 2019{\natexlab{b}},
  \aap, 632, A65

\bibitem[{{Lammer} {et~al.}(2003){Lammer}, {Selsis}, {Ribas}, {Guinan},
  {Bauer}, \& {Weiss}}]{lammer03}
{Lammer}, H., {Selsis}, F., {Ribas}, I., {et~al.} 2003, \apjl, 598, L121

\bibitem[{{Lissauer} {et~al.}(2011){Lissauer}, {Fabrycky}, {Ford}, {Borucki},
  {Fressin}, {Marcy}, {Orosz}, {Rowe}, {Torres}, {Welsh}, {Batalha}, {Bryson},
  {Buchhave}, {Caldwell}, {Carter}, {Charbonneau}, {Christiansen}, {Cochran},
  {Desert}, {Dunham}, {Fanelli}, {Fortney}, {Gautier}, {Geary}, {Gilliland},
  {Haas}, {Hall}, {Holman}, {Koch}, {Latham}, {Lopez}, {McCauliff}, {Miller},
  {Morehead}, {Quintana}, {Ragozzine}, {Sasselov}, {Short}, \&
  {Steffen}}]{lissauer11}
{Lissauer}, J.~J., {Fabrycky}, D.~C., {Ford}, E.~B., {et~al.} 2011, \nat, 470,
  53

\bibitem[{{Lissauer} {et~al.}(2013){Lissauer}, {Jontof-Hutter}, {Rowe},
  {Fabrycky}, {Lopez}, {Agol}, {Marcy}, {Deck}, {Fischer}, {Fortney}, {Howell},
  {Isaacson}, {Jenkins}, {Kolbl}, {Sasselov}, {Short}, \& {Welsh}}]{lissauer13}
{Lissauer}, J.~J., {Jontof-Hutter}, D., {Rowe}, J.~F., {et~al.} 2013, \apj,
  770, 131

\bibitem[{{Lopez} \& {Fortney}(2014)}]{lopez14}
{Lopez}, E.~D. \& {Fortney}, J.~J. 2014, \apj, 792, 1

\bibitem[{{Loyd} {et~al.}(2020){Loyd}, {Shkolnik}, {Schneider},
  {Richey-Yowell}, {Barman}, {Peacock}, \& {Pagano}}]{loyd20}
{Loyd}, R.~O.~P., {Shkolnik}, E.~L., {Schneider}, A.~C., {et~al.} 2020, \apj,
  890, 23

\bibitem[{{Lozovsky} {et~al.}(2021){Lozovsky}, {Helled}, {Pascucci}, {Dorn},
  {Venturini}, \& {Feldmann}}]{lozovsky2021}
{Lozovsky}, M., {Helled}, R., {Pascucci}, I., {et~al.} 2021, \aap, 652, A110

\bibitem[{{MacDonald}(2019)}]{macdonald2019}
{MacDonald}, M.~G. 2019, \mnras, 487, 5062

\bibitem[{{Mamajek} \& {Hillenbrand}(2008)}]{mamajek08}
{Mamajek}, E.~E. \& {Hillenbrand}, L.~A. 2008, \apj, 687, 1264

\bibitem[{{Marcy} {et~al.}(2014){Marcy}, {Isaacson}, {Howard}, {Rowe},
  {Jenkins}, {Bryson}, {Latham}, {Howell}, {Gautier}, {Batalha}, {Rogers},
  {Ciardi}, {Fischer}, {Gilliland}, {Kjeldsen}, {Christensen-Dalsgaard},
  {Huber}, {Chaplin}, {Basu}, {Buchhave}, {Quinn}, {Borucki}, {Koch}, {Hunter},
  {Caldwell}, {Van Cleve}, {Kolbl}, {Weiss}, {Petigura}, {Seager}, {Morton},
  {Johnson}, {Ballard}, {Burke}, {Cochran}, {Endl}, {MacQueen}, {Everett},
  {Lissauer}, {Ford}, {Torres}, {Fressin}, {Brown}, {Steffen}, {Charbonneau},
  {Basri}, {Sasselov}, {Winn}, {Sanchis-Ojeda}, {Christiansen}, {Adams},
  {Henze}, {Dupree}, {Fabrycky}, {Fortney}, {Tarter}, {Holman}, {Tenenbaum},
  {Shporer}, {Lucas}, {Welsh}, {Orosz}, {Bedding}, {Campante}, {Davies},
  {Elsworth}, {Handberg}, {Hekker}, {Karoff}, {Kawaler}, {Lund}, {Lundkvist},
  {Metcalfe}, {Miglio}, {Silva Aguirre}, {Stello}, {White}, {Boss}, {Devore},
  {Gould}, {Prsa}, {Agol}, {Barclay}, {Coughlin}, {Brugamyer}, {Mullally},
  {Quintana}, {Still}, {Thompson}, {Morrison}, {Twicken}, {D{\'e}sert},
  {Carter}, {Crepp}, {H{\'e}brard}, {Santerne}, {Moutou}, {Sobeck}, {Hudgins},
  {Haas}, {Robertson}, {Lillo-Box}, \& {Barrado}}]{marcy14}
{Marcy}, G.~W., {Isaacson}, H., {Howard}, A.~W., {et~al.} 2014, \apjs, 210, 20

\bibitem[{{Masuda} {et~al.}(2013){Masuda}, {Hirano}, {Taruya}, {Nagasawa}, \&
  {Suto}}]{masuda13}
{Masuda}, K., {Hirano}, T., {Taruya}, A., {Nagasawa}, M., \& {Suto}, Y. 2013,
  \apj, 778, 185

\bibitem[{{Maxted} {et~al.}(2015){Maxted}, {Serenelli}, \&
  {Southworth}}]{maxted15}
{Maxted}, P.~F.~L., {Serenelli}, A.~M., \& {Southworth}, J. 2015, \aap, 577,
  A90

\bibitem[{{Mazeh} {et~al.}(2016){Mazeh}, {Holczer}, \& {Faigler}}]{mazeh16}
{Mazeh}, T., {Holczer}, T., \& {Faigler}, S. 2016, \aap, 589, A75

\bibitem[{{McDonald} {et~al.}(2019){McDonald}, {Kreidberg}, \&
  {Lopez}}]{mcdonald19}
{McDonald}, G.~D., {Kreidberg}, L., \& {Lopez}, E. 2019, \apj, 876, 22

\bibitem[{{Meibom} {et~al.}(2009){Meibom}, {Mathieu}, \& {Stassun}}]{meibom09}
{Meibom}, S., {Mathieu}, R.~D., \& {Stassun}, K.~G. 2009, \apj, 695, 679

\bibitem[{{Micela} {et~al.}(1985){Micela}, {Sciortino}, {Serio}, {Vaiana},
  {Bookbinder}, {Golub}, {Harnden}, \& {Rosner}}]{micela85}
{Micela}, G., {Sciortino}, S., {Serio}, S., {et~al.} 1985, \apj, 292, 172

\bibitem[{{Mills} {et~al.}(2019){Mills}, {Howard}, {Weiss}, {Steffen},
  {Isaacson}, {Fulton}, {Petigura}, {Kosiarek}, {Hirsch}, \&
  {Boisvert}}]{mills19}
{Mills}, S.~M., {Howard}, A.~W., {Weiss}, L.~M., {et~al.} 2019, \aj, 157, 145

\bibitem[{Montmerle {et~al.}(2010)Montmerle, Ehrenreich, Lagrange, \&
  Matsuyama}]{montmerle10}
Montmerle, T., Ehrenreich, D., Lagrange, A.-M., \& Matsuyama, I. 2010, in EAS
  Publications Series, Vol.~41, Physics and Astrophysics of Planetary Systems
  (EDP Sciences), 171–175

\bibitem[{{Noyes} {et~al.}(1984){Noyes}, {Hartmann}, {Baliunas}, {Duncan}, \&
  {Vaughan}}]{noyes84}
{Noyes}, R.~W., {Hartmann}, L.~W., {Baliunas}, S.~L., {Duncan}, D.~K., \&
  {Vaughan}, A.~H. 1984, \apj, 279, 763

\bibitem[{{Otegi} {et~al.}(2020){Otegi}, {Bouchy}, \& {Helled}}]{otegi20}
{Otegi}, J.~F., {Bouchy}, F., \& {Helled}, R. 2020, \aap, 634, A43

\bibitem[{{Owen} \& {Lai}(2018)}]{owen18}
{Owen}, J.~E. \& {Lai}, D. 2018, \mnras, 479, 5012

\bibitem[{{Owen} {et~al.}(2020){Owen}, {Shaikhislamov}, {Lammer}, {Fossati}, \&
  {Khodachenko}}]{owen2020}
{Owen}, J.~E., {Shaikhislamov}, I.~F., {Lammer}, H., {Fossati}, L., \&
  {Khodachenko}, M.~L. 2020, \ssr, 216, 129

\bibitem[{{Owen} \& {Wu}(2017)}]{owen17}
{Owen}, J.~E. \& {Wu}, Y. 2017, \apj, 847, 29

\bibitem[{{Pallavicini} {et~al.}(1981){Pallavicini}, {Golub}, {Rosner},
  {Vaiana}, {Ayres}, \& {Linsky}}]{pallavicini81}
{Pallavicini}, R., {Golub}, L., {Rosner}, R., {et~al.} 1981, \apj, 248, 279

\bibitem[{{Palle} {et~al.}(2019){Palle}, {Nowak}, {Luque}, {Hidalgo},
  {Barrag{\'a}n}, {Prieto-Arranz}, {Hirano}, {Fridlund}, {Gandolfi},
  {Livingston}, {Dai}, {Morales}, {Lafarga}, {Albrecht}, {Alonso}, {Amado},
  {Caballero}, {Cabrera}, {Cochran}, {Csizmadia}, {Deeg}, {Eigm{\"u}ller},
  {Endl}, {Erikson}, {Fukui}, {Guenther}, {Grziwa}, {Hatzes}, {Korth},
  {K{\"u}rster}, {Kuzuhara}, {Monta{\~n}es Rodr{\'\i}guez}, {Murgas}, {Narita},
  {Nespral}, {P{\"a}tzold}, {Persson}, {Quirrenbach}, {Rauer}, {Redfield},
  {Reiners}, {Ribas}, {Smith}, {Van Eylen}, {Winn}, \& {Zechmeister}}]{palle19}
{Palle}, E., {Nowak}, G., {Luque}, R., {et~al.} 2019, \aap, 623, A41

\bibitem[{{Petigura} {et~al.}(2018){Petigura}, {Benneke}, {Batygin}, {Fulton},
  {Werner}, {Krick}, {Gorjian}, {Sinukoff}, {Deck}, {Mills}, \&
  {Deming}}]{petigura18}
{Petigura}, E.~A., {Benneke}, B., {Batygin}, K., {et~al.} 2018, \aj, 156, 89

\bibitem[{{Petigura} {et~al.}(2016){Petigura}, {Howard}, {Lopez}, {Deck},
  {Fulton}, {Crossfield}, {Ciardi}, {Chiang}, {Lee}, {Isaacson}, {Beichman},
  {Hansen}, {Schlieder}, \& {Sinukoff}}]{petigura16}
{Petigura}, E.~A., {Howard}, A.~W., {Lopez}, E.~D., {et~al.} 2016, \apj, 818,
  36

\bibitem[{{Pizzolato} {et~al.}(2003){Pizzolato}, {Maggio}, {Micela},
  {Sciortino}, \& {Ventura}}]{pizzolato2003}
{Pizzolato}, N., {Maggio}, A., {Micela}, G., {Sciortino}, S., \& {Ventura}, P.
  2003, \aap, 397, 147

\bibitem[{{Rauer} {et~al.}(2014){Rauer}, {Catala}, {Aerts}, {Appourchaux},
  {Benz}, {Brandeker}, {Christensen-Dalsgaard}, {Deleuil}, {Gizon}, {Goupil},
  {G{\"u}del}, {Janot-Pacheco}, {Mas-Hesse}, {Pagano}, {Piotto}, {Pollacco},
  {Santos}, {Smith}, {Su{\'a}rez}, {Szab{\'o}}, {Udry}, {Adibekyan}, {Alibert},
  {Almenara}, {Amaro-Seoane}, {Eiff}, {Asplund}, {Antonello}, {Barnes},
  {Baudin}, {Belkacem}, {Bergemann}, {Bihain}, {Birch}, {Bonfils}, {Boisse},
  {Bonomo}, {Borsa}, {Brand{\~a}o}, {Brocato}, {Brun}, {Burleigh}, {Burston},
  {Cabrera}, {Cassisi}, {Chaplin}, {Charpinet}, {Chiappini}, {Church},
  {Csizmadia}, {Cunha}, {Damasso}, {Davies}, {Deeg}, {D{\'{\i}}az}, {Dreizler},
  {Dreyer}, {Eggenberger}, {Ehrenreich}, {Eigm{\"u}ller}, {Erikson}, {Farmer},
  {Feltzing}, {de Oliveira Fialho}, {Figueira}, {Forveille}, {Fridlund},
  {Garc{\'{\i}}a}, {Giommi}, {Giuffrida}, {Godolt}, {Gomes da Silva},
  {Granzer}, {Grenfell}, {Grotsch-Noels}, {G{\"u}nther}, {Haswell}, {Hatzes},
  {H{\'e}brard}, {Hekker}, {Helled}, {Heng}, {Jenkins}, {Johansen},
  {Khodachenko}, {Kislyakova}, {Kley}, {Kolb}, {Krivova}, {Kupka}, {Lammer},
  {Lanza}, {Lebreton}, {Magrin}, {Marcos-Arenal}, {Marrese}, {Marques},
  {Martins}, {Mathis}, {Mathur}, {Messina}, {Miglio}, {Montalban}, {Montalto},
  {Monteiro}, {Moradi}, {Moravveji}, {Mordasini}, {Morel}, {Mortier},
  {Nascimbeni}, {Nelson}, {Nielsen}, {Noack}, {Norton}, {Ofir}, {Oshagh},
  {Ouazzani}, {P{\'a}pics}, {Parro}, {Petit}, {Plez}, {Poretti}, {Quirrenbach},
  {Ragazzoni}, {Raimondo}, {Rainer}, {Reese}, {Redmer}, {Reffert},
  {Rojas-Ayala}, {Roxburgh}, {Salmon}, {Santerne}, {Schneider}, {Schou},
  {Schuh}, {Schunker}, {Silva-Valio}, {Silvotti}, {Skillen}, {Snellen}, {Sohl},
  {Sousa}, {Sozzetti}, {Stello}, {Strassmeier}, {{\v S}vanda}, {Szab{\'o}},
  {Tkachenko}, {Valencia}, {Van Grootel}, {Vauclair}, {Ventura}, {Wagner},
  {Walton}, {Weingrill}, {Werner}, {Wheatley}, \& {Zwintz}}]{rauer14}
{Rauer}, H., {Catala}, C., {Aerts}, C., {et~al.} 2014, Experimental Astronomy,
  38, 249

\bibitem[{{Ricker} {et~al.}(2015){Ricker}, {Winn}, {Vanderspek}, {Latham},
  {Bakos}, {Bean}, {Berta-Thompson}, {Brown}, {Buchhave}, {Butler}, {Butler},
  {Chaplin}, {Charbonneau}, {Christensen-Dalsgaard}, {Clampin}, {Deming},
  {Doty}, {De Lee}, {Dressing}, {Dunham}, {Endl}, {Fressin}, {Ge}, {Henning},
  {Holman}, {Howard}, {Ida}, {Jenkins}, {Jernigan}, {Johnson}, {Kaltenegger},
  {Kawai}, {Kjeldsen}, {Laughlin}, {Levine}, {Lin}, {Lissauer}, {MacQueen},
  {Marcy}, {McCullough}, {Morton}, {Narita}, {Paegert}, {Palle}, {Pepe},
  {Pepper}, {Quirrenbach}, {Rinehart}, {Sasselov}, {Sato}, {Seager},
  {Sozzetti}, {Stassun}, {Sullivan}, {Szentgyorgyi}, {Torres}, {Udry}, \&
  {Villasenor}}]{ricker2015}
{Ricker}, G.~R., {Winn}, J.~N., {Vanderspek}, R., {et~al.} 2015, Journal of
  Astronomical Telescopes, Instruments, and Systems, 1, 014003

\bibitem[{{Rogers} {et~al.}(2011){Rogers}, {Bodenheimer}, {Lissauer}, \&
  {Seager}}]{rogers2011}
{Rogers}, L.~A., {Bodenheimer}, P., {Lissauer}, J.~J., \& {Seager}, S. 2011,
  \apj, 738, 59

\bibitem[{{Sandoval} {et~al.}(2021){Sandoval}, {Contardo}, \&
  {David}}]{sandoval21}
{Sandoval}, A., {Contardo}, G., \& {David}, T.~J. 2021, \apj, 911, 117

\bibitem[{{Sanz-Forcada} {et~al.}(2011){Sanz-Forcada}, {Micela}, {Ribas},
  {Pollock}, {Eiroa}, {Velasco}, {Solano}, \&
  {Garc{\'\i}a-{\'A}lvarez}}]{sanzForcada11}
{Sanz-Forcada}, J., {Micela}, G., {Ribas}, I., {et~al.} 2011, \aap, 532, A6

\bibitem[{{Skumanich}(1972)}]{skumanich72}
{Skumanich}, A. 1972, \apj, 171, 565

\bibitem[{{St{\"o}kl} {et~al.}(2015){St{\"o}kl}, {Dorfi}, \&
  {Lammer}}]{stokl2015}
{St{\"o}kl}, A., {Dorfi}, E., \& {Lammer}, H. 2015, \aap, 576, A87

\bibitem[{{Sun} {et~al.}(2019){Sun}, {Ioannidis}, {Gu}, {Schmitt}, {Wang}, \&
  {Kouwenhoven}}]{sun19}
{Sun}, L., {Ioannidis}, P., {Gu}, S., {et~al.} 2019, \aap, 624, A15

\bibitem[{{Szab{\'o}} \& {Kiss}(2011)}]{szabo11}
{Szab{\'o}}, G.~M. \& {Kiss}, L.~L. 2011, \apjl, 727, L44

\bibitem[{{Tu} {et~al.}(2015){Tu}, {Johnstone}, {G{\"u}del}, \&
  {Lammer}}]{tu15}
{Tu}, L., {Johnstone}, C.~P., {G{\"u}del}, M., \& {Lammer}, H. 2015, \aap, 577,
  L3

\bibitem[{{van der Walt} {et~al.}(2011){van der Walt}, {Colbert}, \&
  {Varoquaux}}]{vanderWaltEtal2011numpy}
{van der Walt}, S., {Colbert}, S.~C., \& {Varoquaux}, G. 2011, Computing in
  Science and Engineering, 13, 22

\bibitem[{{Van Eylen} {et~al.}(2018){Van Eylen}, {Agentoft}, {Lundkvist},
  {Kjeldsen}, {Owen}, {Fulton}, {Petigura}, \& {Snellen}}]{vanEylen18}
{Van Eylen}, V., {Agentoft}, C., {Lundkvist}, M.~S., {et~al.} 2018, \mnras,
  479, 4786

\bibitem[{{van Saders} {et~al.}(2016){van Saders}, {Ceillier}, {Metcalfe},
  {Silva Aguirre}, {Pinsonneault}, {Garc{\'\i}a}, {Mathur}, \&
  {Davies}}]{vanSaders16}
{van Saders}, J.~L., {Ceillier}, T., {Metcalfe}, T.~S., {et~al.} 2016, \nat,
  529, 181

\bibitem[{{Vanderburg} {et~al.}(2017){Vanderburg}, {Becker}, {Buchhave},
  {Mortier}, {Lopez}, {Malavolta}, {Haywood}, {Latham}, {Charbonneau},
  {L{\'o}pez-Morales}, {Adams}, {Bonomo}, {Bouchy}, {Collier Cameron},
  {Cosentino}, {Di Fabrizio}, {Dumusque}, {Fiorenzano}, {Harutyunyan},
  {Johnson}, {Lorenzi}, {Lovis}, {Mayor}, {Micela}, {Molinari}, {Pedani},
  {Pepe}, {Piotto}, {Phillips}, {Rice}, {Sasselov}, {S{\'e}gransan},
  {Sozzetti}, {Udry}, \& {Watson}}]{vanderburg17}
{Vanderburg}, A., {Becker}, J.~C., {Buchhave}, L.~A., {et~al.} 2017, \aj, 154,
  237

\bibitem[{{Vissapragada} {et~al.}(2020){Vissapragada}, {Jontof-Hutter},
  {Shporer}, {Knutson}, {Liu}, {Thorngren}, {Lee}, {Chachan}, {Mawet},
  {Millar-Blanchaer}, {Nilsson}, {Tinyanont}, {Vasisht}, \&
  {Wright}}]{vissapragada20}
{Vissapragada}, S., {Jontof-Hutter}, D., {Shporer}, A., {et~al.} 2020, \aj,
  159, 108

\bibitem[{{Watson} {et~al.}(1981){Watson}, {Donahue}, \& {Walker}}]{watson81}
{Watson}, A.~J., {Donahue}, T.~M., \& {Walker}, J.~C.~G. 1981, \icarus, 48, 150

\bibitem[{{Weiss} {et~al.}(2013){Weiss}, {Marcy}, {Rowe}, {Howard}, {Isaacson},
  {Fortney}, {Miller}, {Demory}, {Fischer}, {Adams}, {Dupree}, {Howell},
  {Kolbl}, {Johnson}, {Horch}, {Everett}, {Fabrycky}, \& {Seager}}]{weiss13}
{Weiss}, L.~M., {Marcy}, G.~W., {Rowe}, J.~F., {et~al.} 2013, \apj, 768, 14

\bibitem[{{Wright} {et~al.}(2011){Wright}, {Drake}, {Mamajek}, \&
  {Henry}}]{wright11}
{Wright}, N.~J., {Drake}, J.~J., {Mamajek}, E.~E., \& {Henry}, G.~W. 2011,
  \apj, 743, 48

\end{thebibliography}
